\documentclass[12pt]{article}
\usepackage{amsfonts,amsmath,graphicx,amssymb,geometry,arydshln,textcomp,amsthm,hyperref,framed,color,appendix,lscape}
\usepackage[round]{natbib}
\usepackage[onehalfspacing]{setspace}
\usepackage{ragged2e}
\usepackage{xcolor,xspace,colortbl,rotating}
\usepackage{tabulary}
\usepackage{booktabs}
\usepackage{multirow}
\usepackage{bm}
\usepackage{cleveref}
\usepackage{threeparttable}
\usepackage{threeparttablex}
\usepackage{subfig}
\usepackage{appendix}
\usepackage{etoc}
\usepackage{enumerate,enumitem}
\usepackage{longtable}
\usepackage{rotating}
\usepackage{pdflscape}          
\usepackage{mathtools}
\DeclareGraphicsExtensions{.pdf,.svg,.eps,.ps,.png,.jpg,.jpeg}
\theoremstyle{definition}
\newtheorem {theorem}[]{Theorem}

\newtheorem{assumption}{Assumption}

\newtheorem {lemma}{Lemma}

\newtheorem {remark}[]{Remark}

\crefname{subequation}{subequation}{subequations}
\Crefname{subequation}{Assumption}{assumptions}
\crefname{assumption}{Assumption}{Assumptions}
\hypersetup{colorlinks,linkcolor={blue},citecolor={blue},urlcolor={blue}}

\begin{document}

	\author{Xiao Huang\thanks{The author is grateful for valuable comments from Zheng Fang, Daniel Henderson, Junsoo Lee, Xiaochun Liu, Whitney Newey, Alejandro Sanchez-Becerra, Ruoxuan Xiong, Zhaoguo Zhan, and seminar participants at the University of Alabama and the 3rd Georgia Econometrics Workshop at Emory University.   The author also thanks the Office of Research at the Kennesaw State University for computation support. Correspondence address: Department of Economics, Finance, and Quantitative Analysis, Coles College of Business, Kennesaw State University, GA 30144, USA. Email: xhuang3@kennesaw.edu.}}
	\title{\Large Composite Quantile Factor Model}
	\date{ \today}
	\maketitle
	
	\doublespace
	
	\begin{abstract}
		This paper introduces the method of composite quantile factor model for factor analysis in high-dimensional panel data. We propose to estimate the factors and factor loadings across multiple quantiles of the data, allowing the estimates to better adapt to features of the data at different quantiles while still modeling the mean of the data.  We  develop the limiting distribution of the estimated factors and factor loadings, and an information criterion for consistent factor number selection  is also discussed. Simulations show that the proposed estimator and the information criterion have good finite sample properties for several non-normal distributions under consideration. We also consider an empirical study on the factor analysis for $246$ quarterly macroeconomic variables.  A companion R package \texttt{cqrfactor} is  developed.
	\end{abstract}
	
	\bigskip
	
	
	\bigskip
	
	
	\newpage  
	
	\normalsize
	
	\doublespace
	
	\section{Introduction} \label{intro}
	
	Factor model is a useful statistical tool to describe data with unobserved and systematic components. Detailed textbook discussions on classical factor analysis for data with a fixed or small number of variables can be found in \cite{lawleymaxwell1971factoranalysis,anderson2003introduction}. Following the important work in \cite{stockandwatson1998nber,stockandwatson2002JASA,stockandwatson2002jbes,baiandng2002ecma,bai2003ecma} on high-dimensional panel data, the research on factor analysis in panel data has been extended to many directions, and there now exists a large body of literature on factor analysis for high-dimensional panel data. See \cite{baiandwang2016are} for a review on recent developments in this burgeoning field.
	
	At the heart of factor analysis for high-dimensional panel data is the use of principal component analysis (PCA) method. The estimates for factors are chosen to be the normalized eigenvectors of the sample covariance matrix of the data, based on which we can estimate the factor loadings using the least-squares (LS) method.  These eigenvectors coincide with the solutions  in PCA, and the procedure's simplicity  greatly contributes to its wide popularity as a research tool in empirical macroeconomics and finance.
	
	Despite its simplicity, the PCA-based procedure typically imposes some higher-order moment conditions on the factor and error terms (see, e.g., the assumptions in \cite{stockandwatson2002JASA,bai2003ecma}) in order to obtain desirable asymptotic results. However,  the sample covariance matrix on which PCA operates may be irrelevant for data with infinite (or very large) variances and PCA becomes invalid. Weakening or even removing these conditions will be appealing since many data in applications such as finance are either heavy tailed or of unknown nature. Two approaches for robust factor analysis have emerged in recent literature.  \cite{chenetal2021qfm} introduce the quantile factor models (QFM), where both the factors and factor loadings are quantile-dependent. At the quantile position $ 0.5 $, QFM estimator can be interpreted as the least absolute deviation (LAD) estimator that may be robust to certain error distributions.  \cite{andoandbai2020jasa} provide a more general framework that adds a regression component with heterogeneous coefficients. By assuming the factors and errors follow a joint elliptical distribution, \cite{heetal2022JBESnomoment} propose a second approach to replacing the sample covariance matrix in PCA with the spatial Kendall's tau matrix that can handle non-normal distributions with large or infinite variances, a situation in which the standard PCA fails.
	
	This paper studies another approach to robust factor analysis and we term it composite quantile factor models (CQFM). Our approach is inspired by the interesting work in \cite{zouandyuan2008cqr}. \cite{zouandyuan2008cqr} notice that, in a linear regression with infinite error variance, the parameter estimator will no longer have root-$ n $ consistency or asymptotic normality; a robust procedure such as LAD can be used but its relative efficiency to the LS estimator can be very small. The authors propose to estimate the regression coefficients by simultaneously minimizing the standard quantile regression objective function at multiple quantile positions and call this procedure composite quantile regression (CQR). \cite{zouandyuan2008cqr} demonstrate the good finite sample properties of the CQR estimator for several non-normal error distributions.
	
	Although factor analysis is different from linear regression, they are intrinsically connected (see, e.g., \cite{stockandwatson1998nber} for the use of LS method in deriving the solution to the factor model). If CQR works in linear regression, we conjecture a variant of it  will also work in factor model. This paper studies the extension of CQR  to the estimation of factor model by simultaneously minimizing the objective function at multiple quantiles. The resulting estimates are shown to have good finite sample properties under various non-normal error distributions. Because the CQFM visits different quantiles of data during estimation, the estimated factors can usually pick up more skewness (and kurtosis) information about the data, often resulting a better fit of the model. It is important to point out that, although CQFM uses the method of quantile regression, its estimates  are for the mean factor model. This sets our paper apart from the work in \cite{andoandbai2020jasa,chenetal2021qfm}, where the goal is to estimate parameters at a specific quantile position.
	
	We make the following contributions to the growing literature on panel factor analysis. First, we introduce CQFM as a new method to perform factor analysis on the mean factor model that can capture features of data at different quantiles. Second, we develop the asymptotic distribution results for the estimated factors and factor loadings; an information criterion is also developed to consistently select the factor number. Third, we provide extensive simulation evidence to show that CQFM works well for several non-normal error distributions. A special case of CQFM is when one chooses to optimize the objective function at a single quantile position, say $ 0.5 $. This reduces CQFM to the QFM in \cite{chenetal2021qfm}. In several simulation examples, we demonstrate the advantage of CQFM as a result of using information at multiple quantiles. We also develop an R package \texttt{cqrfactor} that implements the CQFM method in this paper. The \texttt{cqrfactor} package can be downloaded from \url{https://github.com/xhuang20/cqrfactor}.
	
	The rest of the paper is organized as follows. Section 2 sets up the objective function for CQFM and discusses the estimation procedure and the asymptotic results. Section 3 discusses the information criterion for the selection of factor numbers. Section 4 presents all simulation results. Section 5 applies CQFM method to the modeling of the quarterly macroeconomic data in \cite{mccrackenandng2020FREDQD}. Section 6 concludes.  The online supplement contains all proofs, additional figures and tables.
	
	\section{Model estimation and the asymptotic results}
	
	\subsection{The model and the algorithm}
	
	Let $ Y_{it} $ be the observation at time $ t $ for the $ i $th cross-section unit. Consider the following factor model:
	\begin{equation} \label{eq: factor model}						
		Y_{it} = \lambda_{0i}' F_{0t} + \varepsilon_{it}, \text{ for } i = 1, \cdots, N, t = 1,\cdots,T,
	\end{equation}
	where $F_{0t}$ is an $r \times 1$ vector of factors, $\lambda_{0i}$ is an $r \times 1$ vector of factor loading, and $\varepsilon_{it}$ is the error term. Both $F_{0t}$ and $\lambda_{0i}$ are unobserved, and the goal is to estimate them jointly. We assume the number of factors ($r$) is known. An information criterion will be developed to estimate $r$ consistently in \Cref{sec:factor number}. Rewrite \cref{eq: factor model} in matrix form to have
	\begin{equation} \label{eq: factor model matrix form}
		\underset{T \times N}{Y \vphantom{\Lambda_0'}} = \underset{T \times r}{F_0  \vphantom{\Lambda_0'}} \underset{r \times N}{\Lambda_0'} +  \underset{T \times N}{\varepsilon  \vphantom{\Lambda_0'}}.
	\end{equation}
	where $Y_{it}$ and $\varepsilon_{it}$ are elements of $Y$ and $\varepsilon$, respectively, and
	\begin{equation}
		F_0 = \big[F_{01}, \cdots,F_{0t},\cdots,	F_{0T}\big]', \qquad
		\Lambda_0 = \big[\lambda_{01},\cdots,	\lambda_{0i}, 	\cdots,	\lambda_{0N}\big]'.
	\end{equation}
	
	Let $ \tau$ be a quantile position with $0 < \tau < 1$ and $b_{0\tau}$ be the $ 100\tau\% $ quantile of $\varepsilon_{it}$. For the quantile factor model, we seek the estimates for $b_{0\tau}$, $F_{0t}$, and $\lambda_{0i}$ that minimize the following QFM objective function:
	\begin{equation} \label{eq: QFM obj}
		\frac{1}{NT} \sum_{i=1}^{N} \sum_{t=1}^{T} \rho_{\tau}\left(Y_{it} - b_{\tau} - \lambda_i'F_t\right),
	\end{equation}
	where $\rho_{\tau}(u) = u(\tau - \mathbf{I}(u \leq 0))$ is the check function in quantile regression. The estimates from \cref{eq: QFM obj} depend on $\tau$ and can be written as $\hat{F}_t(\tau)$ and $\hat{\lambda}_i(\tau)$, an approach adopted in \cite{andoandbai2020jasa,chenetal2021qfm}.
	
	In CQFM, instead of estimating the model at a single quantile position $\tau$, we estimate the model simultaneously at multiple quantiles by choosing a sequence of $K$ quantiles, $0 < \tau_1<\tau_2<\cdots<\tau_K<1$, and minimizing the following objection function:
	\begin{equation} \label{eq: CQFM obj}
		\frac{1}{NT} \sum_{k=1}^{K}\sum_{i=1}^{N}\sum_{t=1}^{T} \rho_{\tau_k}(Y_{it} -b_{\tau_{k}}- \lambda_i'F_t),
	\end{equation}
	where $b_{\tau_{k}}$ estimates $b_{0\tau_{k}}$, the $100\tau_{k}\%$ quantile of $\varepsilon_{it}$.  Let $\hat{\lambda}_i$ and $\hat{F}_t$ be the estimators for $\lambda_{0i}$ and $F_{0t}$ in \cref{eq: CQFM obj}. Unlike the solutions to \cref{eq: QFM obj}, $\hat{\lambda}_i$ and $\hat{F}_t$ are not dependent on any specific quantile position, and they estimate parameters in the mean factor model. By minimizing the objective function across multiple quantiles, the estimators can adapt to data features at different quantiles while still giving estimates for the mean of the process. We usually select equally spaced quantiles with $\tau_{k} = \frac{k}{K+1}$ for $k = 1,2,\cdots,K$ and $K$ is an odd number such as $5$ or $7$.  This will always include the $50\%$ quantile in estimation, but an even number of quantiles also works for \cref{eq: CQFM obj}. The number $K$ can be viewed as a tuning parameter of CQFM. In the special case of $K=1$, i.e., when a single quantile position is used in \cref{eq: CQFM obj}, CQFM reduces to the quantile factor model. Our R package can estimate both CQFM and QFM.
	
	There is no closed-form solution to the minimization exercise in \cref{eq: CQFM obj}, $\hat{b}_{\tau_{k}}$, $\hat{\lambda}_i$, and $\hat{F}_t$ need to be obtained through an iterative algorithm. Since $\lambda_i$ and $F_t$ appear as a product in \cref{eq: CQFM obj}, they are not separately identifiable. We use the following normalization for identification purposes:
	\begin{equation} \label{eq: normalization}
		\begin{aligned}
			\frac{1}{T} \sum_{t=1}^{T} \hat{F}_t \hat{F}_t' &= I_r, \quad \text{an identity matrix of dimension $r$},\\
			\frac{1}{N} \sum_{i=1}^{N} \hat{\lambda}_i \hat{\lambda}_i' &= \Sigma_{\hat{\lambda}}, \quad \text{a diagonal matrix with decreasing diagonal elements}.
		\end{aligned}
	\end{equation}
	We describe the steps of the algorithm below. Let $s$ denote the iteration step and the ranges of the subscripts $i,t,k$ are the same as those appear in \cref{eq: CQFM obj}.
	
	\begin{enumerate}[nosep]
		\item[] \textit{Step 1}. Choose a random starting value for $F_t^{(0)}$ for all $t$. Use $F_t^{(0)}$ to get the initial estimates for $\lambda_{i}^{(0)}$ and $b_{\tau_{k}}^{(0)}$.
		\item[] \textit{Step 2}. Given $b_{\tau_{k}}^{(s-1)}$ and $\lambda_i^{(s-1)}$ , obtain $F_t^{(s)} $ that minimizes \cref{eq: CQFM obj}. 
		\item[] \textit{Step 3}. Given $b_{\tau_{k}}^{(s-1)}$ and $F_t^{(s)} $, obtain $\lambda_i^{(s)}$ that minimizes \cref{eq: CQFM obj}.
		\item[] \textit{Step 4}. Given $\lambda_i^{(s)}$ and $F_t^{(s)} $, obtain $b_{\tau_{k}}^{(s)}$ that minimizes \cref{eq: CQFM obj}.
		\item[] \textit{Step 5}. Repeat Steps 2 to 4 for $s = 1, 2, \cdots$ until estimates converge. Normalize the final solution according to \cref{eq: normalization}.
	\end{enumerate}
	
	A few remarks follow.
	
	\begin{remark}
		The minimization exercise in \cref{eq: CQFM obj} is non-convex in the parameters. However, it is convex, for example, when we solve  $\lambda_i^{(s)}$ given $b_{\tau_{k}}^{(s-1)}$ and $F_t^{(s-1)}$. Our simulation experience indicates the final solution is not sensitive to the starting values of $F_t^{(0)}$. Our R package \texttt{cqrfactor} allows the user to supply different random seeds to initialize $F_t^{(0)}$, making it easy to check the solution's sensitivity to starting values. This iterative strategy is also used in several other papers such as \cite{bai2009ecma,chenetal2021qfm}.
	\end{remark}
	
	\begin{remark}
		When solving \cref{eq: CQFM obj} in steps 2 to 4,  we use the majorization-minimization (MM) algorithm for quantile regression described in \cite{hunterandlange2000JCGSMM}. This is one of the several popular methods for solving a quantile regression problem.
	\end{remark}
	
	\begin{remark}
		In step 5, there is no unique way to define the convergence of the algorithm. Between steps $s$ and $s+1$, one can check the difference in the loss function \cref{eq: CQFM obj} to see if it is small enough; alternatively, one can check the (average) absolute change in parameter estimates between steps $s$ and $s+1$. 
	\end{remark}
	
	\subsection{Asymptotic results of the estimators}
	
	We make the following assumptions to derive the asymptotic results.
	
	\begin{assumption} \label{asump: factor and facor loading}
		The factors $F_{0t}$ are random with $\frac{1}{T}\sum_{t=1}^{T}F_{0t} \rightarrow E(F_{0t}) = 0$ and $\frac{1}{T}\sum_{t=1}^{T}F_{0t} F_{0t}' \rightarrow \Sigma_{F_0} = I_r$ as $T \rightarrow \infty$.  The factor loadings have the limit $\frac{1}{N} \sum_{i=1}^{N} \lambda_{0i}\lambda_{0i}' \rightarrow \Sigma_{\lambda_0} $,  a diagonal matrix with $\sigma_{ii} > \sigma_{jj} > 0$ if $i < j$.
	\end{assumption}
	
	\begin{assumption} \label{asump: error density}
		The distribution of the error term $\varepsilon_{it}$ has an absolutely continuous cumulative function $F_{\varepsilon}$ with a continuous density function $f_{\varepsilon}$ that is uniformly bounded away from $0$ and $\infty$.
	\end{assumption}
	
	\begin{assumption} \label{asump: iid}
		The error terms $\varepsilon_{it}$  are i.i.d. and are independent of the factors $F_{0t}$ across all $i$ and $t$.
	\end{assumption}
	
	\Cref{asump: factor and facor loading} is almost identical to \citet[Assumption~F1]{stockandwatson2002JASA}
	and \citet[Assumption~1(i)]{chenetal2021qfm} and can help identify both $F_0$ and $\Lambda_0$. See \cite{baing2013pcrfactor} for a more detailed discussion on the identification in factor models. \Cref{asump: error density} is a standard one in quantile regression. This assumption is made for the unconditional distribution of $\varepsilon_{it}$. If we consider the conditional distribution of $\varepsilon_{it}$ given $F_{0t}$ in \Cref{asump: error density}, all expectations in the proof will be conditional. The i.i.d. requirement in \Cref{asump: iid} is strong. However, this assumption simplifies the presentation of the asymptotic results and allows us to make direct comparison of our asymptotic results to the cross-section regression result in \cite{zouandyuan2008cqr}; in addition, the simple form of our asymptotic results facilitates the efficiency comparison between the CQFM-based factors and the PCA-based factors in \cite{bai2003ecma} (see a remark following \Cref{thm: asymptotic distribution} for a discussion). We can modify \Cref{asump: iid} so that it is conditional on $F_{0t}$, and the asymptotic covariance in \Cref{thm: asymptotic distribution} will have the standard sandwich form. We discuss this in a remark following \Cref{thm: asymptotic distribution}.  Our simulation section includes results for errors with heteroskedasticity and AR(1) structure, and CQFM continues to give good results especially when the sample size is large. 
	
	The following theorem gives the asymptotic distribution of the estimated factors and factor loadings.
	
	\begin{theorem} \label{thm: asymptotic distribution}
		Under \cref{asump: factor and facor loading,asump: error density,asump: iid},  the asymptotic distribution of $\sqrt{N}(\hat{F}_t - F_{0t})$ is $N(0, \Sigma_{\text{CQFM,}F})$ with
		\begin{equation*}
			\Sigma_{\text{CQFM,}F} = \frac{\sum_{k_1  = 1}^{K}\sum_{k_2=1}^{K} \min(\tau_{k_1}, \tau_{k_2}) (1 -\max(\tau_{k_1}, \tau_{k_2})}{\Big(\sum_{k=1}^{K} f_{\varepsilon}(b_{0\tau_{k}}) \Big)^2} \Sigma_{\lambda_0}^{-1};
		\end{equation*}
		the asymptotic distribution of $\sqrt{T}(\hat{\lambda}_{i} - \lambda_{0i})$ is $N(0, \Sigma_{\text{CQFM,}\lambda})$ with
		\begin{equation*}
			\Sigma_{\text{CQFM,}\lambda} = \frac{\sum_{k_1  = 1}^{K}\sum_{k_2=1}^{K} \min(\tau_{k_1}, \tau_{k_2}) (1 -\max(\tau_{k_1}, \tau_{k_2})}{\Big(\sum_{k=1}^{K} f_{\varepsilon}(b_{0\tau_{k}}) \Big)^2} \Sigma_{F_0}^{-1}.
		\end{equation*}
	\end{theorem}
	
	The format of the limiting distributions in \Cref{thm: asymptotic distribution} resembles the result for linear regression coefficients in  \citet[Theorem~2.1]{zouandyuan2008cqr}. 
	\begin{remark}
		When $K=1$, CQFM reduces to the quantile factor model. Results in \Cref{thm: asymptotic distribution} are comparable to those in \cite{andoandbai2020jasa}. Use the asymptotic distribution for factor loadings as an example. At quantile position $\tau$, its asymptotic variance is
		\begin{equation} \label{eq:AndoBai lambda var}
			\text{ \citet[Theorem~2]{andoandbai2020jasa}: }\tau(1-\tau)\Gamma_{i,0,\tau}^{-1}V_{i,0,\tau}\Gamma_{i,0,\tau}^{-1},
		\end{equation}
		 where both $ V_{i,0,\tau}$ and $ \Gamma_{i,0,\tau}$ are defined in their theorem and ``$V_{i,0,\tau}$" is similar to $\Sigma_{F_0}$ in \Cref{thm: asymptotic distribution}.  This sandwich estimator for covariance is commonly found in other papers on quantile regression with panel data such as \cite{katoandgalvao2012panelquantile,galvaoandkato2016joesmoothedquantile,chenetal2021qfm}.  In \Cref{thm: asymptotic distribution} with $K=1$, based on \cref{eq:lambda diff equation,eq:lambda distribution}, we have
		\begin{equation} \label{eq:sigma lambda k=1}
			\Sigma_{\text{CQFM,}\lambda} = \tau(1-\tau) \Big(f_{\varepsilon}(b_{0\tau}) \Sigma_{F_0}\Big)^{-1} \Sigma_{F_0} \Big(f_{\varepsilon}(b_{0\tau}), \Sigma_{F_0}\Big)^{-1} =  \frac{\tau(1-\tau)}{f_{\varepsilon}(b_{0\tau})^2 }\Sigma_{F_0}^{-1},
		\end{equation}
		which matches the result in \cite{andoandbai2020jasa}. Our result is made simpler by the i.i.d. errors in \Cref{asump: iid} that allow us to separate $f_{\varepsilon}(b_{0\tau_{k}})$ from $\Sigma_{F_0}$ in the term $f_{\varepsilon}(b_{0\tau_{k}}) \Sigma_{F_0}$; other papers typically consider the distribution of $\varepsilon_{it}$ conditional on either some regressors or the factors, see, for example, the term ``$\Gamma_{i,0,\tau} =T^{-1} \sum_{t=1}^{T}g_{it}(0|\cdot)z_{it,0,\tau}z_{it,0,\tau}$'' in \citet[Theorem~2]{andoandbai2020jasa}, where the conditional density function $g_{it}(0|\cdot)$ cannot be taken out of the summation sign as $T \rightarrow \infty$.  This simplification can also be found  in \citet[Theorem~4.1]{koenker_2005} for the linear quantile regression with i.i.d. errors. If the distribution of $\varepsilon_{it}$ is conditional on $F_{0t}$, $\Sigma_{\text{CQFA,}\lambda}$ will have a format similar to \cref{eq:AndoBai lambda var}.
	\end{remark}
	
	\begin{remark}
		If the true factor and factor loading, $F_{0t}$ and $\lambda_{0i}$, do not meet the normalization conditions in \cref{eq: normalization}, $\hat{F}_t$ and $\hat{\lambda}_{i}$ estimate a rotation of the corresponding true values. Our proof can be adapted to incorporate a rotation matrix. To simplify the presentation of the asymptotic results, we assume factors and loadings are identifiable under the normalization assumptions and omit the rotation matrix in \Cref{thm: asymptotic distribution}, similar to \cite{andoandbai2020jasa}. 
 	\end{remark}
 	
 	\begin{remark}
 		Although the asymptotic results in \Cref{thm: asymptotic distribution} are developed for the panel mean factor model while those in \cite{andoandbai2020jasa,chenetal2021qfm} are for panel quantile factor model, all proofs are related to techniques in quantile regression. \cite{andoandbai2020jasa} give a proof based on the uniform consistency of parameter estimates and higher-order moment conditions on the error term; \cite{chenetal2021qfm} derive the asymptotic results based on a smoothed quantile objective function by replacing the indicator function with a differentiable kernel function. In our proof, we replace the objective function with an asymptotic quadratic form of the parameters and solve $\hat{\lambda}_{i} - \lambda_{i}$ and $\hat{F}_t - F_{0t}$ directly from the first-order conditions, similar to the proof strategy in \cite{zouandyuan2008cqr} for CQR and \cite{koenker_2005} for quantile regression.
 	\end{remark}
 	
 	\begin{remark}
 		To compare the relative efficiency between CQFM and PCA-based solutions, we compute the asymptotic relative efficiency (ARE) of CQFM relative to PCA --- the ratio of their asymptotic variances.  Consider the estimator for $F_0$. In CQFM, its variance is given in \Cref{thm: asymptotic distribution}; for PCA-based factor analysis, the variance is given in \citet[Theorem~1(i)]{bai2003ecma}. We will simplify the variance expression for $\hat{F}_t$ in \citet[Theorem~1(i)]{bai2003ecma} to facilitate the comparison. The notation for factor estimator is ``$\tilde{F}_t$" in \cite{bai2003ecma}, while we use $\hat{F}_t^{\text{PCA}}$ to denote the same estimator. An $r \times r$ rotation matrix, $H = (\Lambda_0'\Lambda_0/N)(F_0'\hat{F}^{\text{PCA}}/T)V_{NT}^{-1}$, is introduced in \citet[p.~158]{bai2003ecma} to describe the indeterminacy of the solutions, where $V_{NT}$ is a diagonal matrix that contains the eigenvalues of $(NT)^{-1}YY'$. For our purpose, it will be desirable to set $H = I_r$ so that $\sqrt{N}(\hat{F}_t^{\text{PCA}} - H'F_{0t})$ in \citet[Theorem~1(i)]{bai2003ecma} becomes $\sqrt{N}(\hat{F}_t^{\text{PCA}} - F_{0t})$, matching the format in \Cref{thm: asymptotic distribution}. Replacing $F_0$ in $H$ with the estimator $\hat{F}^{\text{PCA}}$ and using the normalization $ F^{\text{PCA}\prime} F^{\text{PCA}} / T= I_r$, we obtain $V_{NT} = \Lambda_0'\Lambda_0/N \rightarrow \Sigma_{\lambda}$, where the convergence result follows Assumption B in \cite{bai2003ecma}. This result, combined with equation (7) in \citet[p.~150]{bai2003ecma}, suggests the variance of $\sqrt{N}(\hat{F}_t^{\text{PCA}} - H'F_{0t})$ in \citet[Theorem~1(i)]{bai2003ecma} can be written as $\sigma_{\varepsilon}^2 \Sigma_{\lambda}^{-1}$ if $\varepsilon_{it}$ is i.i.d. and independent of $F_{0t}$, where $\sigma_{\varepsilon}^2$ is the variance of $\varepsilon_{it}$ and is assumed to be a finite number. This result greatly simplifies the efficiency comparison between CQFM and PCA-based factor analysis.  Define the ARE of CQFM relative to the PCA-based factor analysis as
 		\begin{equation} \label{eq:are factor}
 			\text{ARE}(K)_F = \frac{\sigma_{\varepsilon}^2\Big(\sum_{k=1}^{K} f_{\varepsilon}(b_{0\tau_{k}}) \Big)^2}{\sum_{k_1  = 1}^{K}\sum_{k_2=1}^{K} \min(\tau_{k_1}, \tau_{k_2}) (1 -\max(\tau_{k_1}, \tau_{k_2})},
 		\end{equation} 
 		which is identical to equation (3.1) in \cite{zouandyuan2008cqr}. As a result, we can apply \cite[Theorem~3.1]{zouandyuan2008cqr} to show that the ARE for the factor estimator from CQFM in \cref{eq:are factor} has a relative efficiency of at least $0.7026$ with respect to that of the PCA-based factor analysis when $K \rightarrow \infty$. This result suggests that, compared to PCA, CQFM factors will have about $30\%$ efficiency loss in the worst scenario. This is a conservative theoretical result. In our simulations (see \Cref{tab:mse asym error} and \Cref{tab:mse sym error} in the supplement), the mean squared error (MSE) of the estimated component $\hat{\lambda}_{i}' \hat{F}_t$ are mostly smaller or much smaller than that of PCA-based estimate. Efficiency loss, if any, is small based on our simulation study.
 		
 	\end{remark}
 	
 	\begin{remark}
 		To compute $\Sigma_{\text{CQFM,}F} $ and $\Sigma_{\text{CQFM,}\lambda} $, we first note that quantities such as $\tau_{k_1}$ and $\tau_{k_2}$, along with $K$, are chosen beforehand by the researcher. $\Sigma_{\lambda_0}$ can be replaced with the normalized diagonal matrix $ \hat{\Lambda}'\hat{\Lambda}/N$ while $\Sigma_{F_0}$ is $I_r$.  There is no unique way to estimate the density $f_{\varepsilon}(b_{0\tau_{k}})$ (and its inverse). Since $f_{\varepsilon}(b_{0\tau_{k}})$ is the density of $\varepsilon$ at $100\tau_{k}\%$ quantile and the estimate for $\hat{b}_{\tau_{k}}$ is available, we can first obtain the residuals $\hat{\varepsilon}_{it}$, and use a consistent nonparametric density estimator for the residuals to obtain $\hat{f}_{\varepsilon}(\hat{b}_{\tau_{k}})$. Because of the i.i.d. error assumption, this simple estimate for $\Sigma_{\text{CQFM,}F} $ and $\Sigma_{\text{CQFM,}\lambda} $ is always positive definite.
 	\end{remark}
 	
 	\begin{remark}
 		In a quantile factor model, the conditional quantile at $\tau_{k}$ can be written as
 		\begin{equation} \label{eq:qfm}
 			Q_{Y_{it}}(\tau_{k}) = \lambda_{0i}(\tau_{k})' F_{0t}(\tau_{k}) + b_{0\tau_{k}},
 		\end{equation}
 		where the factors and factor loadings vary with $\tau_{k}$. But our factor model in \cref{eq: factor model}, when used inside \cref{eq: CQFM obj}, have constant factors and factor loadings across selected quantiles. This is not a misspecification since our goal is to estimate the $\lambda_{0i}$ and $F_{0t}$ in the mean of \cref{eq: factor model} but not $ \lambda_{0i}(\tau_{k})$ and $F_{0t}(\tau_{k})$ in \cref{eq:qfm}. Much like in a standard linear regression, in addition to the least-squares method,  one can use the lasso, principal components regression, LAD, Huber loss regression, CQR, \textit{etc.}, for estimation, there are several ways to estimate the mean factor model, and CQFM is one of the alternatives. While still permitting the quantile model in \cref{eq:qfm} for the data, CQFM combines the mean factor model in \cref{eq: factor model} with the composite quantile loss in \cref{eq: CQFM obj}. A single quantile loss in \cref{eq: QFM obj} gives estimates that adapt to data at a particular quantile. By using multiple quantiles, CQFM is designed to give the mean estimates that can adapt to data at multiple quantiles. Our simulation results demonstrate that this approach works well for several examples of data with asymmetry, heteroskedasticity and time series correlation.
 	\end{remark}
 	
 	\section{Factor number selection} \label{sec:factor number}
 	
 	The number of factors is assumed to be known in \Cref{thm: asymptotic distribution}. We discuss the selection of factor number in this section. Since the important work in \cite{baiandng2002ecma} on consistent factor number selection, there has been continued development of new methods in the literature. See \cite{baing2007jbesprimitive,amengualandwatson2007JBES,hallinandliska2007jasa,onatski2009ecma,lamandyao2012AOS} for panel mean regression models and \cite{andoandbai2020jasa,chenetal2021qfm} for panel quantile regression models.
 	
 	Denote $r$ the estimated number of factors. To work with \cref{eq: CQFM obj}, we propose the following information criterion (IC):
 	\begin{align} \label{eq:IC}
 		IC(r) &= \log \Bigg[\frac{1}{NT} \sum_{k}^{K}\sum_{i=1}^{N}\sum_{t=1}^{T} \rho_{\tau_k}(Y_{it} -\hat{b}_{\tau_{k}}(r)- \hat{\lambda}_i(r)'\hat{F}_t(r))\Bigg]  \nonumber \\
 		&\quad + r \times q(N,T),
 	\end{align}
 	where
 	\begin{equation} \label{eq:qnt}
 		q(N,T) = \left(\frac{N+T}{NT}\right) \log\left(\frac{NT}{N+T}\right),
 	\end{equation}
 	and we use $\hat{b}_{\tau_{k}}(r), \hat{\lambda}_i(r)$ and $\hat{F}_t(r)$ to denote estimates based on $r$ number of factors. This information criterion is similar to the one used in \cite{andoandbai2020jasa} and $\textit{IC}_{p1}$ in \citet[p.~201]{baiandng2002ecma}. \Cref{thm:factor number} shows the consistency of $IC(r)$. Let $C_{NT} = \min(N,T)$.
 	\begin{theorem} \label{thm:factor number}
 		Under \Cref{asump: factor and facor loading,asump: error density,asump: iid}, as $N,T \rightarrow \infty$, if  $q(N,T) \rightarrow 0$,  
 		the information criterion in \cref{eq:IC} selects the number of factors consistently.
 	\end{theorem}
 	See the online supplement for the proof.
 	\begin{remark}
 		The condition for $q(N,T)$ in \Cref{thm:factor number} defines a class of penalty functions, and \cref{eq:qnt} is an example of possibly many other penalty functions. The IC with \cref{eq:qnt} works quite well for most of the simulation examples in our study. However, it fails  when the error term follows a \textit{t} distribution with $1$ degree of freedom ($t_1$). In this case, we propose another $q(N,T)$ function that works well with the $t_1$ distribution
 		\begin{equation} \label{eq:qnt 2}
 			q(N,T) = \log \left(\log\left(\frac{NT}{N+T}\right)\right)  \left(\frac{N+T}{NT}\right).
 		\end{equation}
 		This penalty function also meets the requirement for $q(N,T)$ in \Cref{thm:factor number}, but it converges to $0$ faster  and, consequently, imposes less penalty than \cref{eq:qnt} . Its performance for the $t_1$ error distribution is reported in \Cref{tab:factor number sym error} in the online supplement.
 	\end{remark}
	
	\section{Monte Carlo simulation}
	
	In this section, we use Monte Carlo simulation to study the finite sample properties of the CQFM method. To compare CQFM to other methods, we use the R code in \cite{heetal2022JBESnomoment}  to compute the robust two-step (RTS) factors and the matlab code in \cite{chenetal2021qfm} to compute the QFM factors at quantile position $0.5$ (QFM(0.5)) and also the estimated factor numbers. When space permits, we also add the PCA results.
	
	The number of quantiles in CQFM is an additional tuning parameter, and we choose $K = 5$ for demonstration purposes, which corresponds to the quantiles of $0.17, 0.33, 0.5, 0.67$ and $0.83$.  A convergence criterion of $ 10^{-3}$ is used in the MM algorithm.
	
	\subsection{Data simulation}
	
	Consider the following 3-factor data generating process (DGP):
	\begin{equation*}
		Y_{it} = \sum_{j=1}^{3} \lambda_{0i,j} F_{0t,j} + \varepsilon_{it},
	\end{equation*}
	where $F_{0t,1} = 0.8 F_{0t-1,1} + e_{1t}$, $F_{0t,2} = 0.5 F_{0t-1,2} + e_{2t}$, $F_{0t,3} = 0.2 F_{0t-1,3} + e_{3t}$, and both $e$ and $\lambda_{0i,j}$ are i.i.d. $N(0,1)$. This is identical to the DGP in \citet[Section~5.1]{chenetal2021qfm} except that we consider several asymmetrical i.i.d. errors. They are summarized in \Cref{tab:error_asym_description}. Let $\gamma_1$ and $\gamma_2$ be the skewness and excess kurtosis coefficient, respectively.
	
	\begin{table}[htp]
		\centering
		\caption{Description of the $5$ asymmetric error distributions}
		\begin{tabular}{ll}
			\toprule
			\text{Error distribution}& parameter setting \\
			\midrule
			1. skewed normal (\textit{sn}) & $\mu_{\varepsilon} = 0, \sigma_{\varepsilon}=1, \gamma_1=0.99$ \\
			2. skewed \textit{t}   & $\mu_{\varepsilon} = 0, \sigma_{\varepsilon}=1, \gamma_1=0.99, \gamma_2=3$ \\
			3. asymmetric Laplace&  location$=0$, scale$=0.5$, asymmetry$=4$\\
			4. log-normal & $\mu = 0, \sigma = 1.5$ \\
			5. mixture of skewed normal& $0.9\cdot sn(0,1,0.99) + 0.1\cdot sn(0,9,0.99)$ \\
			\bottomrule
		\end{tabular}%
		\label{tab:error_asym_description}%
	\end{table}%
	
	The R package \texttt{sn} is used to simulate the skewed normal and skewed \textit{t} distributions in \Cref{tab:error_asym_description}. If one specifies the skewness parameter directly, the \texttt{sn} package restricts $|\gamma_1| < 0.99527$; we set $\gamma_1 = 0.99$ for the first two error distributions. The asymmetric Laplace error term is generated using the \texttt{rlaplace} function in the R package \texttt{LaplacesDemon}. The three numbers, $0,0.5,4$ correspond to the location, scale, and kappa parameter in the \texttt{rlaplace} function in R. For the asymmetric Laplace distribution, a kappa value of $4$ implies a skewness of about $-1.99$. Next, we consider a more skewed log-normal distribution with mean and s.d. equal to $0$ and $1.5$, respectively. These are the parameter values for the log-normal density, which implies the error term $\varepsilon_{it}$ has its mean, s.d., and skewness equal to $ 3.08, 8.97$ and $33.47$. For both the asymmetric Laplace distribution and the log-normal distribution, we subtract the theoretical mean from the simulated errors so that all error terms have zero mean. Finally, we consider a mixture of skewed normal distribution, where $sn(0,1,0.99)$ and $sn(0,9,0.99)$ denote the skewed normal distribution with $\mu_{\varepsilon} = 0, \sigma_{\varepsilon}=1, \gamma_1=0.99$ and $\mu_{\varepsilon} = 0, \sigma_{\varepsilon}=3, \gamma_1=0.99$, respectively. The weights for the mixture normal are $0.9$ and $0.1$.
	
	We consider five different sample sizes: $\left(N,T\right) = (50,100), (100,50), (100,200), (200,100)$ and $(300,300)$. For each error distribution and sample size, we report the value of an evaluation metric based on $100$ replications for every estimation method. 
	
	\Cref{tab:adj R2 sym error with pca,tab:mse sym error,tab:factor number sym error} in the online supplement report the results for $5$ symmetric error distributions, including $N(0,1)$, \textit{t} distribution with $1$ degree of freedom, \textit{etc}.  \Cref{tab:adj R2 asym error with pca heterodasticity,tab:mse asym error heteroskedasticity,tab:factor number asym error heterskedasticity,tab:adj R2 asym ar1 error with pca,tab:mse asym ar1 error,tab:factor number asym ar1 error} report the results for the following  heteroskedasticity and AR(1) asymmetric errors:
	\begin{align} 
		\text{heteroskedasticity } & Y_{it} = \sum_{j=1}^{3} \lambda_{0i,j} F_{0t,j} + \left[2+\cos(2\pi\times\lambda_{0i,4}F_{0t,4})\right]\times\varepsilon_{it}, \label{eq:heter}\\
		\text{AR(1) error }              & Y_{it} = \sum_{j=1}^{3} \lambda_{0i,j} F_{0t,j} + \varepsilon_{it} \text{ with } \varepsilon_{it} = 0.5\varepsilon_{i,t-1} + u_{it}, \label{eq:ar1}
	\end{align}
	where $\lambda_{0i,4}$ and $F_{0t,4}$ are i.i.d. $N(0,1)$ and $\varepsilon_{it}$ and $u_{it}$ are asymmetric errors defined in \Cref{tab:error_asym_description}. CQFM is found to have good finite properties in these cases too.
	\subsection{Estimation of the factor and factor loading}
	Similar to \citet[Table~1]{chenetal2021qfm}, \Cref{tab:adj R2 asym error} reports the average adjusted $R^2$ from regressing $F_{0t,1}, F_{0t,2}$, and $F_{0t,3}$ on the $3$ estimated factors from the RTS, QFM(0.5), and CQFM methods. Results in \Cref{tab:adj R2 asym error} assess how well the estimated factors span the space spanned by the true factors. \Cref{tab:adj R2 asym error with pca} in the supplement expands \Cref{tab:adj R2 asym error} to include the PCA results.
	
		\begin{table}[htp] \centering
		
		\begin{center}
			\caption{Adj. $R^2$ of regressing  $3$ true factors on the estimated factors} 
			\label{tab:adj R2 asym error} 
			\begin{threeparttable}
				\begin{tabular}{lrrrrrrrrr}    
					\cmidrule(lr){1-10}
					\multirow{2}{*}{(T,N)} & $R^2_{1,\text{RTS}}$  & $R^2_{2,\text{RTS}}$ & $R^2_{3,\text{RTS}}$ & $R^2_{1,\text{QFM}}$ & $R^2_{2,\text{QFM}}$  & $R^2_{3,\text{QFM}}$ & $R^2_{1,\text{CQFM}}$ & $R^2_{2,\text{CQFM}}$  & $R^2_{3,\text{CQFM}}$ \\
					\cmidrule(lr){2-4} \cmidrule(lr){5-7} \cmidrule(lr){8-10}
					&   \multicolumn{9}{c}{$\varepsilon_{it} \sim$ skewed normal}      \\    
					\cmidrule(lr){2-10}
					(50,100) & 0.9950 & 0.9914 & 0.9893 & 0.9915 & 0.9857 & 0.9821 & 0.9951 & 0.9917 & 0.9898 \\    
					(100,50) & 0.9909 & 0.9828 & 0.9786 & 0.9847 & 0.9712 & 0.9645 & 0.9911 & 0.9832 & 0.9790 \\    
					(100,200) & 0.9978 & 0.9961 & 0.9949 & 0.9960 & 0.9928 & 0.9908 & 0.9979 & 0.9962 & 0.9952 \\    
					(200,100) & 0.9959 & 0.9921 & 0.9898 & 0.9926 & 0.9856 & 0.9816 & 0.9961 & 0.9924 & 0.9903 \\      
					(300,300) & 0.9986 & 0.9975 & 0.9967 & 0.9974 & 0.9953 & 0.9938 & 0.9987 & 0.9976 & 0.9969 \\    
					&  \multicolumn{9}{c}{$\varepsilon_{it} \sim$ skewed t}      \\   
					\cmidrule(lr){2-10}
					(50,100) & 0.9950 & 0.9916 & 0.9895 & 0.9936 & 0.9895 & 0.9866 & 0.9954 & 0.9923 & 0.9903 \\    
					(100,50) & 0.9908 & 0.9828 & 0.9790 & 0.9885 & 0.9783 & 0.9737 & 0.9914 & 0.9840 & 0.9804 \\    
					(100,200) & 0.9978 & 0.9960 & 0.9949 & 0.9972 & 0.9950 & 0.9936 & 0.9981 & 0.9964 & 0.9954 \\    
					(200,100) & 0.9959 & 0.9921 & 0.9899 & 0.9947 & 0.9900 & 0.9871 & 0.9962 & 0.9928 & 0.9908 \\       
					(300,300) & 0.9986 & 0.9975 & 0.9967 & 0.9983 & 0.9968 & 0.9958 & 0.9988 & 0.9978 & 0.9971 \\    
					&   \multicolumn{9}{c}{$\varepsilon_{it} \sim$ asymmetric Laplace}      \\  
					\cmidrule(lr){2-10}  
					(50,100) & 0.9569 & 0.9254 & 0.9085 & 0.9482 & 0.9108 & 0.8682 & 0.9759 & 0.9590 & 0.9518 \\    
					(100,50) & 0.9277 & 0.8698 & 0.8438 & 0.9151 & 0.8443 & 0.8062 & 0.9573 & 0.9221 & 0.9057 \\    
					(100,200) & 0.9826 & 0.9673 & 0.9577 & 0.9777 & 0.9566 & 0.9341 & 0.9911 & 0.9834 & 0.9784 \\    
					(200,100) & 0.9664 & 0.9375 & 0.9204 & 0.9571 & 0.9143 & 0.8900 & 0.9823 & 0.9663 & 0.9572 \\    
					(300,300) & 0.9891 & 0.9797 & 0.9739 & 0.9843 & 0.9704 & 0.9613 & 0.9946 & 0.9900 & 0.9874 \\    
					&   \multicolumn{9}{c}{$\varepsilon_{it} \sim$ log-normal}      \\   
					\cmidrule(lr){2-10}
					(50,100) & 0.5958 & 0.3578 & 0.2543 & 0.9541 & 0.8272 & 0.6834 & 0.9889 & 0.9823 & 0.9769 \\    
					(100,50) & 0.5637 & 0.3286 & 0.2245 & 0.9216 & 0.7739 & 0.5421 & 0.9781 & 0.9598 & 0.9512 \\    
					(100,200) & 0.8045 & 0.5902 & 0.4469 & 0.9754 & 0.8402 & 0.5698 & 0.9964 & 0.9932 & 0.9914 \\    
					(200,100) & 0.7470 & 0.5382 & 0.4320 & 0.9687 & 0.8033 & 0.4895 & 0.9931 & 0.9863 & 0.9826 \\        
					(300,300) & 0.8950 & 0.7967 & 0.7229 & 0.9881 & 0.8448 & 0.4473 & 0.9980 & 0.9963 & 0.9952 \\    
					&   \multicolumn{9}{c}{$\varepsilon_{it} \sim$ mixture of skewed normal}      \\   
					\cmidrule(lr){2-10} 
					(50,100) & 0.9908 & 0.9851 & 0.9807 & 0.9899 & 0.9835 & 0.9788 & 0.9937 & 0.9900 & 0.9870 \\    
					(100,50) & 0.9829 & 0.9694 & 0.9609 & 0.9816 & 0.9670 & 0.9586 & 0.9880 & 0.9785 & 0.9731 \\   
					(100,200) & 0.9960 & 0.9929 & 0.9908 & 0.9954 & 0.9920 & 0.9893 & 0.9974 & 0.9954 & 0.9941 \\    
					(200,100) & 0.9926 & 0.9858 & 0.9818 & 0.9914 & 0.9837 & 0.9789 & 0.9952 & 0.9908 & 0.9880 \\     
					(300,300) & 0.9976 & 0.9955 & 0.9941 & 0.9971 & 0.9946 & 0.9929 & 0.9985 & 0.9972 & 0.9964 \\   
					\hline
				\end{tabular}
				\begin{tablenotes}[flushleft]
					\item[] \textit{Notes}: Each number is the average of adjusted $R^2$ over $100$ replications of regressing one of the three true factors on the estimated factors based on the RTS, QFM(0.5), and CQFM method, respectively. We choose $\tau=0.5$ for the QFM method and $K=5$ for the CQFM method.
				\end{tablenotes}
			\end{threeparttable}
		\end{center} 
	\end{table}
	
	For the first two error distributions with small skewness in \Cref{tab:adj R2 asym error}, all three methods perform well. Their differences in the adjusted $R^2$ mostly appear in the third digit. Still, we see CQFM performs slightly better than RTS and QFM. In the case of asymmetric Laplace error, the difference between CQFM and the other two methods start to grow larger. For example, for the sample size $(50,100)$, the adj. $R^2$ associated with $F_{0t,3}$ is $0.8682$ for QFM, while it is $0.9518$ for CQFM. The log-normal error distribution poses the greatest challenge to the other methods, as the regression yields much lower adj. $R^2$s compared to those of CQFM, and increasing sample size from $(50,100)$ to $(300,300)$ does not seem to help. 
	
	To further investigate the accuracy of the estimates, we compute the mean squared error (MSE) of the estimated components and report them in \Cref{tab:mse asym error}. The MSE is defined as
	\begin{equation*}
		\text{MSE} = \frac{1}{NT}\sum_{i=1}^{N}\sum_{t=1}^{T} \big(\lambda_{0i}'F_{0t} - \hat{\lambda}_i'\hat{F}_{t} \big)^2.
	\end{equation*}
	\Cref{tab:mse asym error} clearly indicates that CQFM always yields the smallest MSE for the $5$ error distributions. Depending on the error distribution, CQFM's reduction in MSE can be huge -- its MSE can be a fraction of that of PCA.
	
	We can draw several conclusions based on \Cref{tab:adj R2 asym error,tab:mse asym error}. First, compared to QFM at a single quantile position $\tau = 0.5$, the higher adj. $R^2$ and smaller MSE of CQFM suggests that there is some benefit in performing the estimation at multiple quantile positions simultaneously. Second, CQFM continues to work well in cases such as the asymmetric Laplace and log-normal errors, implying that CQFM can be a useful alternative to PCA in certain cases.
	
	The online supplement also includes the adj. $R^2$ and MSE results (\Cref{tab:adj R2 sym error with pca,tab:mse sym error}) for $5$ symmetric error distributions. Overall, CQFM continues to provide robust estimates.
		
		\begin{table}[ht] \centering
		\begin{center}
			\caption{MSE under asymmetric errors} 
			\label{tab:mse asym error} 
			\begin{threeparttable}
				\begin{tabular}{lrrrrrrrr}    
					\cmidrule(lr){1-9}
                    \multirow{2}{*}{(T,N)}  &   \multicolumn{4}{c}{$\varepsilon_{it} \sim$ skewed normal}  &   \multicolumn{4}{c}{$\varepsilon_{it} \sim$ skewed t}         \\  
					  & RTS & QFM & CQFM & PCA & RTS & QFM & CQFM & PCA \\
					\cmidrule(lr){2-5} \cmidrule(lr){6-9}
					(50,100) & 0.101 & 0.157 & 0.097 & 0.090 & 0.099 & 0.114 & 0.090 & 0.089\\
					(100,50) & 0.092 & 0.155 & 0.088 & 0.089 & 0.092 & 0.114 & 0.084 & 0.089\\
					(100,200) & 0.048 & 0.085 & 0.048 & 0.045 & 0.048 & 0.058 & 0.044 & 0.045\\
					(200,100) & 0.046 & 0.084 & 0.043 & 0.045 & 0.046 & 0.058 & 0.041 & 0.045\\
					(300,300) & 0.021 & 0.040 & 0.020 & 0.020 & 0.021 & 0.026 & 0.019 & 0.020\\
					&   \multicolumn{4}{c}{$\varepsilon_{it} \sim$ asymmetric Laplace}    &   \multicolumn{4}{c}{$\varepsilon_{it} \sim$ log-normal}    \\  
					\cmidrule(lr){2-5}  \cmidrule(lr){6-9}
					(50,100) & 0.836 & 1.139 & 0.458 & 0.801 & 19.888 & 4.124 & 0.214 & 36.456 \\
					(100,50) & 0.794 & 1.134 & 0.439 & 0.799 & 14.344 & 4.157 & 0.211 & 36.752\\
					(100,200) & 0.390 & 0.747 & 0.198 & 0.379 & 6.727 & 4.294 & 0.084 & 26.414\\
					(200,100) & 0.382 & 0.744 & 0.196 & 0.381 & 4.865 & 4.283 & 0.076 & 26.531\\
					(300,300) & 0.167 & 0.440 & 0.080 & 0.165 & 1.726 & 4.377 & 0.038 & 17.464\\
					&   \multicolumn{4}{c}{$\varepsilon_{it} \sim$ mixture of skewed normal}   &   \\   
					\cmidrule(lr){2-5} 
					(50,100) & 0.176 & 0.182 & 0.120 & 0.161 &&&& \\
					(100,50) & 0.168 & 0.182 & 0.113 & 0.164 &&&&\\
					(100,200) & 0.086 & 0.097 & 0.058 & 0.081 &&&& \\
					(200,100) & 0.083 & 0.097 & 0.053 & 0.081  &&&&\\
					(300,300) & 0.037 & 0.046 & 0.025 & 0.036 &&&&\\ 
					\hline
				\end{tabular}
				\begin{tablenotes}[flushleft]
					\item[] \textit{Notes}: Each number is the average MSE over $100$ replications for the RTS, QFM(0.5), CQFM, and PCA method, respectively. We choose $\tau=0.5$ for the QFM method and $K=5$ for the CQFM method.
				\end{tablenotes}
			\end{threeparttable}
		\end{center} 
	\end{table}
	
	\subsection{Estimation of the factor number}
	Next, we study the performance of the information criterion in \cref{eq:IC,eq:qnt}. \Cref{tab:factor number} reports the average estimated factor number and the frequency of correct factor number estimation. Both CQFM and PCA perform well in majority of the cases. For log-normal error, all methods fail in small sample. However, CQFM yields better results when sample size is large  with $\text{Prob}(\hat{r} = 3) = 82\%$ in the case of $(300,300)$. 
	
		\begin{table}[htp] \centering
		\begin{center}
			\caption{Average estimated factor number and frequency of correct estimation} 
			\label{tab:factor number} 
			\begin{threeparttable}
				\renewcommand{\TPTminimum}{\linewidth}
				\makebox[\linewidth] { 
					\begin{tabular}{lrrrrrr}
						\toprule
						(T,N) & QFM      &   CQFM    &    PCA   &  QFM     &  CQFM     &   PCA \\
						\hline
						&   \multicolumn{3}{c}{avg. $\hat{r}$}  &   \multicolumn{3}{c}{$\text{Prob}(\hat{r} = 3)$} \\
						\cmidrule(lr){2-4}  \cmidrule(lr){5-7}
						&  \multicolumn{6}{c}{$\varepsilon_{it} \sim$ skewed normal}        \\
						\cmidrule(lr){2-7} 
						(50,100) & 2.52  & 3     & 3     & 0.61  & 1     & 1 \\
						(100,50) & 2.55  & 3     & 3     & 0.6   & 1     & 1 \\
						(100,200) & 2.94  & 3     & 3     & 0.95  & 1     & 1 \\
						(200,100) & 2.92  & 3     & 3     & 0.92  & 1     & 1 \\
						
						(300,300) & 3     & 3     & 3     & 1     & 1     & 1 \\
						&  \multicolumn{6}{c}{$\varepsilon_{it} \sim$ skewed t}        \\
						
						\cmidrule(lr){2-7}  
						(50,100) & 2.52  & 3     & 3     & 0.59  & 1     & 1 \\
						(100,50) & 2.55  & 3     & 3     & 0.61  & 1     & 1 \\
						(100,200) & 2.94  & 3     & 3     & 0.95  & 1     & 1 \\
						(200,100) & 2.92  & 3     & 3     & 0.92  & 1     & 1 \\
						
						(300,300) & 3     & 3     & 3     & 1     & 1     & 1 \\
						&  \multicolumn{6}{c}{$\varepsilon_{it} \sim$ asymmetric Laplace}        \\
						
						\cmidrule(lr){2-7}  
						(50,100) & 2.66  & 1.8   & 2.9   & 0.66  & 0.14  & 0.9 \\
						(100,50) & 2.75  & 1.74  & 2.91  & 0.71  & 0.12  & 0.91 \\
						(100,200) & 3.31  & 3     & 3     & 0.64  & 1     & 1 \\
						(200,100) & 3.37  & 3     & 3     & 0.57  & 1     & 1 \\
						
						(300,300) & 3.95  & 3     & 3     & 0.05  & 1     & 1 \\
						&  \multicolumn{6}{c}{$\varepsilon_{it} \sim$ log-normal}        \\
						
						\cmidrule(lr){2-7}  
						(50,100) & 2.71  & 1.21  & 3.51  & 0.57  & 0.02  & 0.16 \\
						(100,50) & 2.95  & 1.21  & 3.49  & 0.56  & 0.01  & 0.2 \\
						(100,200) & 3.46  & 2.42  & 3.38  & 0.42  & 0.27  & 0.16 \\
						(200,100) & 3.57  & 2.3   & 3.21  & 0.37  & 0.28  & 0.23 \\
						
						(300,300) & 4     & 3.19  & 3.83  & 0     & 0.82  & 0.21 \\
						&  \multicolumn{6}{c}{$\varepsilon_{it} \sim$ mixture of skewed normal}        \\
						
						\cmidrule(lr){2-7}  
						(50,100) & 2.54  & 3     & 3     & 0.6   & 1     & 1 \\
						(100,50) & 2.61  & 3     & 3     & 0.64  & 1     & 1 \\
						(100,200) & 2.95  & 3     & 3     & 0.96  & 1     & 1 \\
						(200,100) & 2.92  & 3     & 3     & 0.92  & 1     & 1 \\
						
						(300,300) & 3     & 3     & 3     & 1     & 1     & 1 \\
						\bottomrule
					\end{tabular} 
				}   
				\begin{tablenotes}[flushleft]
					\item[] \textit{Notes}: To estimate $r$, we use the rank minimization method in \cite{chenetal2021qfm} for QFM at $\tau = 0.5$, the IC in \cref{eq:IC,eq:qnt} for CQFM, and the $IC_{p1}$ in \cite[p.~201]{baiandng2002ecma} for the PCA method. Avg. $\hat{r}$ is based on $100$ replications.
				\end{tablenotes}
			\end{threeparttable}
		\end{center} 
	\end{table}
	
	For the $5$ symmetric error distributions in \Cref{tab:factor number sym error}, our proposed IC with \cref{eq:qnt} continues to work well except for the $t_1$ error. In this case, the rank-based approach in \cite{chenetal2021qfm} gives good results when sample size is large. After standardizing the data and using \cref{eq:qnt 2} in \cref{eq:IC}, CQFM also gives satisfactory results when the sample size is large. 

	\section{Empirical application}
	
	In this section, we use the quarterly macroeconomic data set, FRED-QD, in \cite{mccrackenandng2020FREDQD} to study the properties of CQFM factors. We use the version ``2023-06.csv", which contains $258$ quarterly observations from 1959/3/1 to 2023/3/1 for $246$ macroeconomic variables. The data link is: \url{https://research.stlouisfed.org/econ/mccracken/fred-databases/}. We use the matlab code in \cite{mccrackenandng2020FREDQD} to prepare the data, including transforming all variables to stationary time series based on the \texttt{tcode} in \cite{mccrackenandng2020FREDQD}, removing outliers, and using the EM algorithm to fill in missing values. The final data set has $255$ quarterly observations and $246$ variables ($T = 255, N = 246$). 
	
	The number of estimated factors varies across different methods. For example, the CQFM estimate is $1$ ($3$ if \cref{eq:qnt 2} is used); the rank minimization method in \cite{chenetal2021qfm} reports $4$ factors at $\tau = 0.5$; the $IC_{p1}$ and $IC_{p2}$ in \citet[p.~201]{baiandng2002ecma} give $12$ and $8$ factors, respectively. Since our focus is on the property of the estimated factor, we follow \cite{stockandwaston2012nberrecession} and choose the number $6$ across different methods. The scree plot in \Cref{fig:screeplot} reveals why the proposed IC with \cref{eq:qnt} selects only one factor: the first eigenvalue is $54$ and explains about $22\%$ of the variation in the (standardized) data while the second eigenvalue is $19$ and explains about $7.8\%$ of the variance in the data.
	
	\begin{figure}[th!]
		\centering
		\includegraphics[width=1.0\textwidth,keepaspectratio=TRUE]{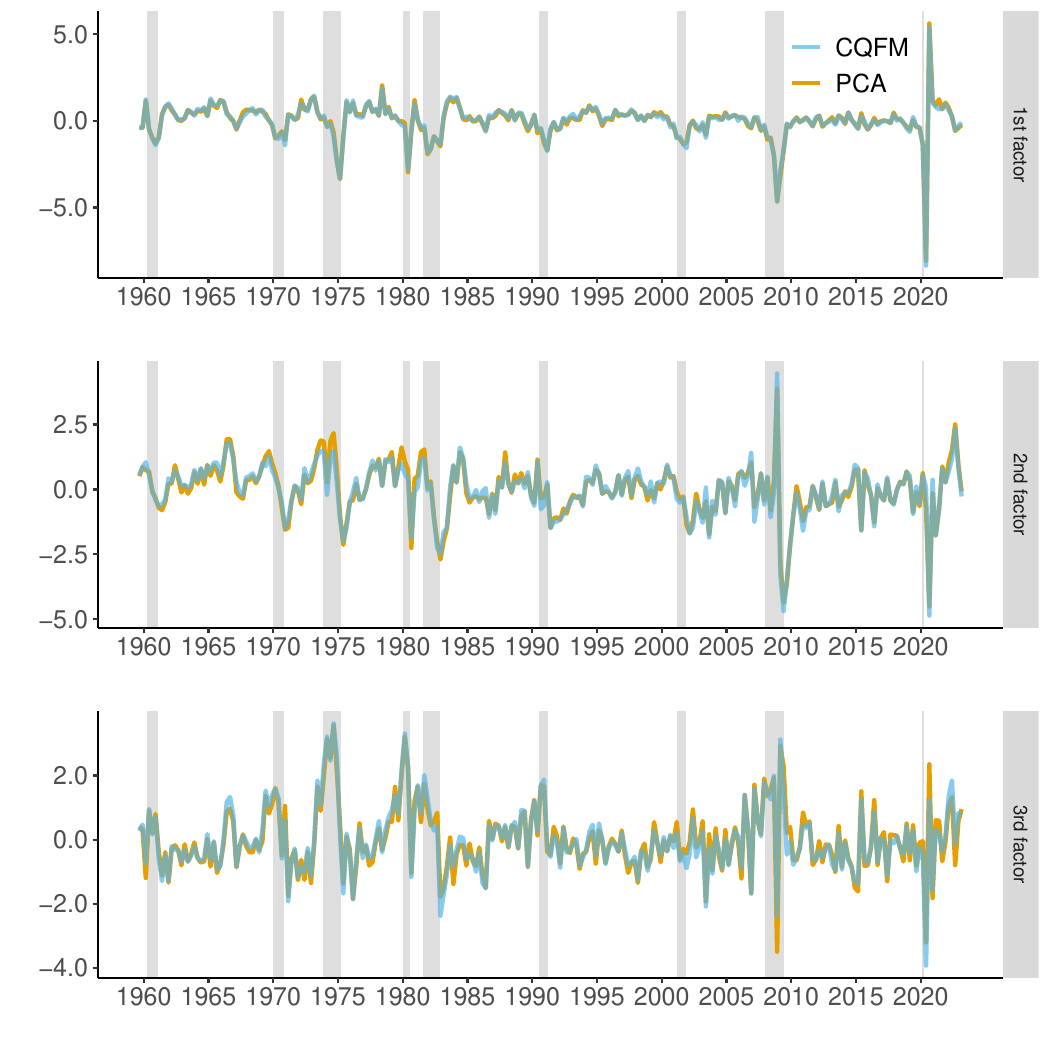}%
		\caption{The first three CQFM and PCA factors from 1959/3/1 to 2023/3/1 }%
		\label{fig:factors1-3}%
	\end{figure}
	
	\Cref{fig:factors1-3} plots the first three CQFM and PCA  factors (factors 4 to 6 are plotted in \Cref{fig:factors4-6}). The first three factors from CQFM and PCA are very similar to each other. Despite this visual similarity, the estimated factors exhibit different moment properties. \Cref{tab:factor moments} summarizes the skewness and kurtosis of the six estimated factors from the four methods. We make a few observations. First, the CQFM-based factors tend to have larger skewness and kurtosis in the first few factors. This means, if the data have large skewness and/or kurtosis, the CQFM-based factors will likely give a better fit for the component ($\lambda_{0i}' F_{0t}$). Second, even if other methods such as PCA-based factors exhibit  larger skewness and/or kurtosis in later factors -- for example, the 5th PCA factor exhibits larger skewness than CQFM, these larger value will unlikely be helpful in capturing the skewness and kurtosis in the data since it is typically the first few factors that determines the overall variability of the data. Third, compared to CQFM-based factors, the QFM-based factors exhibit less skewness and kurtosis, suggesting that estimation done at a single quantile position such as $\tau = 0.5$ may not be effective in capturing certain features of the data; the composite quantile approach is more effective in this regard.  The small MSEs for CQFM in \Cref{tab:mse asym error} attest to the above arguments.  
	
	\begin{table}[t] \centering
		\begin{center}
			\caption{Skewness and kurtosis of the $6$ estimated factors} 
			\label{tab:factor moments} 
			\begin{threeparttable}
				\renewcommand{\TPTminimum}{\linewidth}
				\makebox[\linewidth] {
					\begin{tabular}{lrrrrrr}
						\toprule
						method & $\hat{F}_1$ & $\hat{F}_2$ & $\hat{F}_3$ & $\hat{F}_4$ & $\hat{F}_5$& $\hat{F}_6$ \\
						\hline
						& \multicolumn{6}{c}{skewness ($\gamma_1$)}\\
						\cmidrule(lr){2-7}
						\text{RTS} & 1.85  & -0.33 & -0.38 & -0.27 & -0.52 & 0.17 \\
						\text{QFM} & -1.95 & 0.70  & -0.33 & 0.34  & 0.35  & 0.43 \\
						\text{CQFM} & -2.51 & -0.93 & 0.59 & -0.52  & -0.02 & -0.16 \\
						\text{PCA} & -2.23 & -0.76 & 0.51  & -0.18 & -0.12 & 0.00 \\
						& \multicolumn{6}{c}{kurtosis ($\gamma_2$)}\\
						\cmidrule(lr){2-7}
						\text{RTS} & 14.55 & 4.43  & 13.57 & 2.76  & 3.68  & 3.47 \\
						\text{QFM} & 17.62 & 6.54  & 5.25  & 2.80  & 3.08  & 7.50 \\
						\text{CQFM} & 26.03 & 7.98  & 4.93  & 15.58 & 2.71  & 3.32 \\
						\text{PCA} & 24.36 & 6.62  & 4.51  & 16.70 & 2.88  & 4.87 \\
						\bottomrule
					\end{tabular}%
				}
				\begin{tablenotes}[flushleft]
					\item[] \textit{Notes}: This table reports the skewness and kurtosis of the estimated six factors for different methods based on the FRED-QD data between 1959Q1 and 2023Q1. We focus on the magnitude of $\gamma_1$  since factors have sign indeterminacy.
				\end{tablenotes}
			\end{threeparttable}
		\end{center} 
	\end{table}
	Next, following \cite{stockandwatson2002jbes}, we use the diffusion indexes to forecast one-quarter-ahead macroeconomic variables. The forecasting function is given by 
	\begin{equation} \label{eq:diffusion_indexes_forecast}
		y_{i,t+1} = \beta_i + \sum_{j=0}^{3} \beta_j y_{i,t - j}  + \beta_F' \hat{F}_{t} + \epsilon_{i,t+1}, \text{ for } i = 1, \cdots, 246,
	\end{equation}
	where $y_{it}$ is the original data transformed according to the \texttt{tcode} in FRED-QD ``2023-06.csv" and $\hat{F}_t$ is the estimated $6$ factors at time $t$.  The forecast period starts from $2000 Q1$ to $2023Q1$, a total of $93$ forecasts for each of the $246$ macroeconomic variables, and this forecast period covers three NBER-determined recessions, including the one induced by the recent pandemic.  For each rolling forecast, we use a rolling window of $120$ quarters to estimate factors and the coefficients $\beta_j$ and $\beta_F$. We forecast the data that are transformed using the \texttt{tcode} in \cite{mccrackenandng2020FREDQD} and convert the forecast back to data in their original levels.
	
	\begin{table}[t] \centering
		\begin{center}
			\caption{Forecast RMSE of GDP, Unemployment rate, and Inflation} 
			\label{tab:forecast_rmse} 
			\begin{threeparttable}
				\renewcommand{\TPTminimum}{\linewidth}
				\makebox[\linewidth] {
					\begin{tabular}{lrrrrr}
						\toprule
						variable & \multicolumn{1}{l}{RTS} & \multicolumn{1}{l}{QFM} & \multicolumn{1}{l}{CQFM} & \multicolumn{1}{l}{PCA} & \multicolumn{1}{l}{AR(4)} \\
						\midrule
						\texttt{GDPC1} & 322.131 & 303.498 & \textbf{258.293} & 355.526 & 299.965 \\   
						\texttt{UNRATE} & 1.227 & 1.110 & \textbf{1.093} & 1.246 & 1.203 \\    
						\texttt{CPIAUCSL} & \textbf{3.616} & 4.173 & 3.832 & 3.798 & 3.739 \\    
						\texttt{avg RMSE} & 8269.7 & 8302.2 & \textbf{8053.6} & 8578.6 & 8219.2 \\ 
						\bottomrule
					\end{tabular}%
				}
				\begin{tablenotes}[flushleft]
					\item[] \textit{Notes}: This table reports the average forecast RMSE over $93$ forecasts from $2000Q1$ to $2023Q1$. \texttt{GDPC1} is the real GDP in chained 2012 dollars; \texttt{UNRATE} is the civilian unemployment rate (percent); \texttt{CPIAUCSL} is the CPI for all urban consumers. Results for columns 1 to 4 are based on an AR(4) model with six factors as additional regressors. The last column reports the forecast RMSE of the AR(4) model with no augmented factors. The variable \texttt{avg RMSE} reports the average of RMSE for all the $246$ macroeconomic time series for each of the $5$ methods.
				\end{tablenotes}
			\end{threeparttable}
		\end{center} 
	\end{table}
	\Cref{tab:forecast_rmse} reports the forecast root-MSE (RMSE) for the three most common macroeconomic variables, real gross domestic product (\texttt{GDPC1}), civilian unemployment rate (\texttt{UNRATE}), and consumer price index for all urban consumers (\texttt{CPIAUCSL}) in the FRED-QD data set. CQFM gives good results, but its performance is not the best for the CPI data. It's also somewhat surprising that the AR(4) model can sometimes do better than factor-augmented methods. In the last row, we compute the average of RMSE over the $246$ macroeconomic variables for each of the $5$ models, and CQFM gives the smallest average RMSE. Notice that the results for the $4$ factor-based models are obtained by simply choosing $6$ factors without any additional tuning of the model. \cite{mccrackenandng2020FREDQD} consider $7$ factors, and, for the regression in \cref{eq:diffusion_indexes_forecast}, they try $2^7-1 = 127$ different combinations of the $7$ factors. Similar approach can also be used here to possibly improve the performance of the factor-based models.  In addition, many other aspects of diffusion index modeling can be tuned to yield a favorable model, which includes, but not limited to, the number lags of the factors (we consider only $1$ in our regression), the forecast horizon (3-month, 6-month, one-year, \textit{etc}.), the inclusion of lag variables in the $Y$ matrix in \cref{eq: factor model matrix form} in factor analysis, the use of balanced panel data vs. unbalance panel data with EM-algorithm-generated data, the types of data transformation used, whether to split the data before and after a recession, among others. In the case of CQFM, we can also tune the parameter $K$ to possibly improve its performance. \Cref{tab:forecast_rmse} is a simple demonstration of the use of CQFM-based factors. A more comprehensive study is needed to further study the properties of different factor-based models.
	
	\section{Conclusions}
	
	In this paper, we develop the method of composite quantile factor model. We demonstrate in both simulations and an empirical study that, compared to PCA and several other methods, CQFM can be more effective in modeling asymmetric data due to its capability of adapting to data at multiple quantile positions. Asymptotic distributional theory and an information criterion for consistent factor number selection are also discussed. PCA-based method is popular for factor analysis, and CQFM will be a useful addition to a researcher's toolkit when handling non-normal data.
	
	Many extensions of the current research are possible, and we give two examples that are highly relevant to data modeling. One is the creation of sparsity in CQFM. Adding penalty functions to \cref{eq: CQFM obj} gives
	\begin{equation} \label{eq: CQFM obj with penalty}
		\frac{1}{NT} \sum_{k}^{K}\sum_{i=1}^{N}\sum_{t=1}^{T} \rho_{\tau_k}(Y_{it} -b_{\tau_{k}}- \lambda_i'F_t) + \text{penalty}(F) + \text{penalty}(\Lambda).
	\end{equation}
	\cite{zouandyuan2008cqr} use the adaptive lasso in \cite{zou2006jasaadplasso} to induce sparsity in linear regression, and many other penalty functions are available for $F$ and $\Lambda$. The other example, following the work in \cite{bai2009ecma}, is to add a regression component to \cref{eq: CQFM obj} so that it becomes the panel data model with interactive fixed effects 
	\begin{equation} \label{eq: CQFM obj with regression}
		\frac{1}{NT} \sum_{k}^{K}\sum_{i=1}^{N}\sum_{t=1}^{T} \rho_{\tau_k}(Y_{it} -b_{\tau_{k}}- X_{it}'\beta - \lambda_i'F_t),
	\end{equation}
	where  $X_{it}$ is a vector of regressors. The model in \cref{eq: CQFM obj with regression} is a hybrid of CQR in \cite{zouandyuan2008cqr} and CQFM in the current paper. Given the good finite sample properties of CQR and CQFM under certain non-normal data, we expect estimators from \cref{eq: CQFM obj with regression} will also show some robustness to non-normal data. We leave these topics for future research.

	
	\spacing{1.45}
	\bibliographystyle{ecca}
	\bibliography{reference}
	
\newpage
\setcounter{page}{1}
\spacing{1.42}

	\begin{appendices}
		{\centering \title{\large\MakeUppercase{Supplementary Material to \\``Composite Quantile Factor Model"}\footnote{Email: xhuang3@kennesaw.edu.}\\[10pt]} }
		\begin{center}
			\large
			\author{\textsc{Xiao Huang}}  \\
			\date{\today}
		\end{center}
		\maketitle
		
		\bigskip    
		
		This supplement contains all lemmas and proofs for the theorems, as well as additional figures and tables. 
		
		\setstretch{2}
		\bigskip
		
		\localtableofcontents
		
		\newpage 
		\setstretch{1.35}
		\setcounter{section}{19}
		\setcounter{equation}{0}
		\setcounter{figure}{0}
		\setcounter{table}{0}
		\renewcommand{\theequation}{S.\arabic{equation}}
		\renewcommand\thefigure{S.\arabic{figure}} 
		\renewcommand\thetable{S.\arabic{table}} 
		
		\subsection{Lemmas} \label{supp: lemmas}
		
		This section discusses two lemmas that are used in proving the asymptotic distribution of the estimated factors and factor loadings. Let the parameter vectors be $\theta_0 = (b_{0\tau_{1}}, \cdots, b_{0\tau_{K}}, \lambda_{01}',\\ \cdots,\lambda_{0N}', F_{01}',\cdots,F_{0T}')'$ and $\hat{\theta} = (\hat{b}_{\tau_{1}}, \cdots, \hat{b}_{\tau_{k}}, \hat{\lambda}_1',\cdots,\hat{\lambda}_N', \hat{F}_1',\cdots,\hat{F}_T')'$. Let $\left\Vert\cdot\right\Vert$ be the $\ell_2$ norm. We will repeatedly apply the identity in \cite{knight1998aosl1}: for two variables $x$ and $y$ and a quantile position $\tau_k$, we have $ \rho_{\tau_k}(x-y) - \rho_{\tau_k}(x) = y(\mathbf{I}(x<0)-\tau_{k}) + \int_{0}^{y}[\mathbf{I}(x \leq s) - \mathbf{I}(x \leq 0)]ds$. Define
		\begin{align} \label{eq: metric}
			d(\hat{\theta},\theta_0) = \sqrt{\frac{1}{NT}\sum_{i=1}^{T}\sum_{t=1}^{T}\sum_{k=1}^{K}( \hat{b}_{\tau_{k}}+ \hat{\lambda}_i'\hat{F}_t  -b_{\tau_{0k}}-\lambda_{0i}'F_{0t} )^2}
		\end{align}
		and 
		\begin{align} \label{eq: Wnt}
			W_{NT} &= \frac{1}{NT}\sum_{i=1}^{N}\sum_{t=1}^{T} \sum_{k=1}^{K} \left[\rho_{\tau_k}(Y_{it} - \hat{b}_{\tau_{k}} - \hat{\lambda}_i'\hat{F}_t) - \rho_{\tau_k}(Y_{it} - b_{0\tau_{k}} - \lambda_{0i}'F_{0t}) \right] \nonumber\\
			&\quad-\frac{1}{NT}\sum_{i=1}^{N}\sum_{t=1}^{T}\sum_{k=1}^{K} E\left[\rho_{\tau_k}(Y_{it} - \hat{b}_{\tau_{k}} - \hat{\lambda}_i'\hat{F}_t) - \rho_{\tau_k}(Y_{it} - b_{0\tau_{k}} - \lambda_{0i}'F_{0t}) \right].
		\end{align}
		We will show both $d(\hat{\theta},\theta_0)$ and $W_{NT}$ are $o_p(1)$. Let $c$ be a positive constant.
		
		\begin{lemma} \label{lemma: distance}
			Under \Cref{asump: factor and facor loading,asump: error density,asump: iid},  $d(\hat{\theta},\theta_0) = o_p(1)$ and $ W_{NT} = o_p(1)$ as $N,T \rightarrow \infty$.
		\end{lemma}
		
		\begin{proof}[Proof of \Cref{lemma: distance}]
			Consider the expansion of the term $E[\rho_{\tau_k}(Y_{it} - \hat{b}_{\tau_{k}} - \hat{\lambda}_i'\hat{F}_t) - \rho_{\tau_k}(Y_{it} - $ $b_{0\tau_{k}} - \lambda_{0i}'F_{0t}) ] $ in \cref{eq: Wnt} around the value $c_{0,it} = b_{0\tau_{k}} + \lambda_{0i}'F_{0t}$. An application of the identity in \cite{knight1998aosl1} and the mean value theorem gives
			\begin{align} \label{eq:mvexpansion}
				&E\left[\rho_{\tau_k}(Y_{it} - \hat{b}_{\tau_{k}} - \hat{\lambda}_i'\hat{F}_t) - \rho_{\tau_k}(Y_{it} - b_{0\tau_{k}} - \lambda_{0i}'F_{0t}) \right] = \frac{1}{2} f_{\varepsilon}(c_{it,k}^*)(\hat{b}_{\tau_{k}} + \hat{\lambda}_i'\hat{F}_t -  b_{0\tau_{k}} - \lambda_{0i}'F_{0t})^2 \nonumber\\
				& \geq c (\hat{b}_{\tau_{k}} + \hat{\lambda}_i'\hat{F}_t -  b_{0\tau_{k}} - \lambda_{0i}'F_{0t})^2 \geq 0.
			\end{align}
			where $ c_{it,k}^*$ is between $b_{0\tau_{k}} + \lambda_{0i}'F_{0t} $ and $ \hat{b}_{\tau_{k}} + \hat{\lambda}_i'\hat{F}_t$. Rearrange terms in \cref{eq: Wnt} to have
			\begin{align} \label{eq: Wnt inequality}
				&W_{NT} + \frac{1}{NT}\sum_{i=1}^{N}\sum_{t=1}^{T}\sum_{k=1}^{K} E\left[\rho_{\tau_k}(Y_{it} - \hat{b}_{\tau_{k}} - \hat{\lambda}_i'\hat{F}_t) - \rho_{\tau_k}(Y_{it} - b_{0\tau_{k}} - \lambda_{0i}'F_{0t}) \right] \nonumber\\
				&=\frac{1}{NT}\sum_{i=1}^{N}\sum_{t=1}^{T} \sum_{k=1}^{K} \left[\rho_{\tau_k}(Y_{it} - \hat{b}_{\tau_{k}} - \hat{\lambda}_i'\hat{F}_t) - \rho_{\tau_k}(Y_{it} - b_{0\tau_{k}} - \lambda_{0i}'F_{0t}) \right] \leq 0,
			\end{align}
			where the inequality in \cref{eq: Wnt inequality} holds because $( \hat{b}_{\tau_{k}}, \hat{\lambda}_i', \hat{F}_t')$ is the minimizer of \cref{eq: CQFM obj}.
			Combining \cref{eq:mvexpansion} and \cref{eq: Wnt inequality} gives
			\begin{equation*}
				0 \leq d^2(\hat{\theta},\theta_0) \leq \underset{\hat{\theta }\in \Theta}{\sup} |W_{NT}(\hat{\theta})|,
			\end{equation*}
			which is essentially the same as the last inequality in \citet[p.~895]{chenetal2021qfm}. The proof of $\underset{\hat{\theta} \in \Theta}{\sup} |W_{NT}(\hat{\theta})| = o_p(1)$ follows the same steps in \cite{chenetal2021qfm} and is omitted.
		\end{proof}

		\begin{lemma} \label{lemma: factor product convergence}
			Under \Cref{asump: factor and facor loading,asump: error density,asump: iid}, as $N,T \rightarrow \infty$,
			\begin{equation*}
				\frac{1}{\sqrt{N}} \left \Vert \hat{\Lambda} - \Lambda_0 \right \Vert = o_p(1) \text{ and }\frac{1}{\sqrt{T}}\left \Vert \hat{F} -  F_0 \right \Vert = o_p(1).
			\end{equation*}
		\end{lemma}
		
		\begin{proof}[Proof of \Cref{lemma: factor product convergence}]
			\Cref{lemma: distance} proves that $ d^2(\hat{\theta},\theta_0) = o_p(1)$, which implies that, for every $k$,
			
			\begin{equation} \label{eq: fitted distance}
				\frac{1}{NT}\sum_{i=1}^{N}\sum_{t=1}^{T}( \hat{b}_{\tau_{k}}+ \hat{\lambda}_i'\hat{F}_t  -b_{\tau_{0k}}-\lambda_{0i}'F_{0t} )^2 = o_p(1).
			\end{equation}
			
			Use the inequality $\frac{1}{2}(x^2 + y^2) \leq (x+y)^2$, \cref{eq: fitted distance} gives that, for each $k$,
			\begin{align} 
				&\frac{1}{2} \left[ \frac{1}{NT}\sum_{i=1}^{N}\sum_{t=1}^{T}\left(\hat{b}_{\tau_k}-b_{\tau_{k}}\right)^2 +  \frac{1}{NT}\sum_{i=1}^{N}\sum_{t=1}^{T} \left(\hat{\lambda}_i'\hat{F}_t - \lambda_{0i}'F_{0t}\right)^2 \right] \nonumber \\
				&\leq \frac{1}{NT}\sum_{i=1}^{N}\sum_{t=1}^{T}( \hat{b}_{\tau_{k}}+ \hat{\lambda}_i'\hat{F}_t  -b_{\tau_{0k}}-\lambda_{0i}'F_{0t} )^2 = o_p(1). \label{eq: fitted distance inequality}
			\end{align}
			It follows that
			\begin{equation*}
				\frac{1}{NT}\sum_{i=1}^{N}\sum_{t=1}^{T} \left(\hat{\lambda}_i'\hat{F}_t - \lambda_{0i}'F_{0t}\right)^2 = o_p(1),
			\end{equation*}
			which is equivalent to 
			\begin{equation} \label{eq: fitted op1}
				\frac{1}{\sqrt{NT} } \left \Vert \hat{F} \hat{\Lambda}' - F_0 \Lambda_0 \right \Vert = o_p(1).
			\end{equation}
			Define $P_{\hat{\Lambda}} = \hat{\Lambda}(\hat{\Lambda}'\hat{\Lambda})^{-1}\hat{\Lambda}'$, $M_{\hat{\Lambda}} = I_N - P_{\hat{\Lambda}}$, $P_{\hat{F}} = \hat{F}(\hat{F}'\hat{F})^{-1}\hat{F}'$, and $M_{\hat{F}} = I_T - P_{\hat{F}}$.
			Since multiplying $M_{\hat{\Lambda}}$ shrinks its $\ell_2$ norm, we have
			\begin{equation} \label{eq: M shrink}
				\frac{1}{\sqrt{NT}} \left \Vert \left(\hat{F}\hat{\Lambda}' - F_0\Lambda_0\right) M_{\hat{\Lambda}} \right \Vert \leq \frac{1}{\sqrt{NT}} \left \Vert \hat{F}\hat{\Lambda}' - F_0\Lambda_0 \right \Vert = o_p(1).
			\end{equation}
			With $ \hat{F}\hat{\Lambda}' M_{\hat{\Lambda}} = 0$, the above result implies that $\frac{1}{\sqrt{NT}} \left \Vert F_0 \Lambda_0' M_{\hat{\Lambda}} \right \Vert = o_p(1)$, i.e.,
			\begin{equation*} 
				\frac{1}{NT} \text{tr}\left(M_{\hat{\Lambda}} \Lambda_0 F_0' F_0 \Lambda_0' M_{\hat{\Lambda}}\right) = \text{tr}\left(\frac{F_0' F_0}{T} \frac{\Lambda_0'M_{\hat{\Lambda}}\Lambda_0}{N}\right) = o_p(1).
			\end{equation*}
			Given the normalization condition $F_0'F_0/T = I_r$, we have 
			\begin{equation} \label{eq: trace op 1}
				\text{tr}\left(\frac{\Lambda_0'M_{\hat{\Lambda}}\Lambda_0}{N}\right) = o_p(1) \text{ or } \frac{1}{N} \left\Vert  M_{\hat{\Lambda}} \Lambda_0\right\Vert^2 = o_p(1).
			\end{equation}
			Expanding $ M_{\hat{\Lambda}}$ gives
			\begin{equation} \label{eq: trace op 2}
				\frac{\Lambda_0'M_{\hat{\Lambda}}\Lambda_0}{N} = \frac{\Lambda_0' \Lambda_0}{N} -  \frac{\Lambda_0'P_{\hat{\Lambda}} \Lambda_0}{N} =  \frac{\Lambda_0' \Lambda_0}{N} -  \frac{\Lambda_0' \hat{\Lambda} \hat{\Lambda}' \Lambda_0}{N} \frac{\Sigma_{\hat{\lambda}}^{-1}}{N}.
			\end{equation}
			Combining \cref{eq: trace op 1} and \cref{eq: trace op 2} gives 
			\begin{equation} \label{eq: trace op 3}
				\text{tr}(\Lambda_0'\hat{\Lambda}\hat{\Lambda}' \Lambda_0) \overset{p}{\rightarrow} \text{tr}(\Lambda_0'\Lambda_0 \cdot N\Sigma_{\hat{\lambda}}) =N^2 \text{tr}(\Sigma_{\lambda_0} \Sigma_{\hat{\lambda}}).
			\end{equation}
			Consider the following.
			\begin{align} 
				\left \Vert P_{\hat{\Lambda}} - P_{\Lambda_0}\right \Vert^2 &= \text{tr}\left(  (P_{\hat{\Lambda}} - P_{\Lambda_0})^2   \right) \nonumber \\
				&=\frac{1}{N} \text{tr}(\hat{\Lambda} \hat{\Lambda}' \Sigma_{\hat{\lambda}}^{-1} + \Lambda_0 \Lambda_0' \Sigma_{\lambda_0}^{-1}) - \frac{2}{N^2} \text{tr}(\hat{\Lambda} \hat{\Lambda}' \Lambda_0 \Lambda_0' \Sigma_{\hat{\lambda}}^{-1} \Sigma_{\lambda_0}^{-1} ) \nonumber \\
				&\overset{p}{\rightarrow} 2 \text{tr}(I_r) - 2 \text{tr}(I_r) \tag*{\text{(Using \cref{eq: trace op 3}})} \nonumber \\
				&= 0. \label{eq: P lambda mat convergence}
			\end{align}
			Thus, we have
			\begin{align}
				\frac{1}{\sqrt{N}} \left \Vert  M_{\Lambda_0} \hat{\Lambda}  \right \Vert &= \frac{1}{\sqrt{N}} \left \Vert  (M_{\Lambda_0} - M_{\hat{\Lambda}})\hat{\Lambda}  \right \Vert = \frac{1}{\sqrt{N}} \left \Vert  (P_{\Lambda_0} - P_{\hat{\Lambda}})\hat{\Lambda}  \right \Vert \nonumber \\
				&\leq 	\left \Vert P_{\hat{\Lambda}} - P_{\Lambda_0}\right \Vert \frac{1}{\sqrt{N}} \left \Vert \hat{\Lambda} \right \Vert = o_p(1) O(1) = o_p(1),
			\end{align}
			where the result $\left \| \hat{\Lambda} \right \| /\sqrt{N}= O(1)$ follows the normalization condition $\hat{\Lambda}' \hat{\Lambda}/N = \Sigma_{\hat{\lambda}}$.
			It follows that
			\begin{equation*}
				\frac{1}{\sqrt{N}} \left \Vert  M_{\Lambda_0} \hat{\Lambda}  \right \Vert = \frac{1}{\sqrt{N}} \left \| \hat{\Lambda} - \Lambda_0 (\Lambda_0'\Lambda_0)^{-1} \Lambda_0' \hat{\Lambda} \right \| = \frac{1}{\sqrt{N}} \left \| \hat{\Lambda} - \Lambda_0 G \right \| = o_p(1),
			\end{equation*}
			where $G$ is the rotation matrix. The normalization conditions in \cref{eq: normalization} and \cref{asump: factor and facor loading} suggest the factor loading can be identified, and we can ignore the rotation matrix for simplicity purposes, and the above result becomes $\left \| \hat{\Lambda} - \Lambda_0 \right \| / \sqrt{N} = o_p(1)$. 
			
			Next, we establish a similar result for the factors. Similar to \cref{eq: M shrink}, we have
 			\begin{equation} \label{eq: M shrink factor}
 				\frac{1}{\sqrt{NT}} \left \Vert  M_{\hat{F}}   \left(\hat{F}\hat{\Lambda}' - F_0\Lambda_0\right)  \right \Vert \leq \frac{1}{\sqrt{NT}} \left \Vert \hat{F}\hat{\Lambda}' - F_0\Lambda_0 \right \Vert = o_p(1).
 			\end{equation}
 			Because $ M_{\hat{F}} \hat{F} \hat{\Lambda}' = 0$, we have $\frac{1}{\sqrt{NT}} \left \Vert M_{\hat{F}}  F_0 \Lambda_0' \right \Vert = o_p(1)$, which is equivalent to 
		\begin{equation*}
				\frac{1}{T} \text{tr}\left(M_{\hat{F}} F_0 \frac{\Lambda_0' \Lambda_0}{N} F_0' M_{\hat{F}}\right) = \text{tr} \left(\frac{\Lambda_0'\Lambda_0}{N} \frac{F_0' M_{\hat{F}} F_0}{T}\right) =  o_p(1).
		\end{equation*}
		Given the matrix $\frac{\Lambda_0' \Lambda_0}{N}$ is diagonal, we have $\frac{1}{T}\left\Vert M_{\hat{F}} F_0 \right\Vert^2 = o_p(1)$. Since 
		\begin{equation} \label{eq: MF decomp}
			\frac{F_0' M_{\hat{F}} F_0 }{T} = \frac{F_0' F_0}{T} - \frac{F_0'\hat{F}}{T} \frac{\hat{F}'F_0}{T},
		\end{equation}
		and we conclude that $\text{tr}(\frac{F_0'\hat{F}}{T} \frac{\hat{F}'F_0}{T}) \overset{p}{\rightarrow} \text{tr}(\frac{F_0' F_0}{T} ) = r$. Using an argument similar to that in \cref{eq: P lambda mat convergence}, we obtain  $\left \Vert P_{\hat{F}} - P_{F_0} \right \Vert = o_p(1)$ (also see \citet[p.~1265]{bai2009ecma} for a similar proof).  Consequently, we have
		\begin{align} \label{eq: MFhat}
			\frac{1}{\sqrt{T}} \left\Vert M_{F_0} \hat{F} \right\Vert &= \frac{1}{\sqrt{T}} \left\Vert \left(M_{F_0} - M_{\hat{F}}  \right)\hat{F} \right\Vert = \frac{1}{\sqrt{T}} \left\Vert \left(P_{F_0} - P_{\hat{F}}  \right)\hat{F} \right\Vert \nonumber\\
			& \leq  \left\Vert \left(P_{F_0} - P_{\hat{F}}  \right)\right\Vert \frac{1}{\sqrt{T}} \left\Vert \hat{F} \right\Vert = o_p(1) \cdot O_p(1) = o_p(1),
		\end{align}
		where the result $  \frac{1}{\sqrt{T}} \left\Vert \hat{F} \right\Vert = O_p(1)$ follows the normalization in \cref{eq: normalization}. Rewrite \cref{eq: MFhat} to have
		\begin{equation*}
			\frac{1}{\sqrt{T}} \left\Vert M_{F_0} \hat{F} \right\Vert = \frac{1}{\sqrt{T}} \left\Vert \hat{F} -  F_0\left(F_0'F_0\right)^{-1}F_0' \hat{F} \right\Vert = \frac{1}{\sqrt{T}} \left\Vert \hat{F} -  F_0 H\right\Vert = o_p(1),
		\end{equation*} 
		where the rotation matrix is $H = \left(F_0'F_0\right)^{-1}F_0' \hat{F} $. Since the normalization condition indicates that factors are identifiable, similar to the approach in \cite{andoandbai2020jasa}, we can omit the rotation matrix $H$ for simplicity purposes and have
		$\frac{1}{\sqrt{T}} \left\Vert \hat{F} -  F_0 \right\Vert = o_p(1)$. 
	\end{proof}
	
%
		
		\subsection{Proof of \Cref{thm: asymptotic distribution}} \label{supp: theroem 1 proof}    
		
		\begin{proof}[Proof of \Cref{thm: asymptotic distribution}]
			Define $w_{NT,k} = \sqrt{NT}(\hat{b}_{\tau_{k}} - b_{0\tau_{k}})$, $u_{T,i} = \sqrt{T}(\hat{\lambda}_i - \lambda_{0i})$ and $v_{N,t} = \sqrt{N}(\hat{F}_t - F_{0t})$.  Ignoring the scaling factor $\frac{1}{NT}$, the objective function in \cref{eq: CQFM obj} can be modified as follows:
			\begin{align} \label{eq:loss func diff 0}
				\mathcal{L}_{NT} &= \sum_{k=1}^{K}\sum_{i=1}^{N}\sum_{t=1}^{T} \left[\rho_{\tau_k}\left(\varepsilon_{it} -b_{0\tau_{k}} - \frac{w_k + u_i'v_t + \sqrt{N}u_i'F_{0t} + \sqrt{T}\lambda_{0i}'v_t}{\sqrt{NT}} \right) -  \rho_{\tau_k}(\varepsilon_{it} -b_{0\tau_{k}})\right] ,
			\end{align}
			whose minimizer is $\left\{w_{NT,k}, u_{T,i},v_{N,t} \right\}$, and it can be verified by their substitution in \cref{eq:loss func diff 0}. Using the identity in \cite{knight1998aosl1} gives
			\begin{align}
				\mathcal{L}_{NT} &= \sum_{k=1}^{K} \frac{1}{\sqrt{NT}} \sum_{i=1}^{N}\sum_{t=1}^{T}\left(I(\varepsilon_{it} < \tau_{0k}) - \tau_{k}\right) w_k +\frac{1}{\sqrt{NT}} \sum_{i=1}^{N}\sum_{t=1}^{T} u_i'v_t \sum_{k=1}^{K} \left(I(\varepsilon_{it} < \tau_{0k}) - \tau_{k}\right) \nonumber \\
				&\quad+ \sum_{i=1}^{N}u_i' \frac{1}{\sqrt{T}} \sum_{t=1}^{T} F_{0t} \left[\sum_{k=1}^{K}I(\varepsilon_{it} < \tau_{0k}) - \tau_{k}\right] \nonumber\\
				&\quad+ \sum_{t=1}^{T} v_t' \frac{1}{\sqrt{N}}\sum_{i=1}^{N}\lambda_{0i} \left[\sum_{k=1}^{K}I(\varepsilon_{it} < \tau_{0k}) - \tau_{k}\right] + \sum_{k=1}^{K} \mathcal{B}_{NT,k} \label{eq:loss func diff 01}\\
				&= \sum_{k=1}^{K}z_{NT,k}w_k + \frac{1}{\sqrt{NT}} \sum_{i=1}^{N}\sum_{t=1}^{T} u_i'v_t \sum_{k=1}^{K} \left(I(\varepsilon_{it} < \tau_{0k}) - \tau_{k}\right) + \sum_{i=1}^{N}u_i'z_{T,F_0} \nonumber \\
				&\quad + \sum_{t=1}^{T}v_t' z_{N,\lambda_0} + \sum_{k=1}^{K}\mathcal{B}_{NT,k},  \label{eq:loss func diff 02}
			\end{align}
			where
			\begin{align} 
				z_{NT,k} &= \frac{1}{\sqrt{NT}} \sum_{i=1}^{N}\sum_{t=1}^{T}\left(I(\varepsilon_{it} < \tau_{0k}) - \tau_{k}\right), \\
				z_{T,F_0} &= \frac{1}{\sqrt{T}} \sum_{t=1}^{T} F_{0t} \left[I(\varepsilon_{it} < \tau_{0k}) - \tau_{k}\right],\\
				z_{N,\lambda_0} &= \frac{1}{\sqrt{N}}\sum_{i=1}^{N}\lambda_{0i} \left[I(\varepsilon_{it} < \tau_{0k}) - \tau_{k}\right],\\
				\mathcal{B}_{NT,k} &= \sum_{i=1}^{N}\sum_{t=1}^{T} \int_{0}^{\frac{w_k + u_i'v_t + \sqrt{N}u_i'F_{0t} + \sqrt{T}\lambda_{0i}'v_t}{\sqrt{NT}}  } \big(I(\varepsilon_{it} - b_{0\tau_{k}} \leq s) - I(\varepsilon_{it} - b_{0\tau_{k}} \leq 0)\big)ds. \label{eq:Bnt 0}
			\end{align}
			Under the i.i.d. error assumption and the implied moment conditions from the normalization conditions, $\left\{z_{NT,k},z_{T,F_0}', z_{N,\lambda_0}'\right\} \overset{d}{\rightarrow} \left\{z_k, z_{F_0}', z_{\lambda_0}'\right\}$ with a multivariate normal distribution as $N,T\rightarrow \infty$. The second term in \cref{eq:loss func diff 02}, $\frac{1}{\sqrt{NT}} \sum_{i=1}^{N}\sum_{t=1}^{T} u_i'v_t \sum_{k=1}^{K} \left(I(\varepsilon_{it} < \tau_{0k}) - \tau_{k}\right)$, has a smaller probability order than, for example, the fourth term in \cref{eq:loss func diff 02}. Both $u_i$ and $v_t$ are $O_p(1)$ since they are parameter values in optimization; $\lambda_{0i}$ is either $O(1)$ or $O_p(1)$ if we treat it as a random variable. With an extra $\sqrt{T}$ on the denominator, the second term has a smaller order in probability and can be ignored. See a related argument in our later proof that uses the results in \Cref{lemma: factor product convergence}. Let $\mathbf{u} = (u_1',\cdots,u_N')'$, $\mathbf{v} = (v_1',\cdots,v_T')'$ and $\tilde{\theta} = \left(w_1,\cdots,w_k, \mathbf{u}',\mathbf{v}'\right)$. The value of $\tilde{\theta}$ at $(b_{0\tau_{k}}, \lambda_{0i}, F_{0t})$ is $\tilde{\theta}_0 = \mathbf{0}$. The second-order Taylor series expansion of $E(\mathcal{B}_{NT,k})$ at $\tilde{\theta}_0$ becomes
			\begin{align}
				E\left(B_{NT,k}\right) &=\sum_{i=1}^{N}\sum_{t=1}^{T} \int_{0}^{\frac{w_k + u_i'v_t + \sqrt{N}u_i'F_{0t} + \sqrt{T}\lambda_{0i}'v_t}{\sqrt{NT}}  } \big(F(b_{0\tau_{k}} + s) - F(b_{0\tau_{k}} \leq 0)\big)ds \nonumber \\
				&=\frac{1}{2} f_{\varepsilon}(b_{0\tau_{k}})\sum_{i=1}^{N}\sum_{t=1}^{T} (w_k,u_i',v_t')
				\begin{bmatrix}
					\frac{1}{NT} & \frac{1}{\sqrt{NT}} \frac{1}{\sqrt{T}} F_{0t}' & \frac{1}{\sqrt{NT}}\lambda_{0i}'\\
					\frac{1}{\sqrt{T}} F_{0t} & \frac{1}{T} F_{0t}F_{0t}'  & \frac{1 }{\sqrt{T}} F_{0t} \lambda_{0i}'\\
					\frac{1}{\sqrt{NT}} \frac{1}{\sqrt{N}} \lambda_{0i} & \frac{1}{\sqrt{NT}} \lambda_{0i} F_{0t}' & \frac{1}{N} \lambda_{0i} \lambda_{0i}' 
				\end{bmatrix}
				(w_k,u_i',v_t')' \label{eq:EB1} \\
				&= \frac{1}{2} f_{\varepsilon}(b_{0\tau_{k}})(w_k,u_i',v_t')
				\begin{bmatrix}
					1 & 0 & 0\\
					0 & I_r & 0\\
					0 & 0 &\Sigma_{\lambda_0} 
				\end{bmatrix}
				(w_k,u_i',v_t')' + o(1),  \label{eq:EB2}
			\end{align}
			where the second equality holds because the first derivative of $E(\mathcal{B}_{NT}^{(k)})$ w.r.t. $\tilde{\theta}$ is $0$ when evaluated at $\tilde{\theta}_0$. The third equality follows the normalization conditions. For example, since $\left\| F_0 \right\|/\sqrt{T} = O_p(1)$, each element of $F_{0t}$ is $O_p(1)$ and $F_{0t}/\sqrt{T} \rightarrow 0$ as $T \rightarrow \infty$.
			
			The variance of $E(B_{NT}^{(k)})$ is given by
				\begin{align*}
				\text{Var}(B_{NT,k}) &= \sum_{i=1}^{N}\sum_{t=1}^{T} E\bigg(\int_{0}^{\frac{w_k + u_i'v_t + \sqrt{N}u_i'F_{0t} + \sqrt{T}\lambda_{0i}'v_t}{\sqrt{NT}}  }  \nonumber\\
				&\quad \big(I(\varepsilon_{it} - b_{0\tau_{k}} \leq s) - I(\varepsilon_{it} - b_{0\tau_{k}} \leq 0) - F_{\varepsilon}(b_{0\tau_{k}} + s ) + F_{\varepsilon}(b_{0\tau_{k}})\big)ds \bigg)^2 \nonumber
				\end{align*}
				\begin{align} \label{eq:Bnt0 var}
				&\leq \sum_{i=1}^{N}\sum_{t=1}^{T} E \bigg( \bigg |  \mathrlap{\int_{0}^{\frac{w_k + u_i'v_t + \sqrt{N}u_i'F_{0t} + \sqrt{T}\lambda_{0i}'v_t}{\sqrt{NT}}  }}
				\qquad \big(I(\varepsilon_{it} - b_{0\tau_{k}} \leq s) - I(\varepsilon_{it} - b_{0\tau_{k}} \leq 0) - F_{\varepsilon}(b_{0\tau_{k}} + s ) + F_{\varepsilon}(b_{0\tau_{k}})\big)ds  \bigg | \bigg) \nonumber \\
				&\quad \times 2 \Big | \frac{w_k + u_i'v_t + \sqrt{N}u_i'F_{0t} + \sqrt{T}\lambda_{0i}'v_t}{\sqrt{NT}} \Big | \nonumber \\
				&\quad \leq 2 E(B_{NT,k}) \times 2 \max_{i,t} \left | \frac{w_k}{\sqrt{NT}} + \frac{u_i'v_t}{\sqrt{NT}} + \frac{u_i'F_{0t}}{\sqrt{T}} + \frac{\lambda_{0i}'v_t}{\sqrt{N}} \right | \rightarrow 0 \text{ as } N,T \rightarrow \infty,
			\end{align}
			and we have $B_{NT,k}$ converges in probability to the first term in \cref{eq:EB2}.
			Let $C$ be the constant Hessian in \cref{eq:EB2}. Combining the above results gives
			\begin{equation} \label{eq:limit quadratic form}
				\mathcal{L}_{NT} \overset{d}{\rightarrow} \sum_{k=1}^{K} z_k w_k + \sum_{i=1}^{N} z_{F_0}' u_i + \sum_{t=1}^{T} z_{\lambda_0}' v_t +\frac{1}{2}\sum_{k=1}^{K}f_{\varepsilon}(b_{0\tau_{k}}) \sum_{i=1}^{N}\sum_{t=1}^{T}(w_k,u_i',v_t')C(w_k,u_i',v_t')', 
			\end{equation}
			a quadratic form in $ (w_k,u_i',v_t')$. Hence, the minimizer of $\mathcal{L}_{NT}$ will also converges in distribution to the minimizer of the quadratic form in \cref{eq:limit quadratic form}, based on which we can derive the asymptotic distribution of the minimizer $\left\{w_{NT,k}, u_{T,i},v_{N,t} \right\}$.
			
			In the following, we work with a scaled version of \cref{eq:loss func diff 0} and $\{ \hat{b}_{\tau_{k}} - b_{0\tau_{k}}, \hat{\lambda}_i - \lambda_{0i}, \hat{F}_t - F_{0t}  \}$ directly to see their asymptotic distributions. Consider the following loss function
			\begin{align} \label{eq:loss func diff 1}
				L_{NT} &= \frac{1}{NT} \sum_{k=1}^{K}\sum_{i=1}^{N}\sum_{t=1}^{T} \left[\rho_{\tau_k}(Y_{it} -\hat{b}_{\tau_{k}}- \hat{\lambda}_i'\hat{F}_t) - \rho_{\tau_k}(Y_{it} -b_{0\tau_{k}}- \lambda_{0i}'F_{0t})\right] \nonumber \\
				&= \frac{1}{NT} \sum_{k=1}^{K}\sum_{i=1}^{N}\sum_{t=1}^{T} \left[\rho_{\tau_k}\left(\varepsilon_{it} -b_{0\tau_{k}} - \left(\hat{b}_{\tau_{k}} - b_{0\tau_{k}}\right)- \left(\hat{\lambda}_i'\hat{F}_t - \lambda_{0i}'F_{0t}\right) \right) -  \rho_{\tau_k}(\varepsilon_{it} -b_{0\tau_{k}})\right] 
			\end{align}
			Use the identity in \cite{knight1998aosl1} again to have
			\begin{align} \label{eq:loss func diff 2}
				L_{NT} &= \frac{1}{NT} \sum_{k=1}^{K}\sum_{i=1}^{N}\sum_{t=1}^{T} \Big[ \big((\hat{b}_{\tau_{k}} - b_{0\tau_{k}}) + (\hat{\lambda}_i'\hat{F}_t - \lambda_{0i}'F_{0t}) \big) \cdot \big(I(\varepsilon_{it} - b_{0\tau_{k}} < 0) - \tau_{k}\big) + \nonumber\\
				&\quad \int_{0}^{(\hat{b}_{\tau_{k}} - b_{0\tau_{k}}) + (\hat{\lambda}_i'\hat{F}_t - \lambda_{0i}'F_{0t}) } \big(I(\varepsilon_{it} - b_{0\tau_{k}} \leq s) - I(\varepsilon_{it} - b_{0\tau_{k}} \leq 0)\big)ds  \Big] = \mathrm{I} + \mathrm{II},
			\end{align}
			where
			\begin{align}
				\mathrm{I} &= \frac{1}{NT} \sum_{k=1}^{K}\sum_{i=1}^{N}\sum_{t=1}^{T} \big((\hat{b}_{\tau_{k}} - b_{0\tau_{k}}) + (\hat{\lambda}_i'\hat{F}_t - \lambda_{0i}'F_{0t}) \big) \cdot \big(I(\varepsilon_{it} - b_{0\tau_{k}} < 0) - \tau_{k}\big), \label{eq:I term} 
			\end{align}
			\begin{align}
				\mathrm{II} &= \frac{1}{NT} \sum_{k=1}^{K}\sum_{i=1}^{N}\sum_{t=1}^{T} \int_{0}^{(\hat{b}_{\tau_{k}} - b_{0\tau_{k}}) + (\hat{\lambda}_i'\hat{F}_t - \lambda_{0i}'F_{0t}) } \big(I(\varepsilon_{it} - b_{0\tau_{k}} \leq s) - I(\varepsilon_{it} - b_{0\tau_{k}} \leq 0)\big)ds. \label{eq:II term}
			\end{align}
			
			Using $\hat{\lambda}_i'\hat{F}_t - \lambda_{0i}'F_{0t} = (\hat{\lambda}_i - \lambda_{0i})'(\hat{F}_t - F_{0t}) + (\hat{\lambda}_i - \lambda_{0i})'F_{0t} + \lambda_{0i}'(\hat{F}_t - F_{0t})$,
			\cref{eq:I term,eq:II term} become 
			\begin{align}
					\mathrm{I} &= \frac{1}{NT} \sum_{k=1}^{K}\sum_{i=1}^{N}\sum_{t=1}^{T} \big((\hat{b}_{\tau_{k}} - b_{0\tau_{k}}) + (\hat{\lambda}_i - \lambda_{0i})'(\hat{F}_t - F_{0t}) + (\hat{\lambda}_i - \lambda_{0i})'F_{0t} + \lambda_{0i}'(\hat{F}_t - F_{0t})  \big) \cdot \nonumber\\
					&\quad \big(I(\varepsilon_{it} - b_{0\tau_{k}} < 0) - \tau_{k}\big), \label{eq:I term2} \\
				\mathrm{II} &= \frac{1}{NT} \sum_{k=1}^{K}\sum_{i=1}^{N}\sum_{t=1}^{T} \mathrlap{\int_{0}^{(\hat{b}_{\tau_{k}} - b_{0\tau_{k}}) + (\hat{\lambda}_i - \lambda_{0i})'(\hat{F}_t - F_{0t}) + (\hat{\lambda}_i - \lambda_{0i})'F_{0t} + \lambda_{0i}'(\hat{F}_t - F_{0t})  } } 
				\qquad \big(I(\varepsilon_{it} - b_{0\tau_{k}} \leq s) - I(\varepsilon_{it} - b_{0\tau_{k}} \leq 0)\big)ds. \label{eq:II term2}
			\end{align}
			It is clear that $L_{NT}$ in \cref{eq:loss func diff 1} is identical to \cref{eq:loss func diff 0} except for the scaling factor $1/NT$.
			As $N,T \rightarrow \infty$, \cref{eq:II term2} converges to some expected value. To see this, define
			\begin{align} \label{eq:Bnt}
				B_{NT,k} &= \frac{1}{NT} \sum_{i=1}^{N}\sum_{t=1}^{T} \mathrlap{\int_{0}^{(\hat{b}_{\tau_{k}} - b_{0\tau_{k}}) + (\hat{\lambda}_i - \lambda_{0i})'(\hat{F}_t - F_{0t}) + (\hat{\lambda}_i - \lambda_{0i})'F_{0t} + \lambda_{0i}'(\hat{F}_t - F_{0t})  }} 
				\qquad \big(I(\varepsilon_{it} - b_{0\tau_{k}} \leq s) - I(\varepsilon_{it} - b_{0\tau_{k}} \leq 0)\big)ds,
			\end{align}
			where $B_{NT,k}$ is similar to $\mathcal{B}_{NT,k}$ in \cref{eq:Bnt 0}. The result in \cref{eq:EB1} implies the following expectation of $B_{NT,k}$ and its second-order Taylor series expansion evaluated at $\hat{b}_{\tau_{k}} = b_{0\tau_{k}}, \hat{\lambda}_i = \lambda_{0i},$ and $\hat{F} = F_{0t}$: 
			\begin{align} \label{eq:2nd order expansion}
				&E(B_{NT,k})=\frac{1}{NT} \sum_{i=1}^{N}\sum_{t=1}^{T}  \mathrlap{\int_{0}^{(\hat{b}_{\tau_{k}} - b_{0\tau_{k}}) + (\hat{\lambda}_i - \lambda_{0i})'(\hat{F}_t - F_{0t}) + (\hat{\lambda}_i - \lambda_{0i})'F_{0t} + \lambda_{0i}'(\hat{F}_t - F_{0t})  }} \qquad \big( F_{\varepsilon}(b_{0\tau_{k}}+s) -F_{\varepsilon}(b_{0\tau_{k}})\big)ds \nonumber\\
				& = \frac{1}{NT} \sum_{i=1}^{N}\sum_{t=1}^{T}  \frac{f_{\varepsilon}(b_{0\tau_{k}})  }{2}
				\begin{bmatrix}
					\hat{b}_{\tau_{k}} - b_{0\tau_{k}} \\
					\hat{\lambda}_i - \lambda_{0i} \\
					\hat{F}_t - F_{0t}
				\end{bmatrix}' 
				\begin{bmatrix}
					1 & F_{0t}' & \lambda_{0i}' \\
					F_{0t} & F_{0t}F_{0t}' & F_{0t} \lambda_{0i}' \\
					\lambda_{0i} & \lambda_{0i} F_{0t}' & \hat{\lambda}_i\lambda_{0i}'
				\end{bmatrix}
				\begin{bmatrix}
					\hat{b}_{\tau_{k}} - b_{0\tau_{k}} \\
					\hat{\lambda}_i - \lambda_{0i} \\
					\hat{F}_t - F_{0t}
				\end{bmatrix} + o(1).
			\end{align}
Similar to \cref{eq:Bnt0 var}, we also have $\text{Var}(B_{NT,k}) \rightarrow 0$ as $N,T\rightarrow \infty$.
			Plug \cref{eq:I term2,eq:II term2,eq:2nd order expansion} into \cref{eq:loss func diff 2} and, similar to \cref{eq:loss func diff 01}, we have
			\begin{align*} 
				&L_{NT} = \frac{1}{NT} \sum_{k=1}^{K}\sum_{i=1}^{N}\sum_{t=1}^{T} (\hat{b}_{\tau_{k}} - b_{0\tau_{k}}) \big(I(\varepsilon_{it} - b_{0\tau_{k}} < 0) - \tau_{k}\big) \nonumber 
			\end{align*}
			\begin{align} \label{eq:loss func diff3 }
				&+ \frac{1}{NT} \sum_{k=1}^{K}\sum_{i=1}^{N}\sum_{t=1}^{T} (\hat{\lambda}_i - \lambda_{0i})'(\hat{F}_t - F_{0t}) \big(I(\varepsilon_{it} - b_{0\tau_{k}} < 0) - \tau_{k}\big) \nonumber\\
				&+ \frac{1}{NT} \sum_{k=1}^{K}\sum_{i=1}^{N}\sum_{t=1}^{T} (\hat{\lambda}_i - \lambda_{0i})'F_{0t} \big(I(\varepsilon_{it} - b_{0\tau_{k}} < 0) - \tau_{k}\big) \nonumber \nonumber \\
				&+ \frac{1}{NT} \sum_{k=1}^{K}\sum_{i=1}^{N}\sum_{t=1}^{T} \lambda_{0i}'(\hat{F}_t - F_{0t}) \big(I(\varepsilon_{it} - b_{0\tau_{k}} < 0) - \tau_{k}\big) \nonumber \nonumber \\
				&+ \frac{1}{NT} \sum_{k=1}^{K}\sum_{i=1}^{N}\sum_{t=1}^{T} \frac{ f_{\varepsilon}(b_{0\tau_{k}}) }{2}
				\begin{bmatrix}
					\hat{b}_{\tau_{k}} - b_{0\tau_{k}} \\
					\hat{\lambda}_i - \lambda_{0i} \\
					\hat{F}_t - F_{0t}
				\end{bmatrix}' 
				\begin{bmatrix}
					1 & F_{0t}' & \lambda_{0i}' \\
					F_{0t} & F_{0t}F_{0t}' & F_{0t} \lambda_{0i}' \\
					\lambda_{0i} & \lambda_{0i} F_{0t}' & \hat{\lambda}_i\lambda_{0i}'
				\end{bmatrix}
				\begin{bmatrix}
					\hat{b}_{\tau_{k}} - b_{0\tau_{k}} \\
					\hat{\lambda}_i - \lambda_{0i} \\
					\hat{F}_t - F_{0t}
				\end{bmatrix} + o_p(1).
			\end{align}
			Based on \Cref{lemma: factor product convergence}, the second term involving the product $(\hat{\lambda}_i - \lambda_{0i})'(\hat{F}_t - F_{0t})$ in \cref{eq:loss func diff3 } has smaller order in probability than the third and the fourth terms in \cref{eq:loss func diff3 }, and it can be omitted in the following analysis. Next, we take the partial derivative of \cref{eq:loss func diff3 } w.r.t. $ \hat{b}_{\tau_{k}} - b_{0\tau_{k}}, \hat{\lambda}_i - \lambda_{0i},$ and $ \hat{F}_t - F_{0t}$ and set the first-order conditions to zero. 
			
			Consider the partial derivative of \cref{eq:loss func diff3 } w.r.t. $\hat{b}_{\tau_{k}} - b_{0\tau_{k}}$.
			\begin{align} \label{eq:derivative b}
				\frac{\partial L_{NT}}{\partial  (\hat{b}_{\tau_{k}} - b_{0\tau_{k}})} &= \frac{1}{NT}\sum_{i=1}^{N}\sum_{t=1}^{T} \big(I(\varepsilon_{it} - b_{0\tau_{k}} < 0) - \tau_{k}\big) + f_{\varepsilon}(b_{0\tau_{k}}) (\hat{b}_{\tau_{k}} - b_{0\tau_{k}}) \nonumber \\
				&\quad + f_{\varepsilon}(b_{0\tau_{k}}) \cdot \frac{1}{T}\sum_{t=1}^{T}F_{0t}' \cdot \frac{1}{N}\sum_{i=1}^{N} (\hat{\lambda}_i - \lambda_{0i}) + f_{\varepsilon}(b_{0\tau_{k}}) \cdot \frac{1}{N}\sum_{i=1}^{N} \lambda_{0i}' \cdot \frac{1}{T}\sum_{t=1}^{T}(\hat{F}_t - F_{0t}) \nonumber \\
				&=\frac{1}{NT}\sum_{i=1}^{N}\sum_{t=1}^{T} \big(I(\varepsilon_{it} - b_{0\tau_{k}} < 0) - \tau_{k}\big) + f_{\varepsilon}(b_{0\tau_{k}}) (\hat{b}_{\tau_{k}} - b_{0\tau_{k}}) + o_p(1),
			\end{align}
			where the third term on the right is $0$ because $\sum_{i=1}^{T}F_{0t}/T \rightarrow 0$ under \cref{asump: factor and facor loading} and the fourth term is $ o_p(1)$ because in \cref{lemma: factor product convergence} we establish the result $\frac{1}{\sqrt{T}} \left\Vert \hat{F} -  F_0 \right\Vert = o_p(1)$, which implies $  \frac{1}{T}\sum_{t=1}^{T}(\hat{F}_t - F_{0t}) = o_p(1)$.
		Setting \cref{eq:derivative b} to $0$ gives
			\begin{equation} \label{eq:b diff equation}
				\sqrt{NT} (\hat{b}_{\tau_{k}} - b_{0\tau_{k}}) = -f_{\varepsilon}(b_{0\tau_{k}})^{-1} \frac{1}{\sqrt{NT}} \sum_{i=1}^{N}\sum_{t=1}^{T} \big(I(\varepsilon_{it} - b_{0\tau_{k}} < 0) - \tau_{k}\big).
			\end{equation}
			
			Consider the partial derivative of \cref{eq:loss func diff3 } w.r.t. $ \hat{\lambda}_i - \lambda_{0i}$.
			\begin{equation*}
				\frac{\partial L_{NT}}{\partial  (\hat{\lambda}_i - \lambda_{0i})} =\frac{1}{N} \cdot \frac{1}{T} \sum_{t=1}^{T} F_{0t} \sum_{k=1}^{K}  \big(I(\varepsilon_{it} - b_{0\tau_{k}} < 0) - \tau_{k}\big) 
			\end{equation*}
			\begin{align} \label{eq:derivative lambda}
					&\quad + \frac{1}{N}\sum_{k=1}^{K} f_{\varepsilon}(b_{0\tau_{k}}) \bigg[\frac{1}{T}\sum_{t=1}^{T} F_{0t} (\hat{b}_{\tau_{k}} - b_{0\tau_{k}}) + \frac{1}{T} \sum_{t=1}^{T} F_{0t} F_{0t}' (\hat{\lambda}_i - \lambda_{0i}) + \frac{1}{T} \sum_{t=1}^{T} F_{0t} \lambda_{0i}' (\hat{F}_t - F_{0t})\bigg] \nonumber \\
					&\quad = \frac{1}{NT} \sum_{t=1}^{T} F_{0t} \sum_{k=1}^{K}  \big(I(\varepsilon_{it} - b_{0\tau_{k}} < 0) - \tau_{k}\big) + \frac{1}{N}\sum_{k=1}^{K} f_{\varepsilon}(b_{0\tau_{k}})  \frac{1}{T} \sum_{t=1}^{T} F_{0t} F_{0t}' (\hat{\lambda}_i - \lambda_{0i}) + o_p(1).
			\end{align}
			Under \Cref{asump: factor and facor loading}, we have $\frac{1}{T}\sum_{t=1}^{T} F_{0t} (\hat{b}_{\tau_{k}} - b_{0\tau_{k}}) \rightarrow 0 $  and  we will show $\frac{1}{T} \sum_{t=1}^{T} F_{0t} \lambda_{0i} (\hat{F}_t - F_{0t}) = o_p(1)$ so that \cref{eq:derivative lambda} holds. To see this, we write this term in a more detailed matrix format. Let $\hat{F}_{t,j}$, $F_{0t,j}$ and $\lambda_{0i,j}$ be the $j$th element of the $r \times 1$ vector $\hat{F}_t$, $F_{0t}$ and $\lambda_{0i}$, respectively.
			\begin{align}
				\frac{1}{T} \sum_{t=1}^{T} F_{0t} \lambda_{0i}' (\hat{F}_t - F_{0t})  &= \frac{1 }{T} \sum_{t=1}^{T} \begin{bmatrix}
					F_{0t,1} \\
					F_{0t,2} \\
					\vdots    \\
					F_{0t,r}
				\end{bmatrix}
				\begin{bmatrix}
					\lambda_{0i,1}, \lambda_{0i,2}, \cdots, \lambda_{0i,r}
				\end{bmatrix}
				\begin{bmatrix}
					\hat{F}_{t,1} - F_{0t,1} \\
					\hat{F}_{t,2} - F_{0t,2} \\
					\vdots \\
					\hat{F}_{t,r} - F_{0t,r}
				\end{bmatrix} \nonumber \\
				&= \begin{bmatrix}
					\sum_{j=1}^{r} \lambda_{0i,j} \frac{1}{T} \sum_{t=1}^{T} F_{0t,1}(\hat{F}_{t,j} - F_{0t,j}) \\
					\sum_{j=1}^{r} \lambda_{0i,j} \frac{1}{T} \sum_{t=1}^{T} F_{0t,2}(\hat{F}_{t,j} - F_{0t,j}) \\
					\vdots \\
					\sum_{j=1}^{r} \lambda_{0i,j} \frac{1}{T} \sum_{t=1}^{T} F_{0t,r}(\hat{F}_{t,j} - F_{0t,j}) \label{eq: FlambdaFhat}
				\end{bmatrix},
			\end{align}
			which implies that we need to show all terms such as $\sum_{j=1}^{r} \lambda_{0i,j}\frac{1}{T} \sum_{t=1}^{T} F_{0t,1}(\hat{F}_{t,j} - F_{0t,j}) = o_p(1)$. From \Cref{lemma: factor product convergence}, we have 
			\begin{equation*}
				\frac{1}{T} \left \| \hat{F} - F_0 \right \|^2 = \frac{1}{T} \sum_{t=1}^{T} \sum_{j=1}^{r} (\hat{F}_{t,j} - F_{0t,j})^2 = o_p(1),
			\end{equation*}
			which implies 
			\begin{align*}
				&\sum_{j=1}^{r} \lambda_{0i,j} \frac{1}{T} \sum_{t=1}^{T} F_{0t,1}(\hat{F}_{t,j} - F_{0t,j}) \leq \left | \sum_{j=1}^{r} \lambda_{0i,j} \right | \left |  \frac{1}{T} \sum_{t=1}^{T} F_{0t,1}(\hat{F}_{t,j} - F_{0t,j}) \right | \\
				&\leq \left(r\sum_{j=1}^{r} \lambda_{0i,j}^2\right)^{1/2} \left(\frac{1}{T} \sum_{t=1}^{T} F_{0t,1}^2\right)^{1/2} \left(\frac{1}{T} \sum_{t=1}^{T}(\hat{F}_{t,j} - F_{0t,j})^2\right)^{1/2} = O(1)\cdot 1 \cdot o_p(1) = o_p(1),
			\end{align*}
			where the results for $O(1)$ and $1$ follow the normalization conditions, and the $o_p(1)$ term is the result of \Cref{lemma: factor product convergence}. More specifically, the normalization of $\Lambda_0$ implies $\frac{1}{N}\sum_{i=1}^{N}\sum_{j=1}^{r}\lambda_{0i,j}^2 = O(1)$. Since $\sum_{j=1}^{r}\lambda_{0i,j}^2 \geq 0$ for every $i$, we conclude $\sum_{j=1}^{r}\lambda_{0i,j}^2 = O(1)$.    Hence we conclude every element in the vector in \cref{eq: FlambdaFhat} is $o_p(1)$ and \cref{eq:derivative lambda} holds.
			
			     
			Setting \cref{eq:derivative lambda} to $0$ gives
			\begin{equation} \label{eq:lambda diff equation}
				\sqrt{T}(\hat{\lambda}_i - \lambda_{0i}) = - \Big(\sum_{k=1}^{K} f_{\varepsilon}(b_{0\tau_{k}}) \frac{1}{T}\sum_{t=1}^{T} F_{0t} F_{0t}' \Big)^{-1} \frac{1}{\sqrt{T}} \sum_{t=1}^{T} \Big[ F_{0t} \sum_{k=1}^{K} \big(I(\varepsilon_{it} - b_{0\tau_{k}} < 0) - \tau_{k}\big) \Big]
			\end{equation}
		
		Next, we compute the variance of $\frac{1}{\sqrt{T}} \sum_{t=1}^{T} \Big[ F_{0t} \sum_{k=1}^{K} \big(I(\varepsilon_{it} - b_{0\tau_{k}} < 0) - \tau_{k}\big) \Big]$. 
		Given  the result that $E\Big(\sum_{k=1}^{K} (I(\varepsilon_{it} - b_{0\tau_{k}} < 0) -\tau_{k})\Big) = 0$, and $\varepsilon_{it}$ is i.i.d., we obtain
		\begin{align} \label{eq: lambda equation var}
			&\text{Var}\Big(\frac{1}{\sqrt{T}} \sum_{t=1}^{T} \Big[ F_{0t} \sum_{k=1}^{K} \big(I(\varepsilon_{it} - b_{0\tau_{k}} < 0) - \tau_{k}\big) \Big]\Big) \nonumber\\
			 & = \frac{1}{T} \sum_{1}^{T} \text{Var} \left(
			 \begin{bmatrix}
			 	F_{0t,1} \cdot \sum_{k=1}^{K} \big(I(\varepsilon_{it} - b_{0\tau_{k}} < 0) - \tau_{k}\big)\\
			 	\vdots \\
			 	F_{0t,r} \cdot \sum_{k=1}^{K} \big(I(\varepsilon_{it} - b_{0\tau_{k}} < 0) - \tau_{k}\big)
			 \end{bmatrix}
			 \right) \nonumber \\
			&= \Sigma_{F_0} \sum_{k_1  = 1}^{K}\sum_{k_2=2}^{K} \min(\tau_{k_1}, \tau_{k_2}) (1 -\max(\tau_{k_1}, \tau_{k_2}) ).
		\end{align}
		As $N,T \rightarrow \infty$, \cref{eq:lambda diff equation,eq: lambda equation var} lead to the following asymptotic distribution:
		\begin{equation} \label{eq:lambda distribution}
			\sqrt{T}(\hat{\lambda}_i - \lambda_{0i}) \sim N(0,\Sigma_{\text{CQFA,}\lambda}),
		\end{equation}
		where 
		
		\begin{equation*}
			\Sigma_{\text{CQFA,}\lambda} = \frac{\sum_{k_1  = 1}^{K}\sum_{k_2=1}^{K} \min(\tau_{k_1}, \tau_{k_2}) (1 -\max(\tau_{k_1}, \tau_{k_2})}{\Big(\sum_{k=1}^{K} f_{\varepsilon}(b_{0\tau_{k}}) \Big)^2} \Sigma_{F_0}^{-1}.
		\end{equation*}
		
		Consider the partial derivative of \cref{eq:loss func diff3 } w.r.t. $ \hat{F}_t - F_{0t}$.
		
		\begin{align} \label{eq:derivative F}
			&\frac{\partial L_{NT}}{\partial  (\hat{F}_t - F_{0t})} =\frac{1}{NT} \sum_{i=1}^{N} \lambda_{0i} \sum_{k=1}^{K}  \big(I(\varepsilon_{it} - b_{0\tau_{k}} < 0) - \tau_{k}\big) \nonumber \\
			&\quad + \frac{1}{T}\sum_{k=1}^{K} f_{\varepsilon}(b_{0\tau_{k}}) \bigg[\frac{1}{N}\sum_{i=1}^{N} \lambda_{0i} (\hat{b}_{\tau_{k}} - b_{0\tau_{k}}) + \frac{1}{N} \sum_{i=1}^{N} \lambda_{0i} F_{0t}' (\hat{\lambda}_{i} - \lambda_{0i}) + \frac{1}{N} \sum_{i=1}^{N} \lambda_{0i} \lambda_{0i}' (\hat{F}_t - F_{0t})\bigg] \nonumber \\
			&\quad = \frac{1}{NT} \sum_{i=1}^{N} \lambda_{0i} \sum_{k=1}^{K}  \big(I(\varepsilon_{it} - b_{0\tau_{k}} < 0) - \tau_{k}\big) + \frac{1}{T}\sum_{k=1}^{K} f_{\varepsilon}(b_{0\tau_{k}})  \frac{1}{N} \sum_{i=1}^{N} \lambda_{0i} \lambda_{0i}' (\hat{F}_t - F_{0t}) + o_p(1).
		\end{align}
		To obtain \cref{eq:derivative F}, we note that $\frac{1}{N}\sum_{i=1}^{N} \lambda_{0i} (\hat{b}_{\tau_{k}} - b_{0\tau_{k}}) = o_p(1) $ since $\hat{b}_{\tau_{k}} - b_{0\tau_{k}}= O_p(1/\sqrt{NT})$ in \cref{eq:b diff equation} and each element of the vector $\frac{1}{N} \sum_{i=1}^{N}\lambda_{0i}$ is $O(1)$ due to the normalization in \Cref{asump: factor and facor loading}. For the term $\frac{1}{N} \sum_{i=1}^{N} \lambda_{0i} F_{0t}' (\hat{\lambda}_{i} - \lambda_{0i})$, we have
		\begin{equation} 
			 \frac{1}{N} \sum_{i=1}^{N} \lambda_{0i} F_{0t}' (\hat{\lambda}_{i} - \lambda_{0i}) = \begin{bmatrix}
			 	\sum_{j=1}^{r} F_{0t,j} \frac{1}{N} \sum_{i=1}^{N} \lambda_{0i,1}(\hat{\lambda}_{i,j} - \lambda_{0i,j}) \\
			 	\sum_{j=1}^{r} F_{0t,j} \frac{1}{N} \sum_{i=1}^{N} \lambda_{0i,2}(\hat{\lambda}_{i,j} - \lambda_{0i,j}) \\
			 	\vdots \\
			 	\sum_{j=1}^{r} F_{0t,j} \frac{1}{N} \sum_{i=1}^{N} \lambda_{0i,r}(\hat{\lambda}_{i,j} - \lambda_{0i,j}) \label{eq: lambdaFlambdahat}
			 \end{bmatrix},
		\end{equation}
		The first element in \cref{eq: lambdaFlambdahat} is
		\begin{align*}
			&\sum_{j=1}^{r} F_{0t,j} \frac{1}{N} \sum_{i=1}^{N} \lambda_{0i,1}(\hat{\lambda}_{i,j} - \lambda_{0i,j}) \leq \left | \sum_{j=1}^{r} F_{0t,j}    \right | \left |  \frac{1}{N} \sum_{i=1}^{N} \lambda_{0i,1}(\hat{\lambda}_{i,j} - \lambda_{0i,j}) \right | \\
			&\leq \left(r\sum_{j=1}^{r} F_{0t,j}^2\right)^{1/2} \left(\frac{1}{N} \sum_{i=1}^{N} \lambda_{0i,1}^2\right)^{1/2} \left(\frac{1}{N} \sum_{i=1}^{N}(\hat{\lambda}_{i,j} - \lambda_{0i,j})^2\right)^{1/2}\\
			&= O_p(1)\cdot O(1) \cdot o_p(1) = o_p(1),
		\end{align*}
		where the $O_p(1)$ and $O(1)$ results follow the normalization conditions for $F_0$ and $\Lambda_0$ and the $o_p(1)$ term is the result of \Cref{lemma: factor product convergence}. Thus we conclude that $\frac{1}{N} \sum_{i=1}^{N} \lambda_{0i} F_{0t}' (\hat{\lambda}_{i} - \lambda_{0i}) = o_p(1)$ and \cref{eq:derivative F} holds. Setting \cref{eq:derivative F} to $0$ gives
		\begin{equation} \label{eq:F diff equation}
			\sqrt{N}(\hat{F}_t - F_{0t}) = - \Big(\sum_{k=1}^{K} f_{\varepsilon}(b_{0\tau_{k}}) \frac{1}{T}\sum_{t=1}^{N} \lambda_{0i} \lambda_{0i}' \Big)^{-1} \frac{1}{\sqrt{N}} \sum_{i=1}^{N} \Big[ \lambda_{0i} \sum_{k=1}^{K} \big(I(\varepsilon_{it} - b_{0\tau_{k}} < 0) - \tau_{k}\big) \Big].
		\end{equation}
		The asymptotic distribution is given by 
		\begin{equation} \label{eq:F distribution}
			\sqrt{N}(\hat{F}_t - F_{0t}) \sim N(0,\Sigma_{\text{CQFA,}F}),
		\end{equation}
		where 
		\begin{equation*}
			\Sigma_{\text{CQFA,}F} = \frac{\sum_{k_1  = 1}^{K}\sum_{k_2=1}^{K} \min(\tau_{k_1}, \tau_{k_2}) (1 -\max(\tau_{k_1}, \tau_{k_2})}{\Big(\sum_{k=1}^{K} f_{\varepsilon}(b_{0\tau_{k}}) \Big)^2} \Sigma_{\lambda_0}^{-1},
		\end{equation*}
		and the derivation of $\Sigma_{\text{CQFA,}F} $ is similar to that for $\Sigma_{\text{CQFA,}\lambda} $ in \cref{eq:lambda distribution}.
		
	\end{proof}

		\subsection{Proof of \Cref{thm:factor number}} \label{supp:thm factor number proof}
		In the following proof, let $r$ be the estimated number of factors and $r_0$ be the true number of factors. Write the estimated factor and factor loading as $\hat{F}_t(r)$ and $\hat{\lambda}_i(r)$ when the estimated number of factor is $r$.
		\begin{proof}[Proof of \Cref{thm:factor number}]
			We consider two cases. When $r > r_0$, we use a similar method in Lemma 4 in \cite{baiandng2002ecma}. Let $H_r$ be an $r_0 \times r$ matrix with rank$(H_r)=\min(r,r_0)$, and an example $H_r$ can be found in \citet[Theorem~1]{baiandng2002ecma}. Let $H_r^+$ be the generalized inverse of $H_r$ so that $H_r H_r^+ = I_{r_0}$. Define the following transformed factor and factor loading vectors in the $r$-dimensional space
			\begin{equation} \label{eq:transformed factor and loading}
				F_{0t}(r) = H_r' F_{0t} \text{ and } \lambda_{0i}(r) = H_r^+ \lambda_{0i}.
			\end{equation}
			Since $ \lambda_{0i}(r)'F_{0t}(r) = \lambda_{0i} F_{0t}$, the transform in \cref{eq:transformed factor and loading} can be viewed as a representation of the true factor and factor loading in the $r$-dimensional space. 
			
			Define
			\begin{align} 
				V(r) &= \frac{1}{NT} \sum_{i=1}^{N} \sum_{t=1}^{T} \sum_{k=1}^{K} \rho_{\tau_k}(Y_{it} - b_{0\tau_{k}} - \hat{\lambda}_i(r)'\hat{F}_t(r)),  \label{eq:V r function}\\
				V(r_0) &= \frac{1}{NT} \sum_{i=1}^{N} \sum_{t=1}^{T} \sum_{k=1}^{K} \rho_{\tau_k}(Y_{it} - b_{0\tau_{k}}  - \lambda_{0i}'F_{0t}) \nonumber \\
				&= \frac{1}{NT} \sum_{i=1}^{N} \sum_{t=1}^{T} \sum_{k=1}^{K} \rho_{\tau_k}(Y_{it} - b_{0\tau_{k}}  - \lambda_{0i}(r)'F_{0t}(r)).  \label{eq:V r0 function}
			\end{align}
			Apply the Knight's equality to \cref{eq:V r function} and we have
			
			\begin{align}
				&V(r) = \frac{1}{NT} \sum_{i=1}^{N} \sum_{t=1}^{T} \sum_{k=1}^{K} \rho_{\tau_k}(\varepsilon_{it} - b_{0\tau_{k}} - (\hat{b}_{\tau_{k}} - b_{0\tau_{k}}) -  (\hat{\lambda}_i(r)'\hat{F}_t(r) - \lambda_{0i}(r)' F_{0t}(r))) \nonumber \\
				&= \frac{1}{NT} \sum_{i=1}^{N} \sum_{t=1}^{T} \sum_{k=1}^{K} \Bigl\{\rho_{\tau_k}(\varepsilon_{it} - b_{0\tau_{k}})  + \big[(\hat{b}_{\tau_{k}} - b_{0\tau_{k}}) + (\hat{\lambda}_i(r)'\hat{F}_t(r) - \lambda_{0i}(r)' F_{0t}(r)) \big] \nonumber \\
				&\quad \times \big(I(\varepsilon_{it} - b_{0\tau_{k}} < 0) - \tau_k \big)  \nonumber\\
				&\quad + \int_{0}^{(\hat{b}_{\tau_{k}} - b_{0\tau_{k}}) + \hat{\lambda}_i(r)'\hat{F}_t(r) - \lambda_{0i}(r)' F_{0t}(r)} \big[I(\varepsilon_{it} - b_{0\tau_{k}} < s) - I(\varepsilon_{it} - b_{0\tau_{k}} < 0)\big]ds\Bigr\} \nonumber \\
				&= \frac{1}{NT} \sum_{i=1}^{N} \sum_{t=1}^{T} \sum_{k=1}^{K} \rho_{\tau_k}(\varepsilon_{it} - b_{0\tau_{k}}) \nonumber \\
				&\quad + \frac{1}{NT} \sum_{i=1}^{N} \sum_{t=1}^{T} \sum_{k=1}^{K} \big[(\hat{b}_{\tau_{k}} - b_{0\tau_{k}}) + (\hat{\lambda}_i(r) - \lambda_{0i}(r))' (\hat{F}_t(r) - F_{0t}(r)) +(\hat{\lambda}_i(r) - \lambda_{0i}(r))'  F_{0t}(r) \nonumber \\
				&\quad +  \lambda_{0i}(r)' (\hat{F}_t(r) - F_{0t}(r))\big]  \times \big(I(\varepsilon_{it} - b_{0\tau_{k}} < 0) - \tau_k \big) \nonumber \\
				&\quad +\frac{1}{NT} \sum_{i=1}^{N} \sum_{t=1}^{T} \sum_{k=1}^{K} \int_{0}^{(\hat{b}_{\tau_{k}} - b_{0\tau_{k}}) + \hat{\lambda}_i(r)'\hat{F}_t(r) - \lambda_{0i}(r)' F_{0t}(r)} \big[I(\varepsilon_{it} - b_{0\tau_{k}} < s) - I(\varepsilon_{it} - b_{0\tau_{k}} < 0)\big]ds \nonumber \\
				&= V(r_0) + \mathrm{I} + \mathrm{II}.
			\end{align}
			Consider $\mathrm{I}$. From \Cref{thm: asymptotic distribution}, the implication of \cref{eq:b diff equation} that $\hat{b}_{\tau_{k}} - b_{0\tau_{k}}$ is an $O_p(1/\sqrt{NT})$ term, and the fact that $\varepsilon_{it}$ is i.i.d. and independent of $F_{0t}$ and $\lambda_{0i}$, we have
			\begin{align*}
				\mathrm{I} &=  \frac{1}{NT} \sum_{i=1}^{N} \sum_{t=1}^{T} \sum_{k=1}^{K} \left[O_p\left(\frac{1}{\sqrt{NT}}\right) + O_p(\frac{1}{\sqrt{N}})O_p(\frac{1}{\sqrt{T}}) + O_p\left(\frac{1}{\sqrt{T}} \right)+ O_p\left(\frac{1}{\sqrt{N}}\right)\right]\cdot O_p\left(\frac{1}{\sqrt{NT}}\right) \nonumber\\
				&=O_p\left(\frac{1}{C_{NT}}\right).
			\end{align*}
			Although the true number of factors is $r_0$, \cref{eq:transformed factor and loading} represents the $r_0$ factors in an $r$-dimensional space.  Consequently, we can use $r$ as the number of factors in \cref{eq: factor model matrix form} for estimation, and apply results in \Cref{lemma: factor product convergence} to the terms in $\mathrm{I}$.
			
			For term $\mathrm{II}$, using the similar argument in \cref{eq:2nd order expansion,eq:loss func diff3 }, we can show $\mathrm{II}$ converges to the following quadratic form
			
			\begin{align*}
				\mathrm{II} &\rightarrow \frac{f_{\varepsilon}(b_{0\tau_{k}})}{2NT} \sum_{k=1}^{K}\sum_{i=1}^{N}\sum_{t=1}^{T} 
				\begin{bmatrix}
					\hat{b}_{\tau_{k}}(r) - b_{0\tau_{k}}(r) \\
					\hat{\lambda}_i(r) - \lambda_{0i}(r) \\
					\hat{F}_t(r) - F_{0t}(r)
				\end{bmatrix}' 
				\begin{bmatrix}
					1 & F_{0t}(r)' & \lambda_{0i}(r)' \\
					F_{0t}(r) & F_{0t}(r)F_{0t}(r)' & F_{0t}(r) \lambda_{0i}(r)' \\
					\lambda_{0i}(r) & \lambda_{0i}(r) F_{0t}(r)' & \hat{\lambda}_i(r)\lambda_{0i}(r)'
				\end{bmatrix}
				\begin{bmatrix}
					\hat{b}_{\tau_{k}}(r) - b_{0\tau_{k}}(r) \\
					\hat{\lambda}_i(r) - \lambda_{0i}(r) \\
					\hat{F}_t(r) - F_{0t}(r)
				\end{bmatrix} \nonumber \\
				&= O_p\left(\frac{1}{\sqrt{NT}}\right)  + O_p\left(\frac{1}{\sqrt{T}} \right) + O_p\left(\frac{1}{\sqrt{N}} \right) + O_p\left(\frac{1}{N\sqrt{T}} \right) + O_p\left(\frac{1}{T\sqrt{N}} \right) = O_p\left(\frac{1}{C_{NT}}\right).\nonumber
			\end{align*}
			Combining the results for the terms $\mathrm{I}$ and $\mathrm{II}$, we conclude that $V(r) - V(r_0) = O_p\left(1/C_{NT}\right)$, which implies $V(r)/V(r_0) = 1+O_p(1/C_{NT})$ and $\log(V(r)/V(r_0)) = O_p(1/C_{NT})$. As a result, we have
			\begin{equation*}
				P(IC(r) - IC(r_0) < 0) \leq P(O_p(1/C_{NT})+q(N,T) < 0) \rightarrow 0,
			\end{equation*}
			which proves the probability for $IC(r)$ in \cref{eq:IC} to select $r > r_0$ is $0$ when $N,T \rightarrow \infty$.
			
			Next, consider the case $r < r_0$. Replace $r$ with $r_0$ in \cref{eq:V r0 function}. By the law of large numbers, we know both \cref{eq:V r function,eq:V r0 function} will converge to some expectations. While $\lambda_{0i}$ and $F_{0t}$ will minimize \cref{eq:V r0 function} as $N,T\rightarrow \infty$, $\hat{\lambda}_i(r)$ and $\hat{F}_{t}(r)$ will not attain the same minimum value when plugged into \cref{eq:V r function} since $\hat{\lambda}_i(r)$ and $\hat{F}_{t}(r)$ cannot span the space spanned by $\lambda_{0i}$ and $F_{0t}$ when $r < r_0$. Hence, we conclude that $V(r) - V(r_0) > 0$ as $N,T\rightarrow \infty$. This is similar to the proof in \citet[supplement page~29]{andoandbai2020jasa}. Hence, we have $V(r)/V(r_0) > 1$ and $\log(V(r)/V(r_0)) > c$ for some positive constant $c$. Finally, as $N,T \rightarrow \infty$, we have
			\begin{equation*}
				P(IC(r)-IC(r_0)<0) \leq P(c + (r - r_0)g(N,T) < 0 ) \rightarrow 0,
			\end{equation*}
			where we use the result that $g(N,T) \rightarrow 0$ as $N,T \rightarrow 0$. 
			
			The analysis for the two cases, $r > r_0$ and $r < r_0$ proves that $IC(r)$ will select the correct number of factor $r_0$ asymptotically because the value of the information criterion at $r$ is always lager than or equal to the value of information criterion at $r_0$.
			
		\end{proof}
		
		\subsection{Additional tables}
		
	\begin{landscape}
		\subsubsection{\Cref{tab:adj R2 asym error with pca}: \Cref{tab:adj R2 asym error} with additional adjusted $R^2$s of the PCA method under asymmetric errors}
	\begin{ThreePartTable}
		\begin{TableNotes}[flushleft]\footnotesize
			\item[] \textit{Notes}: Every number is the average of adjusted $R^2$ over $100$ replications of regressing one of the three true factors on the estimated factors based on the RTS, QFM(0.5), CQFM, and PCA method, respectively. We choose $\tau=0.5$ for the QFM method and $K=5$ for the CQFM method. This table is the same as \Cref{tab:adj R2 asym error} except for the addition of the PCA results.
		\end{TableNotes}
		\begin{longtable}{lrrrrrrrrrrrr} 
			\caption{Adj. $R^2$ of regressing  $3$ true factors on the estimated factors  -- \Cref{tab:adj R2 asym error} with the PCA results}\\
			\label{tab:adj R2 asym error with pca}\\
			\toprule
			\multirow{2}{*}{(T,N)} & $R^2_{1,\text{RTS}}$  & $R^2_{2,\text{RTS}}$ & $R^2_{3,\text{RTS}}$ & $R^2_{1,\text{QFM}}$ & $R^2_{2,\text{QFM}}$  & $R^2_{3,\text{QFM}}$ & $R^2_{1,\text{CQFM}}$ & $R^2_{2,\text{CQFM}}$  & $R^2_{3,\text{CQFM}}$ &$R^2_{1,\text{PCA}}$ & $R^2_{2,\text{PCA}}$& $R^2_{3,\text{PCA}}$ \\
			\cmidrule(lr){2-4} \cmidrule(lr){5-7} \cmidrule(lr){8-10}  \cmidrule(lr){11-13}
			\endfirsthead
			\toprule
			{(T,N)} & $F_{1}^{\text{RTS}}$  & $F_2^{\text{RTS}}$ & $F_3^{\text{RTS}}$ & $\hat{F}_1^{\text{QFM}}$ & $F_2^{\text{QFM}}$ & $F_3^{\text{QFM}}$ & $\hat{F}_1^{\text{CQFM}}$ & $F_2^{\text{CQFM}}$ & $F_3^{\text{CQFM}}$ & $\hat{F}_1^{\text{PCA}}$ & $F_2^{\text{PCA}}$ & $F_3^{\text{PCA}}$ \\
			\cmidrule(lr){2-4} \cmidrule(lr){5-7} \cmidrule(lr){8-10}  \cmidrule(lr){11-13}
			\endhead
			\endfoot
			\bottomrule
			\insertTableNotes
			\endlastfoot
			
			&   \multicolumn{12}{c}{$\varepsilon_{it} \sim$ skewed normal}      \\    
			\cmidrule(lr){2-13}
			(50,100) & 0.9950 & 0.9914 & 0.9893 & 0.9915 & 0.9857 & 0.9821 & 0.9951 & 0.9917 & 0.9898 & 0.9950 & 0.9914 & 0.9893 \\    
			(100,50) & 0.9909 & 0.9828 & 0.9786 & 0.9847 & 0.9712 & 0.9645 & 0.9911 & 0.9832 & 0.9790 & 0.9909 & 0.9828 & 0.9787 \\    
			(100,200) & 0.9978 & 0.9961 & 0.9949 & 0.9960 & 0.9928 & 0.9908 & 0.9979 & 0.9962 & 0.9952 & 0.9978 & 0.9961 & 0.9949 \\    
			(200,100) & 0.9959 & 0.9921 & 0.9898 & 0.9926 & 0.9856 & 0.9816 & 0.9961 & 0.9924 & 0.9903 & 0.9959 & 0.9921 & 0.9899 \\    
			
			(300,300) & 0.9986 & 0.9975 & 0.9967 & 0.9974 & 0.9953 & 0.9938 & 0.9987 & 0.9976 & 0.9969 & 0.9986 & 0.9975 & 0.9967 \\    
			&  \multicolumn{12}{c}{$\varepsilon_{it} \sim$ skewed t}      \\   
			\cmidrule(lr){2-13} 
			(50,100) & 0.9950 & 0.9916 & 0.9895 & 0.9936 & 0.9895 & 0.9866 & 0.9954 & 0.9923 & 0.9903 & 0.9951 & 0.9916 & 0.9896 \\    
			(100,50) & 0.9908 & 0.9828 & 0.9790 & 0.9885 & 0.9783 & 0.9737 & 0.9914 & 0.9840 & 0.9804 & 0.9908 & 0.9828 & 0.9790 \\    
			(100,200) & 0.9978 & 0.9960 & 0.9949 & 0.9972 & 0.9950 & 0.9936 & 0.9981 & 0.9964 & 0.9954 & 0.9978 & 0.9960 & 0.9950 \\    
			(200,100) & 0.9959 & 0.9921 & 0.9899 & 0.9947 & 0.9900 & 0.9871 & 0.9962 & 0.9928 & 0.9908 & 0.9959 & 0.9921 & 0.9899 \\    
			
			(300,300) & 0.9986 & 0.9975 & 0.9967 & 0.9983 & 0.9968 & 0.9958 & 0.9988 & 0.9978 & 0.9971 & 0.9986 & 0.9975 & 0.9967 \\    
			&   \multicolumn{12}{c}{$\varepsilon_{it} \sim$ asymmetric Laplace}      \\  
			\cmidrule(lr){2-13}  
			(50,100) & 0.9569 & 0.9254 & 0.9085 & 0.9482 & 0.9108 & 0.8682 & 0.9759 & 0.9590 & 0.9518 & 0.9566 & 0.9255 & 0.9070 \\    
			(100,50) & 0.9277 & 0.8698 & 0.8438 & 0.9151 & 0.8443 & 0.8062 & 0.9573 & 0.9221 & 0.9057 & 0.9274 & 0.8683 & 0.8417 \\    
			(100,200) & 0.9826 & 0.9673 & 0.9577 & 0.9777 & 0.9566 & 0.9341 & 0.9911 & 0.9834 & 0.9784 & 0.9827 & 0.9673 & 0.9577 \\    
			(200,100) & 0.9664 & 0.9375 & 0.9204 & 0.9571 & 0.9143 & 0.8900 & 0.9823 & 0.9663 & 0.9572 & 0.9664 & 0.9375 & 0.9202 \\    
			
			(300,300) & 0.9891 & 0.9797 & 0.9739 & 0.9843 & 0.9704 & 0.9613 & 0.9946 & 0.9900 & 0.9874 & 0.9891 & 0.9797 & 0.9739 \\    
			
			&   \multicolumn{12}{c}{$\varepsilon_{it} \sim$ log-normal}      \\    
			\cmidrule(lr){2-13}
			
			(50,100) & 0.5958 & 0.3578 & 0.2543 & 0.9541 & 0.8272 & 0.6834 & 0.9889 & 0.9823 & 0.9769 & 0.3050 & 0.1109 & 0.0874 \\    
			(100,50) & 0.5637 & 0.3286 & 0.2245 & 0.9216 & 0.7739 & 0.5421 & 0.9781 & 0.9598 & 0.9512 & 0.2241 & 0.0787 & 0.0600 \\    
			(100,200) & 0.8045 & 0.5902 & 0.4469 & 0.9754 & 0.8402 & 0.5698 & 0.9964 & 0.9932 & 0.9914 & 0.4009 & 0.0976 & 0.0648 \\    
			(200,100) & 0.7470 & 0.5382 & 0.4320 & 0.9687 & 0.8033 & 0.4895 & 0.9931 & 0.9863 & 0.9826 & 0.3668 & 0.0642 & 0.0389 \\    
			
			(300,300) & 0.8950 & 0.7967 & 0.7229 & 0.9881 & 0.8448 & 0.4473 & 0.9980 & 0.9963 & 0.9952 & 0.6190 & 0.1077 & 0.0449 \\    
			&   \multicolumn{12}{c}{$\varepsilon_{it} \sim$ mixture of skewed normal}      \\  
			\cmidrule(lr){2-13}    
			(50,100) & 0.9908 & 0.9851 & 0.9807 & 0.9899 & 0.9835 & 0.9788 & 0.9937 & 0.9900 & 0.9870 & 0.9909 & 0.9851 & 0.9810 \\    
			(100,50) & 0.9829 & 0.9694 & 0.9609 & 0.9816 & 0.9670 & 0.9586 & 0.9880 & 0.9785 & 0.9731 & 0.9829 & 0.9693 & 0.9609 \\    
			(100,200) & 0.9960 & 0.9929 & 0.9908 & 0.9954 & 0.9920 & 0.9893 & 0.9974 & 0.9954 & 0.9941 & 0.9960 & 0.9929 & 0.9908 \\    
			(200,100) & 0.9926 & 0.9858 & 0.9818 & 0.9914 & 0.9837 & 0.9789 & 0.9952 & 0.9908 & 0.9880 & 0.9926 & 0.9858 & 0.9818 \\    
			
			(300,300) & 0.9976 & 0.9955 & 0.9941 & 0.9971 & 0.9946 & 0.9929 & 0.9985 & 0.9972 & 0.9964 & 0.9976 & 0.9955 & 0.9941 \\    
		\end{longtable} 
	\end{ThreePartTable}
\end{landscape}

		\begin{landscape}
			\subsubsection{\Cref{tab:adj R2 sym error with pca}: Adjusted $R^2$ under symmetric errors}
			\begin{ThreePartTable}
				\begin{TableNotes}[flushleft]\footnotesize
					\item[] \textit{Notes}: Each number is the average of adjusted $R^2$ over $100$ replications of regressing one of the three true factors on the estimated factors based on the RTS, QFM(0.5), CQFM, and PCA method, respectively. We choose $\tau=0.5$ for the QFM method and $K=5$ for the CQFM method.
				\end{TableNotes}
				\begin{longtable}{lrrrrrrrrrrrr} 
					\caption{Adj. $R^2$ of regressing  $3$ true factors on the estimated factors under symmetric errors}\\
					\label{tab:adj R2 sym error with pca}\\
					\toprule
					\multirow{2}{*}{(T,N)} & $F_{1}^{\text{RTS}}$  & $F_2^{\text{RTS}}$ & $F_3^{\text{RTS}}$ & $F_1^{\text{QFM}}$ & $F_2^{\text{QFM}}$ & $F_3^{\text{QFM}}$ & $F_1^{\text{CQFM}}$ & $F_2^{\text{CQFM}}$ & $F_3^{\text{CQFM}}$ & $F_1^{\text{PCA}}$ & $F_2^{\text{PCA}}$ & $F_3^{\text{PCA}}$ \\
					\cmidrule(lr){2-4} \cmidrule(lr){5-7} \cmidrule(lr){8-10}  \cmidrule(lr){11-13}
					\endfirsthead
					\toprule
					{(T,N)} & $F_{1}^{\text{RTS}}$  & $F_2^{\text{RTS}}$ & $F_3^{\text{RTS}}$ & $F_1^{\text{QFM}}$ & $F_2^{\text{QFM}}$ & $F_3^{\text{QFM}}$ & $F_1^{\text{CQFM}}$ & $F_2^{\text{CQFM}}$ & $F_3^{\text{CQFM}}$ & $F_1^{\text{PCA}}$ & $F_2^{\text{PCA}}$ & $F_3^{\text{PCA}}$ \\
					\cmidrule(lr){2-4} \cmidrule(lr){5-7} \cmidrule(lr){8-10}  \cmidrule(lr){11-13}
					\endhead
					\endfoot
					\bottomrule
					\insertTableNotes
					\endlastfoot
					
					&   \multicolumn{12}{c}{$\varepsilon_{it} \sim N(0,1)$}      \\    
					\cmidrule(lr){2-13}
					 (50,100) & 0.9950 & 0.9916 & 0.9895 & 0.9936 & 0.9895 & 0.9866 & 0.9954 & 0.9923 & 0.9903 & 0.9951 & 0.9916 & 0.9896 \\
					 (100,50) & 0.9908 & 0.9828 & 0.9790 & 0.9885 & 0.9783 & 0.9737 & 0.9914 & 0.9840 & 0.9804 & 0.9908 & 0.9828 & 0.9790 \\
					 (100,200) & 0.9978 & 0.9960 & 0.9949 & 0.9972 & 0.9950 & 0.9936 & 0.9981 & 0.9964 & 0.9954 & 0.9978 & 0.9960 & 0.9950 \\
					 (200,100) & 0.9959 & 0.9921 & 0.9899 & 0.9947 & 0.9900 & 0.9871 & 0.9962 & 0.9928 & 0.9908 & 0.9959 & 0.9921 & 0.9899 \\
					 
					 (300,300) & 0.9986 & 0.9975 & 0.9967 & 0.9983 & 0.9968 & 0.9958 & 0.9988 & 0.9978 & 0.9971 & 0.9986 & 0.9975 & 0.9967 \\
					&  \multicolumn{12}{c}{$\varepsilon_{it} \sim t_1$}      \\   
					\cmidrule(lr){2-13} 
					 (50,100) & 0.0530 & 0.0415 & 0.0436 & 0.9821 & 0.9689 & 0.9609 & 0.9777 & 0.9623 & 0.9523 & 0.0439 & 0.0383 & 0.0404 \\
					 (100,50) & 0.0483 & 0.0326 & 0.0290 & 0.9662 & 0.9391 & 0.9258 & 0.9577 & 0.9226 & 0.8980 & 0.0167 & 0.0213 & 0.0191 \\
					 (100,200) & 0.0212 & 0.0181 & 0.0193 & 0.9937 & 0.9883 & 0.9849 & 0.9919 & 0.9853 & 0.9808 & 0.0189 & 0.0185 & 0.0230 \\
					 (200,100) & 0.0223 & 0.0136 & 0.0105 & 0.9877 & 0.9764 & 0.9697 & 0.9839 & 0.9695 & 0.9617 & 0.0097 & 0.0103 & 0.0124 \\
					 
					 (300,300) & 0.0106 & 0.0069 & 0.0057 & 0.9964 & 0.9932 & 0.9912 & 0.9953 & 0.9912 & 0.9885 & 0.0074 & 0.0069 & 0.0069 \\
					&   \multicolumn{12}{c}{$\varepsilon_{it} \sim$ Laplace, location = 0, scale = 1}      \\  
					\cmidrule(lr){2-13}  
					(50,100) & 0.9896 & 0.9831 & 0.9794 & 0.9919 & 0.9864 & 0.9837 & 0.9920 & 0.9868 & 0.9842 & 0.9896 & 0.9833 & 0.9797 \\
					(100,50) & 0.9813 & 0.9662 & 0.9587 & 0.9851 & 0.9730 & 0.9670 & 0.9854 & 0.9736 & 0.9678 & 0.9813 & 0.9661 & 0.9587 \\
					(100,200) & 0.9958 & 0.9921 & 0.9897 & 0.9970 & 0.9945 & 0.9929 & 0.9969 & 0.9941 & 0.9923 & 0.9958 & 0.9921 & 0.9897 \\
					(200,100) & 0.9916 & 0.9840 & 0.9796 & 0.9941 & 0.9887 & 0.9855 & 0.9938 & 0.9880 & 0.9848 & 0.9916 & 0.9840 & 0.9796 \\
					
					(300,300) & 0.9973 & 0.9950 & 0.9935 & 0.9983 & 0.9968 & 0.9959 & 0.9981 & 0.9964 & 0.9954 & 0.9973 & 0.9950 & 0.9935 \\   
					&   \multicolumn{12}{c}{$\varepsilon_{it} \sim 0.9N(0,1) + 0.1N(0,9)$ }      \\    
					\cmidrule(lr){2-13}
					 (50,100) & 0.9905 & 0.9841 & 0.9802 & 0.9909 & 0.9843 & 0.9795 & 0.9928 & 0.9875 & 0.9841 & 0.9905 & 0.9841 & 0.9803 \\
					(100,50) & 0.9836 & 0.9695 & 0.9615 & 0.9834 & 0.9696 & 0.9622 & 0.9870 & 0.9757 & 0.9700 & 0.9836 & 0.9695 & 0.9615 \\
					(100,200) & 0.9961 & 0.9928 & 0.9908 & 0.9962 & 0.9928 & 0.9910 & 0.9970 & 0.9945 & 0.9929 & 0.9961 & 0.9928 & 0.9909 \\
					(200,100) & 0.9926 & 0.9859 & 0.9812 & 0.9926 & 0.9859 & 0.9814 & 0.9942 & 0.9891 & 0.9856 & 0.9926 & 0.9859 & 0.9812 \\
					
					(300,300) & 0.9976 & 0.9954 & 0.9940 & 0.9976 & 0.9955 & 0.9941 & 0.9981 & 0.9965 & 0.9955 & 0.9976 & 0.9954 & 0.9940 \\
					&   \multicolumn{12}{c}{$\varepsilon_{it} \sim 0.9N(0,1) + 0.1N(0,100)$ }      \\  
					\cmidrule(lr){2-13}    
					 (50,100) & 0.9334 & 0.8849 & 0.8556 & 0.9900 & 0.9828 & 0.9782 & 0.9918 & 0.9858 & 0.9820 & 0.9319 & 0.8740 & 0.8321 \\
					(100,50) & 0.9035 & 0.8233 & 0.7819 & 0.9819 & 0.9670 & 0.9586 & 0.9850 & 0.9719 & 0.9653 & 0.9001 & 0.8076 & 0.7609 \\
					(100,200) & 0.9756 & 0.9537 & 0.9418 & 0.9960 & 0.9924 & 0.9905 & 0.9967 & 0.9938 & 0.9919 & 0.9755 & 0.9529 & 0.9403 \\
					(200,100) & 0.9561 & 0.9165 & 0.8908 & 0.9922 & 0.9850 & 0.9802 & 0.9935 & 0.9878 & 0.9838 & 0.9558 & 0.9152 & 0.8886 \\
					
					(300,300) & 0.9853 & 0.9722 & 0.9638 & 0.9974 & 0.9952 & 0.9938 & 0.9979 & 0.9961 & 0.9950 & 0.9853 & 0.9721 & 0.9637 \\   
				\end{longtable} 
			\end{ThreePartTable}
		\end{landscape}

		\subsubsection{\Cref{tab:mse sym error}: MSE under symmetric errors}

			\begin{table}[htp] \centering
			\begin{center}
				\caption{MSE under symmetric errors} 
				\label{tab:mse sym error} 
				\begin{threeparttable}
					\begin{tabular}{lrrrrrrrr}    
						\cmidrule(lr){1-9}
						\multirow{2}{*}{(T,N)}  &   \multicolumn{4}{c}{$\varepsilon_{it} \sim N(0,1)$}  &   \multicolumn{4}{c}{$\varepsilon_{it} \sim t_1$}         \\  
						& RTS & QFM & CQFM & PCA & RTS & QFM & CQFM & PCA \\
						\cmidrule(lr){2-5} \cmidrule(lr){6-9}
						(50,100) & 0.098 & 0.135 & 0.110 & 0.088 & 16980540.454 & 0.333 & 0.443 & 19809797.706 \\
						(100,50) & 0.091 & 0.135 & 0.102 & 0.088 & 15105565.250 & 0.330 & 0.493 & 19809805.496 \\
						(100,200) & 0.048 & 0.070 & 0.055 & 0.045 & 473589.065 & 0.135 & 0.178 & 5497608.996 \\
						(200,100) & 0.046 & 0.070 & 0.052 & 0.045 & 147530.806 & 0.135 & 0.176 & 5497659.458 \\
						(300,300) & 0.021 & 0.031 & 0.024 & 0.020 & 58500.884 & 0.054 & 0.072 & 5569166.347 \\
						&   \multicolumn{4}{c}{$\varepsilon_{it} \sim$ Laplace(0,1)}    &   \multicolumn{4}{c}{$\varepsilon_{it} \sim 0.9N(0,1) + 0.1N(0,9)$}    \\  
						\cmidrule(lr){2-5}  \cmidrule(lr){6-9}
						(50,100) & 0.196 & 0.143 & 0.152 & 0.180 & 0.180 & 0.165 & 0.141 & 0.165 \\
						(100,50) & 0.185 & 0.141 & 0.142 & 0.181 & 0.167 & 0.162 & 0.129 & 0.164 \\
						(100,200) & 0.096 & 0.063 & 0.075 & 0.090 & 0.086 & 0.081 & 0.068 & 0.082 \\
						(200,100) & 0.093 & 0.064 & 0.068 & 0.091 & 0.083 & 0.081 & 0.063 & 0.082 \\
						(300,300) & 0.041 & 0.025 & 0.030 & 0.040 & 0.037 & 0.036 & 0.028 & 0.036 \\
						&   \multicolumn{4}{c}{$\varepsilon_{it} \sim 0.9N(0,1) + 0.1N(0,100)$}   &   \\   
						\cmidrule(lr){2-5} 
						(50,100) & 1.231 & 0.179 & 0.158 & 1.310 &       &       &       &  \\
						(100,50) & 1.114 & 0.177 & 0.149 & 1.266 &       &       &       &  \\
						(100,200) & 0.536 & 0.086 & 0.078 & 0.545 &       &       &       &  \\
						(200,100) & 0.520 & 0.087 & 0.071 & 0.542 &       &       &       &  \\
						(300,300) & 0.224 & 0.038 & 0.034 & 0.226 &       &       &       &  \\
						\hline
					\end{tabular}
					\begin{tablenotes}[flushleft]
						\item[] \textit{Notes}: Every number is the average  MSE over $100$ replications for the RTS, QFM(0.5), CQFM, and PCA method, respectively. We choose $\tau=0.5$ for the QFM method and $K=5$ for the CQFM method.
					\end{tablenotes}
				\end{threeparttable}
			\end{center} 
		\end{table}
		
		\subsubsection{\Cref{tab:factor number sym error}: Factor number estimation under symmetric errors}

\begin{table}[htp] \centering
	\begin{center}
		\caption{Estimated factor number and frequency of correct estimation for symmetric errors} 
		\label{tab:factor number sym error} 
		\begin{threeparttable}
			\renewcommand{\TPTminimum}{\linewidth}
			\makebox[\linewidth] { 
				\begin{tabular}{lrrrrrr}
					\toprule
					(T,N) & QFM      &   CQFM    &    PCA   &  QFM     &  CQFM     &   PCA \\
					\hline
					&   \multicolumn{3}{c}{avg. $\hat{r}$}  &   \multicolumn{3}{c}{$\text{Prob}(\hat{r} = 3)$} \\
					\cmidrule(lr){2-4}  \cmidrule(lr){5-7}
					&  \multicolumn{6}{c}{$\varepsilon_{it} \sim N(0,1)$}        \\
					\cmidrule(lr){2-7} 
					(50,100) & 2.5   & 3     & 3     & 0.59  & 1     & 1 \\
					(100,50) & 2.58  & 3     & 3     & 0.63  & 1     & 1 \\
					(100,200) & 2.94  & 3     & 3     & 0.95  & 1     & 1 \\
					(200,100) & 2.93  & 3     & 3     & 0.93  & 1     & 1 \\
					(300,300) & 3     & 3     & 3     & 1     & 1     & 1 \\
					&  \multicolumn{6}{c}{$\varepsilon_{it} \sim t_1$}        \\
					\cmidrule(lr){2-7}  
					(50,100) & 2.43  & 2.67 & 5.99   & 0.52  & 0.68   & 0 \\
					(100,50) & 2.7   & 2.6   & 5.96  & 0.66  & 0.37    & 0 \\
					(100,200) & 2.95  & 2.99  & 6     & 0.96  & 0.99    & 0 \\
					(200,100) & 2.93  & 3.44 & 6     & 0.93  & 0.61    & 0 \\
					(300,300) & 3     & 3 & 6     & 1     & 1 & 0 \\
					&  \multicolumn{6}{c}{$\varepsilon_{it} \sim$ Laplace}        \\
					
					\cmidrule(lr){2-7}  
					(50,100) & 2.53  & 3     & 3     & 0.6   & 1     & 1 \\
					(100,50) & 2.63  & 3     & 3     & 0.66  & 1     & 1 \\
					(100,200) & 2.94  & 3     & 3     & 0.95  & 1     & 1 \\
					(200,100) & 2.92  & 3     & 3     & 0.92  & 1     & 1 \\
					(300,300) & 3     & 3     & 3     & 1     & 1     & 1 \\
					&  \multicolumn{6}{c}{$\varepsilon_{it} \sim 0.9N(0,1) + 0.1N(0,9)$}        \\
					\cmidrule(lr){2-7}  
					(50,100) & 2.54  & 3     & 3     & 0.62  & 1     & 1 \\
					(100,50) & 2.59  & 3     & 3     & 0.62  & 1     & 1 \\
					(100,200) & 2.94  & 3     & 3     & 0.95  & 1     & 1 \\
					(200,100) & 2.93  & 3     & 3     & 0.93  & 1     & 1 \\
					(300,300) & 3     & 3     & 3     & 1     & 1     & 1 \\
					&  \multicolumn{6}{c}{$\varepsilon_{it} \sim 0.9N(0,1) + 0.1N(0,100)$}        \\
					\cmidrule(lr){2-7}  
					(50,100) & 2.55  & 2.8   & 2.58  & 0.64  & 0.81  & 0.62 \\
					(100,50) & 2.59  & 2.86  & 2.6   & 0.63  & 0.86  & 0.61 \\
					(100,200) & 2.94  & 3     & 3     & 0.95  & 1     & 1 \\
					(200,100) & 2.92  & 3     & 3     & 0.92  & 1     & 1 \\
					(300,300) & 3     & 3     & 3     & 1     & 1     & 1 \\
					\bottomrule
				\end{tabular} 
			}   
			\begin{tablenotes}[flushleft]
				\item[] \textit{Notes}: Same as that in \Cref{tab:factor number}. Results for CQFM with $t_1$ error are obtained by standardizing the data, using \cref{eq:qnt 2} in \cref{eq:IC}, and letting $K=25$ in CQFM.
			\end{tablenotes}
		\end{threeparttable}
	\end{center} 
\end{table}
		
		\begin{landscape}
			\subsubsection{\Cref{tab:adj R2 asym error with pca heterodasticity}: Adjusted $R^2$ under asymmetric errors with heteroskedasticity}
			\begin{ThreePartTable}
				\begin{TableNotes}[flushleft]\footnotesize
					\item[] \textit{Notes}: Every number is the average of adjusted $R^2$ over $100$ replications of regressing one of the three true factors on the estimated factors based on the RTS, QFM(0.5), and CQFM method, respectively. We choose $\tau=0.5$ for the QFM method and $K=5$ for the CQFM method. The DGP is described in \cref{eq:heter}.
				\end{TableNotes}
				\begin{longtable}{lrrrrrrrrrrrr} 
					\caption{Adj. $R^2$ of regressing  $3$ true factors on the estimated factors under asymmetric errors and heteroskedasticity}\\
					\label{tab:adj R2 asym error with pca heterodasticity}\\
					\toprule
					\multirow{2}{*}{(T,N)} & $F_{1}^{\text{RTS}}$  & $F_2^{\text{RTS}}$ & $F_3^{\text{RTS}}$ & $\hat{F}_1^{\text{QFM}}$ & $F_2^{\text{QFM}}$ & $F_3^{\text{QFM}}$ & $\hat{F}_1^{\text{CQFM}}$ & $F_2^{\text{CQFM}}$ & $F_3^{\text{CQFM}}$ & $\hat{F}_1^{\text{PCA}}$ & $F_2^{\text{PCA}}$ & $F_3^{\text{PCA}}$ \\
					\cmidrule(lr){2-4} \cmidrule(lr){5-7} \cmidrule(lr){8-10}  \cmidrule(lr){11-13}
					\endfirsthead
					\toprule
					{(T,N)} & $F_{1}^{\text{RTS}}$  & $F_2^{\text{RTS}}$ & $F_3^{\text{RTS}}$ & $\hat{F}_1^{\text{QFM}}$ & $F_2^{\text{QFM}}$ & $F_3^{\text{QFM}}$ & $\hat{F}_1^{\text{CQFM}}$ & $F_2^{\text{CQFM}}$ & $F_3^{\text{CQFM}}$ & $\hat{F}_1^{\text{PCA}}$ & $F_2^{\text{PCA}}$ & $F_3^{\text{PCA}}$ \\
					\cmidrule(lr){2-4} \cmidrule(lr){5-7} \cmidrule(lr){8-10}  \cmidrule(lr){11-13}
					\endhead
					\endfoot
					\bottomrule
					\insertTableNotes
					\endlastfoot
					
					&   \multicolumn{12}{c}{$\varepsilon_{it} \sim \text{skewed normal}$}      \\    
					\cmidrule(lr){2-13}
					(50,100) & 0.9727 & 0.9527 & 0.9397 & 0.9632 & 0.9289 & 0.9168 & 0.9755 & 0.9579 & 0.9468 & 0.9727 & 0.9527 & 0.9397 \\
					(100,50) & 0.9508 & 0.9128 & 0.8909 & 0.9363 & 0.8842 & 0.8488 & 0.9545 & 0.9191 & 0.8987 & 0.9510 & 0.9122 & 0.8906 \\
					(100,200) & 0.9887 & 0.9791 & 0.9726 & 0.9846 & 0.9710 & 0.9615 & 0.9898 & 0.9813 & 0.9754 & 0.9887 & 0.9790 & 0.9725 \\
					(200,100) & 0.9789 & 0.9587 & 0.9479 & 0.9712 & 0.9431 & 0.9288 & 0.9808 & 0.9624 & 0.9526 & 0.9789 & 0.9586 & 0.9478 \\
					(300,300) & 0.9930 & 0.9870 & 0.9831 & 0.9901 & 0.9814 & 0.9761 & 0.9938 & 0.9883 & 0.9848 & 0.9930 & 0.9870 & 0.9831 \\
					&  \multicolumn{12}{c}{$\varepsilon_{it} \sim $ skewed $t$}      \\   
					\cmidrule(lr){2-13} 
					(50,100) & 0.9727 & 0.9521 & 0.9376 & 0.9738 & 0.9552 & 0.9388 & 0.9780 & 0.9627 & 0.9507 & 0.9728 & 0.9521 & 0.9369 \\
					(100,50) & 0.9517 & 0.9112 & 0.8912 & 0.9513 & 0.9104 & 0.8868 & 0.9597 & 0.9267 & 0.9104 & 0.9517 & 0.9104 & 0.8902 \\
					(100,200) & 0.9889 & 0.9790 & 0.9725 & 0.9890 & 0.9795 & 0.9736 & 0.9910 & 0.9832 & 0.9791 & 0.9889 & 0.9789 & 0.9723 \\
					(200,100) & 0.9789 & 0.9592 & 0.9476 & 0.9792 & 0.9600 & 0.9481 & 0.9829 & 0.9673 & 0.9578 & 0.9789 & 0.9591 & 0.9474 \\
					(300,300) & 0.9931 & 0.9870 & 0.9830 & 0.9933 & 0.9874 & 0.9835 & 0.9946 & 0.9899 & 0.9867 & 0.9931 & 0.9870 & 0.9829 \\
					&   \multicolumn{12}{c}{$\varepsilon_{it} \sim$ asymmetric Laplace}      \\  
					\cmidrule(lr){2-13}  
					 (50,100) & 0.7231 & 0.5082 & 0.3845 & 0.7540 & 0.5467 & 0.3553 & 0.8388 & 0.7039 & 0.5807 & 0.7155 & 0.4372 & 0.3184 \\
					 (100,50) & 0.6480 & 0.3788 & 0.2830 & 0.6732 & 0.3722 & 0.3023 & 0.7710 & 0.5818 & 0.4495 & 0.6370 & 0.3283 & 0.2338 \\
					 (100,200) & 0.8950 & 0.7946 & 0.7178 & 0.8961 & 0.6998 & 0.4305 & 0.9447 & 0.8925 & 0.8592 & 0.8893 & 0.7679 & 0.6646 \\
					 (200,100) & 0.8381 & 0.6849 & 0.5925 & 0.8629 & 0.6080 & 0.3442 & 0.9063 & 0.8201 & 0.7776 & 0.8337 & 0.6648 & 0.5502 \\
					 (300,300) & 0.9415 & 0.8916 & 0.8588 & 0.9474 & 0.7450 & 0.3809 & 0.9686 & 0.9415 & 0.9237 & 0.9409 & 0.8892 & 0.8538 \\
					&   \multicolumn{12}{c}{$\varepsilon_{it} \sim$ log-normal }      \\    
					\cmidrule(lr){2-13}
					(50,100) & 0.1306 & 0.0610 & 0.0693 & 0.8032 & 0.5195 & 0.3900 & 0.8513 & 0.6820 & 0.5406 & 0.0560 & 0.0409 & 0.0464 \\
					(100,50) & 0.1112 & 0.0465 & 0.0307 & 0.7475 & 0.4645 & 0.2602 & 0.7924 & 0.6177 & 0.4502 & 0.0263 & 0.0247 & 0.0191 \\
					(100,200) & 0.2022 & 0.0496 & 0.0437 & 0.9347 & 0.7346 & 0.2576 & 0.9580 & 0.9157 & 0.8825 & 0.0305 & 0.0223 & 0.0212 \\
					(200,100) & 0.2361 & 0.0447 & 0.0297 & 0.8943 & 0.6118 & 0.2963 & 0.9291 & 0.8581 & 0.8060 & 0.0164 & 0.0137 & 0.0112 \\
					(300,300) & 0.4625 & 0.0969 & 0.0385 & 0.9636 & 0.7342 & 0.2566 & 0.9762 & 0.9520 & 0.9360 & 0.0134 & 0.0079 & 0.0081 \\
					&   \multicolumn{12}{c}{$\varepsilon_{it} \sim $ mixture of skewed normal }      \\  
					\cmidrule(lr){2-13}    
					(50,100) & 0.9473 & 0.9091 & 0.8905 & 0.9579 & 0.9222 & 0.8970 & 0.9693 & 0.9452 & 0.9336 & 0.9458 & 0.9028 & 0.8738 \\
					(100,50) & 0.9096 & 0.8406 & 0.8051 & 0.9260 & 0.8605 & 0.8222 & 0.9430 & 0.8969 & 0.8750 & 0.9079 & 0.8338 & 0.7901 \\
					(100,200) & 0.9792 & 0.9609 & 0.9501 & 0.9820 & 0.9661 & 0.9554 & 0.9876 & 0.9767 & 0.9699 & 0.9792 & 0.9604 & 0.9489 \\
					(200,100) & 0.9620 & 0.9285 & 0.9066 & 0.9664 & 0.9360 & 0.9192 & 0.9766 & 0.9549 & 0.9421 & 0.9618 & 0.9278 & 0.9053 \\
					(300,300) & 0.9874 & 0.9764 & 0.9689 & 0.9887 & 0.9787 & 0.9722 & 0.9925 & 0.9858 & 0.9815 & 0.9874 & 0.9763 & 0.9688 \\
				\end{longtable} 
			\end{ThreePartTable}
		\end{landscape}
		
		\subsubsection{\Cref{tab:mse asym error heteroskedasticity}: MSE under asymmetric errors with heteroskedasticity}
		\begin{table}[htp] \centering
			\begin{center}
				\caption{MSE under asymmetric errors with heteroskedasticity} 
				\label{tab:mse asym error heteroskedasticity} 
				\begin{threeparttable}
					\begin{tabular}{lrrrrrrrr}    
						\cmidrule(lr){1-9}
						\multirow{2}{*}{(T,N)}  &   \multicolumn{4}{c}{$\varepsilon_{it} \sim$ skewed normal}  &   \multicolumn{4}{c}{$\varepsilon_{it} \sim$ skewed t}         \\  
						& RTS & QFM & CQFM & PCA & RTS & QFM & CQFM & PCA \\
						\cmidrule(lr){2-5} \cmidrule(lr){6-9}
						(50,100) & 0.534 & 0.711 & 0.480 & 0.506 & 0.533 & 0.502 & 0.426 & 0.507 \\
						(100,50) & 0.514 & 0.712 & 0.467 & 0.511 & 0.512 & 0.518 & 0.413 & 0.514 \\
						(100,200) & 0.246 & 0.349 & 0.220 & 0.240 & 0.246 & 0.240 & 0.193 & 0.241 \\
						(200,100) & 0.242 & 0.349 & 0.219 & 0.241 & 0.241 & 0.239 & 0.191 & 0.241 \\
						(300,300) & 0.105 & 0.159 & 0.094 & 0.104 & 0.105 & 0.104 & 0.082 & 0.104 \\
						&   \multicolumn{4}{c}{$\varepsilon_{it} \sim$ asymmetric Laplace}    &   \multicolumn{4}{c}{$\varepsilon_{it} \sim$ log-normal}    \\  
						\cmidrule(lr){2-5}  \cmidrule(lr){6-9}
						(50,100) & 6.904 & 6.542 & 3.892 & 7.508 & 131.320 & 16.924 & 10.504 & 196.980 \\
						(100,50) & 6.824 & 6.687 & 3.874 & 7.526 & 101.513 & 17.118 & 15.258 & 205.817 \\
						(100,200) & 2.580 & 4.798 & 1.302 & 2.849 & 71.079 & 16.400 & 1.013 & 147.965 \\
						(200,100) & 2.630 & 4.794 & 1.278 & 2.884 & 54.737 & 16.315 & 1.001 & 132.482 \\
						(300,300) & 0.970 & 4.156 & 0.505 & 1.006 & 29.903 & 16.143 & 0.405 & 92.768 \\
						&   \multicolumn{4}{c}{$\varepsilon_{it} \sim$ mixture of skewed normal}   &   \\   
						\cmidrule(lr){2-5} 
						(50,100) & 1.010 & 0.857 & 0.621 & 1.044 &       &       &       &  \\
						(100,50) & 0.987 & 0.853 & 0.595 & 1.058 &       &       &       &  \\
						(100,200) & 0.451 & 0.405 & 0.273 & 0.455 &       &       &       &  \\
						(200,100) & 0.444 & 0.405 & 0.269 & 0.454 &       &       &       &  \\
						(300,300) & 0.191 & 0.183 & 0.114 & 0.193 &       &       &       &  \\
						\hline
					\end{tabular}
					\begin{tablenotes}[flushleft]
						\item[] \textit{Notes}: Every number is the average MSE over $100$ replications for the RTS, QFM(0.5), CQFM, and PCA method, respectively. We choose $\tau=0.5$ for the QFM method and $K=5$ for the CQFM method. The DGP is described in \cref{eq:heter}.
					\end{tablenotes}
				\end{threeparttable}
			\end{center} 
		\end{table}
		
		\subsubsection{\Cref{tab:factor number asym error heterskedasticity}: Factor number estimation under asymmetric errors with heteroskedasticity}
			\begin{table}[htp] \centering
			\begin{center}
				\caption{Estimated factor number and frequency of correct estimation under heteroskedasticity} 
				\label{tab:factor number asym error heterskedasticity} 
				\begin{threeparttable}
					\renewcommand{\TPTminimum}{\linewidth}
					\makebox[\linewidth] { 
						\begin{tabular}{lrrrrrr}
							\toprule
							(T,N) & QFM      &   CQFM  &    PCA   &  QFM     &  CQFM  &   PCA \\
						 \cmidrule(lr){2-7}
							&   \multicolumn{3}{c}{avg. $\hat{r}$}  &   \multicolumn{3}{c}{$\text{Prob}(\hat{r} = 3)$} \\
							\cmidrule(lr){2-4}  \cmidrule(lr){5-7}
							&  \multicolumn{6}{c}{$\varepsilon_{it} \sim$ skewed normal}        \\
							\cmidrule(lr){2-7} 
							(50,100) & 2.64  & 3.07  & 2.99  & 0.69  & 0.96  & 0.99 \\
							(100,50) & 2.73  & 3.14  & 3     & 0.76  & 0.95  & 1 \\
							(100,200) & 2.95  & 3     & 3     & 0.96  & 1     & 1 \\
							(200,100) & 2.94  & 3     & 3     & 0.94  & 1     & 1 \\
							(300,300) & 3     & 3     & 3     & 1     & 1     & 1 \\
							&  \multicolumn{6}{c}{$\varepsilon_{it} \sim$ skewed t}        \\
							\cmidrule(lr){2-7}  
							(50,100) & 2.62  & 3.11  & 3.01  & 0.66  & 0.92  & 0.99 \\
							(100,50) & 2.73  & 3.07  & 3     & 0.76  & 0.97  & 1 \\
							(100,200) & 2.95  & 3     & 3     & 0.96  & 1     & 1 \\
							(200,100) & 2.94  & 3     & 3     & 0.94  & 1     & 1 \\
							(300,300) & 3     & 3     & 3     & 1     & 1     & 1 \\
							&  \multicolumn{6}{c}{$\varepsilon_{it} \sim$ asymmetric Laplace}        \\
							\cmidrule(lr){2-7} 
							(50,100) & 5.36  & 1.98  & 1     & 0.07  & 0.13  & 0 \\
							(100,50) & 5.6   & 1.84  & 1     & 0.02  & 0.13  & 0 \\
							(100,200) & 4.86  & 2.89  & 1.02  & 0.02  & 0.73  & 0 \\
							(200,100) & 4.97  & 2.81  & 1.03  & 0.02  & 0.78  & 0 \\
							(300,300) & 4.01  & 3     & 2.14  & 0     & 1     & 0.24 \\ 
							&  \multicolumn{6}{c}{$\varepsilon_{it} \sim$ log-normal}        \\
							\cmidrule(lr){2-7}  
							(50,100) & 1.35  & 1.64  & 3.91  & 0     & 0.07  & 0.16 \\
							(100,50) & 1.43  & 1.64  & 3.6   & 0     & 0.07  & 0.19 \\
							(100,200) & 1.21  & 2.12  & 3.36  & 0     & 0.3   & 0.17 \\
							(200,100) & 1.28  & 2.2   & 3.54  & 0     & 0.27  & 0.2 \\
							(300,300) & 1.51  & 3.09  & 3.45  & 0     & 0.91  & 0.16 \\
							&  \multicolumn{6}{c}{$\varepsilon_{it} \sim$ mixture of skewed normal}        \\
							\cmidrule(lr){2-7}  
							(50,100) & 2.65  & 3.22  & 2.82  & 0.7   & 0.81  & 0.82 \\
							(100,50) & 2.72  & 3.25  & 2.77  & 0.71  & 0.83  & 0.77 \\
							(100,200) & 2.95  & 3     & 3     & 0.96  & 1     & 1 \\
							(200,100) & 2.97  & 3.03  & 3     & 0.97  & 0.97  & 1 \\
							(300,300) & 3     & 3     & 3     & 1     & 1     & 1 \\
							\bottomrule
						\end{tabular} 
					}   
					\begin{tablenotes}[flushleft]
						\item[] \textit{Notes}: Same as \Cref{tab:factor number}. Results for CQFM are obtained based on the IC with \cref{eq:qnt 2}.
					\end{tablenotes}
				\end{threeparttable}
			\end{center} 
		\end{table}
		
		\begin{landscape}
			\subsubsection{\Cref{tab:adj R2 asym ar1 error with pca}: Adjusted $R^2$ under asymmetric and AR(1) errors }
			\begin{ThreePartTable}
				\begin{TableNotes}[flushleft]\footnotesize
					\item[] \textit{Notes}: Each number is the average of adjusted $R^2$ over $100$ replications of regressing one of the three true factors on the estimated factors based on the RTS, QFM(0.5), and CQFM method, respectively. We choose $\tau=0.5$ for the QFM method and $K=5$ for the CQFM method. The DGP is described in \cref{eq:ar1}.
				\end{TableNotes}
				\begin{longtable}{lrrrrrrrrrrrr} 
					\caption{Adj. $R^2$ of regressing  $3$ true factors on the estimated factors under asymmetric and AR(1) errors}\\
					\label{tab:adj R2 asym ar1 error with pca}\\
					\toprule
					\multirow{2}{*}{(T,N)} & $F_{1}^{\text{RTS}}$  & $F_2^{\text{RTS}}$ & $F_3^{\text{RTS}}$ & $\hat{F}_1^{\text{QFM}}$ & $F_2^{\text{QFM}}$ & $F_3^{\text{QFM}}$ & $\hat{F}_1^{\text{CQFM}}$ & $F_2^{\text{CQFM}}$ & $F_3^{\text{CQFM}}$ & $\hat{F}_1^{\text{PCA}}$ & $F_2^{\text{PCA}}$ & $F_3^{\text{PCA}}$ \\
					\cmidrule(lr){2-4} \cmidrule(lr){5-7} \cmidrule(lr){8-10}  \cmidrule(lr){11-13}
					\endfirsthead
					\toprule
					{(T,N)} & $F_{1}^{\text{RTS}}$  & $F_2^{\text{RTS}}$ & $F_3^{\text{RTS}}$ & $\hat{F}_1^{\text{QFM}}$ & $F_2^{\text{QFM}}$ & $F_3^{\text{QFM}}$ & $\hat{F}_1^{\text{CQFM}}$ & $F_2^{\text{CQFM}}$ & $F_3^{\text{CQFM}}$ & $\hat{F}_1^{\text{PCA}}$ & $F_2^{\text{PCA}}$ & $F_3^{\text{PCA}}$ \\
					\cmidrule(lr){2-4} \cmidrule(lr){5-7} \cmidrule(lr){8-10}  \cmidrule(lr){11-13}
					\endhead
					\endfoot
					\bottomrule
					\insertTableNotes
					\endlastfoot
					
					&   \multicolumn{12}{c}{$\varepsilon_{it} \sim \text{skewed normal}$}      \\    
					\cmidrule(lr){2-13}
					(50,100) & 0.9936 & 0.9891 & 0.9861 & 0.9899 & 0.9830 & 0.9776 & 0.9934 & 0.9886 & 0.9852 & 0.9937 & 0.9892 & 0.9863 \\
					(100,50)  & 0.9885 & 0.9773 & 0.9726 & 0.9816 & 0.9650 & 0.9575 & 0.9876 & 0.9759 & 0.9710 & 0.9885 & 0.9774 & 0.9726 \\
					(100,200)  & 0.9972 & 0.9949 & 0.9933 & 0.9954 & 0.9917 & 0.9892 & 0.9971 & 0.9948 & 0.9931 & 0.9972 & 0.9950 & 0.9933 \\
					(200,100)  & 0.9948 & 0.9896 & 0.9865 & 0.9914 & 0.9833 & 0.9783 & 0.9945 & 0.9891 & 0.9859 & 0.9948 & 0.9896 & 0.9865 \\
					(300,300)  & 0.9982 & 0.9966 & 0.9956 & 0.9970 & 0.9943 & 0.9927 & 0.9981 & 0.9965 & 0.9954 & 0.9982 & 0.9966 & 0.9956 \\
					&  \multicolumn{12}{c}{$\varepsilon_{it} \sim $ skewed $t$}      \\   
					\cmidrule(lr){2-13} 
					(50,100)  & 0.9938 & 0.9886 & 0.9857 & 0.9916 & 0.9851 & 0.9814 & 0.9939 & 0.9888 & 0.9864 & 0.9938 & 0.9887 & 0.9858 \\
					(100,50)   & 0.9880 & 0.9780 & 0.9726 & 0.9841 & 0.9702 & 0.9628 & 0.9878 & 0.9776 & 0.9724 & 0.9880 & 0.9781 & 0.9726 \\
					(100,200)  & 0.9971 & 0.9950 & 0.9931 & 0.9962 & 0.9931 & 0.9908 & 0.9973 & 0.9951 & 0.9933 & 0.9972 & 0.9950 & 0.9931 \\
					(200,100)  & 0.9947 & 0.9894 & 0.9865 & 0.9927 & 0.9858 & 0.9816 & 0.9948 & 0.9896 & 0.9866 & 0.9947 & 0.9894 & 0.9865 \\
					(300,300)  & 0.9982 & 0.9966 & 0.9956 & 0.9976 & 0.9954 & 0.9941 & 0.9983 & 0.9968 & 0.9957 & 0.9982 & 0.9966 & 0.9956 \\
					&   \multicolumn{12}{c}{$\varepsilon_{it} \sim$ asymmetric Laplace}      \\  
					\cmidrule(lr){2-13}  
					(50,100)  & 0.9340 & 0.8576 & 0.7759 & 0.9150 & 0.8158 & 0.7174 & 0.9497 & 0.8967 & 0.8489 & 0.9357 & 0.8618 & 0.7769 \\
					(100,50)  & 0.9047 & 0.8094 & 0.7557 & 0.8785 & 0.7469 & 0.6797 & 0.9223 & 0.8487 & 0.8081 & 0.9042 & 0.8058 & 0.7503 \\
					(100,200)   & 0.9749 & 0.9500 & 0.9257 & 0.9670 & 0.9327 & 0.8887 & 0.9810 & 0.9624 & 0.9478 & 0.9752 & 0.9505 & 0.9267 \\
					(200,100)   & 0.9569 & 0.9154 & 0.8875 & 0.9426 & 0.8869 & 0.8406 & 0.9662 & 0.9336 & 0.9139 & 0.9569 & 0.9152 & 0.8870 \\
					(300,300)   & 0.9854 & 0.9719 & 0.9626 & 0.9795 & 0.9591 & 0.9387 & 0.9889 & 0.9790 & 0.9720 & 0.9854 & 0.9719 & 0.9626 \\
					&   \multicolumn{12}{c}{$\varepsilon_{it} \sim$ log-normal }      \\    
					\cmidrule(lr){2-13}
					(50,100)   & 0.4852 & 0.2127 & 0.0983 & 0.8541 & 0.6070 & 0.4188 & 0.8798 & 0.6945 & 0.4976 & 0.3534 & 0.1521 & 0.0813 \\
					(100,50)   & 0.4321 & 0.1520 & 0.0877 & 0.8060 & 0.5435 & 0.3568 & 0.8752 & 0.7186 & 0.6046 & 0.2429 & 0.0880 & 0.0444 \\
					(100,200)    & 0.6285 & 0.2365 & 0.0705 & 0.9441 & 0.7487 & 0.3854 & 0.9767 & 0.9481 & 0.9243 & 0.3646 & 0.0980 & 0.0423 \\
					(200,100)    & 0.6333 & 0.2390 & 0.0867 & 0.9230 & 0.7252 & 0.3104 & 0.9606 & 0.9217 & 0.8941 & 0.2885 & 0.0525 & 0.0279 \\
					(300,300)    & 0.8357 & 0.5227 & 0.1730 & 0.9724 & 0.7799 & 0.3441 & 0.9881 & 0.9769 & 0.9693 & 0.4888 & 0.0831 & 0.0216 \\
					&   \multicolumn{12}{c}{$\varepsilon_{it} \sim $ mixture of skewed normal }      \\  
					\cmidrule(lr){2-13}    
					(50,100)   & 0.9878 & 0.9791 & 0.9732 & 0.9869 & 0.9774 & 0.9707 & 0.9902 & 0.9841 & 0.9789 & 0.9881 & 0.9793 & 0.9736 \\
					(100,50)    & 0.9782 & 0.9597 & 0.9492 & 0.9759 & 0.9546 & 0.9448 & 0.9819 & 0.9661 & 0.9585 & 0.9783 & 0.9597 & 0.9490 \\
					(100,200)    & 0.9949 & 0.9902 & 0.9876 & 0.9943 & 0.9890 & 0.9853 & 0.9959 & 0.9922 & 0.9901 & 0.9949 & 0.9902 & 0.9877 \\
					(200,100)    & 0.9905 & 0.9806 & 0.9755 & 0.9889 & 0.9777 & 0.9719 & 0.9923 & 0.9843 & 0.9802 & 0.9906 & 0.9806 & 0.9755 \\
					(300,300)    & 0.9968 & 0.9939 & 0.9923 & 0.9962 & 0.9928 & 0.9908 & 0.9974 & 0.9952 & 0.9938 & 0.9968 & 0.9939 & 0.9923 \\
				\end{longtable} 
			\end{ThreePartTable}
		\end{landscape}
		
		\subsubsection{\Cref{tab:mse asym ar1 error}: MSE under asymmetric and AR(1) errors}
		\begin{table}[htp] \centering
			\begin{center}
				\caption{MSE under asymmetric and AR(1) errors} 
				\label{tab:mse asym ar1 error} 
				\begin{threeparttable}
					\begin{tabular}{lrrrrrrrr}    
						\cmidrule(lr){1-9}
						\multirow{2}{*}{(T,N)}  &   \multicolumn{4}{c}{$\varepsilon_{it} \sim$ skewed normal}  &   \multicolumn{4}{c}{$\varepsilon_{it} \sim$ skewed t}         \\  
						& RTS & QFM & CQFM & PCA & RTS & QFM & CQFM & PCA \\
						\cmidrule(lr){2-5} \cmidrule(lr){6-9}
						(50,100) & 0.186 & 0.245 & 0.184 & 0.173 & 0.187 & 0.213 & 0.175 & 0.176 \\
						(100,50) & 0.152 & 0.219 & 0.156 & 0.147 & 0.151 & 0.188 & 0.147 & 0.148 \\
						(100,200) & 0.091 & 0.128 & 0.091 & 0.087 & 0.092 & 0.108 & 0.087 & 0.087 \\
						(200,100) & 0.076 & 0.114 & 0.077 & 0.074 & 0.077 & 0.096 & 0.073 & 0.075 \\
						(300,300) & 0.038 & 0.056 & 0.039 & 0.036 & 0.038 & 0.046 & 0.037 & 0.036 \\
						&   \multicolumn{4}{c}{$\varepsilon_{it} \sim$ asymmetric Laplace}    &   \multicolumn{4}{c}{$\varepsilon_{it} \sim$ log-normal}    \\  
						\cmidrule(lr){2-5}  \cmidrule(lr){6-9}
						 (50,100) & 1.981 & 2.324 & 1.466 & 1.947 & 42.800 & 10.486 & 12.697 & 49.761 \\
						(100,50) & 1.519 & 1.982 & 1.156 & 1.551 & 37.263 & 7.648 & 5.144 & 48.708 \\
						(100,200) & 0.824 & 1.175 & 0.618 & 0.811 & 28.682 & 7.136 & 0.898 & 36.477 \\
						(200,100) & 0.672 & 0.988 & 0.510 & 0.670 & 23.374 & 6.964 & 0.662 & 37.015 \\
						(300,300) & 0.312 & 0.510 & 0.235 & 0.309 & 13.940 & 6.830 & 0.280 & 24.307 \\
						&   \multicolumn{4}{c}{$\varepsilon_{it} \sim$ mixture of skewed normal}   &   \\   
						\cmidrule(lr){2-5} 
						(50,100) & 0.344 & 0.339 & 0.275 & 0.326 &       &       &       &  \\
						(100,50) & 0.280 & 0.296 & 0.226 & 0.276 &       &       &       &  \\
						(100,200) & 0.167 & 0.172 & 0.131 & 0.161 &       &       &       &  \\
						(200,100) & 0.138 & 0.151 & 0.110 & 0.135 &       &       &       &  \\
						(300,300) & 0.068 & 0.073 & 0.054 & 0.066 &       &       &       &  \\
						\hline
					\end{tabular}
					\begin{tablenotes}[flushleft]
						\item[] \textit{Notes}: Every number is the average MSE over $100$ replications for the RTS, QFM(0.5), CQFM, and PCA method, respectively. We choose $\tau=0.5$ for the QFM method and $K=5$ for the CQFM method. The DGP is described in \cref{eq:ar1}.
					\end{tablenotes}
				\end{threeparttable}
			\end{center} 
		\end{table}
		
		\subsubsection{\Cref{tab:factor number asym ar1 error}: Factor number estimation under asymmetric AR(1) errors }
		\begin{table}[htp] \centering
			\begin{center}
				\caption{Estimated factor number and frequency of correct estimation under asymmetric AR(1) errors} 
				\label{tab:factor number asym ar1 error} 
				\begin{threeparttable}
					\renewcommand{\TPTminimum}{\linewidth}
					\makebox[\linewidth] { 
						\begin{tabular}{lrrrrrr}
							\toprule
							(T,N) & QFM      &   CQFM   &    PCA   &  QFM     & CQFM     &   PCA \\
							\hline
							&   \multicolumn{3}{c}{avg. $\hat{r}$}  &   \multicolumn{3}{c}{$\text{Prob}(\hat{r} = 3)$} \\
							\cmidrule(lr){2-4}  \cmidrule(lr){5-7}
							&  \multicolumn{6}{c}{$\varepsilon_{it} \sim$ skewed normal}        \\
							\cmidrule(lr){2-7} 
							(50,100) & 2.54  & 3     & 3.24  & 0.61  & 1     & 0.79 \\
							(100,50) & 2.61  & 3     & 3     & 0.66  & 1     & 1 \\
							(100,200) & 2.94  & 3     & 3     & 0.95  & 1     & 1 \\
							(200,100) & 2.92  & 3     & 3     & 0.92  & 1     & 1 \\
							(300,300) & 3     & 3     & 3     & 1     & 1     & 1 \\
							&  \multicolumn{6}{c}{$\varepsilon_{it} \sim$ skewed t}        \\
							\cmidrule(lr){2-7}
							(50,100) & 2.5   & 3     & 3.27  & 0.58  & 1     & 0.8 \\
							(100,50) & 2.6   & 3     & 3     & 0.63  & 1     & 1 \\
							(100,200) & 2.94  & 3     & 3.01  & 0.95  & 1     & 0.99 \\
							(200,100) & 2.93  & 3     & 3     & 0.93  & 1     & 1 \\
							(300,300) & 3     & 3     & 3     & 1     & 1     & 1 \\  
							&  \multicolumn{6}{c}{$\varepsilon_{it} \sim$ asymmetric Laplace}        \\
							\cmidrule(lr){2-7}  
							(50,100) & 3.21  & 1.23  & 3.13  & 0.44  & 0     & 0.68 \\
							(100,50) & 3.1   & 1.13  & 2.81  & 0.57  & 0.01  & 0.81 \\
							(100,200) & 3.34  & 2.43  & 3     & 0.62  & 0.48  & 1 \\
							(200,100) & 3.24  & 2.38  & 3     & 0.72  & 0.41  & 1 \\
							(300,300) & 3.87  & 3     & 3     & 0.13  & 1     & 1 \\
							&  \multicolumn{4}{c}{$\varepsilon_{it} \sim$ log-normal}        \\
							\cmidrule(lr){2-7}  
							(50,100) & 3.9   & 5.96  & 5.5   & 0.21  & 0     & 0.04 \\
							(100,50) & 3.68  & 5.45  & 4.4   & 0.32  & 0.04  & 0.09 \\
							(100,200) & 3.37  & 5.49  & 4.87  & 0.51  & 0.01  & 0.13 \\
							(200,100) & 3.35  & 4.91  & 4.03  & 0.53  & 0.06  & 0.17 \\
							(300,300) & 3.96  & 3.77  & 4.6   & 0.04  & 0.41  & 0.12 \\
							&  \multicolumn{6}{c}{$\varepsilon_{it} \sim$ mixture of skewed normal}        \\
							\cmidrule(lr){2-7}  
							(50,100) & 2.58  & 3     & 3.34  & 0.65  & 1     & 0.72 \\
							(100,50) & 2.62  & 3     & 3     & 0.68  & 1     & 1 \\
							(100,200) & 2.94  & 3     & 3     & 0.95  & 1     & 1 \\
							(200,100) & 2.94  & 3     & 3     & 0.94  & 1     & 1 \\
							(300,300) & 3     & 3     & 3     & 1     & 1     & 1 \\
							\bottomrule
						\end{tabular} 
					}   
					\begin{tablenotes}[flushleft]
						\item[] \textit{Notes}: Same as \Cref{tab:factor number}. Results for CQFM under log-normal error are obtained using the IC with \cref{eq:qnt 2}; all other CQFM results are obtained using the IC with \cref{eq:qnt}.
					\end{tablenotes}
				\end{threeparttable}
			\end{center} 
		\end{table}

\subsection{Additional figures}
\begin{figure}[h]
        	\centering
        	\includegraphics[scale=0.8,keepaspectratio=TRUE]{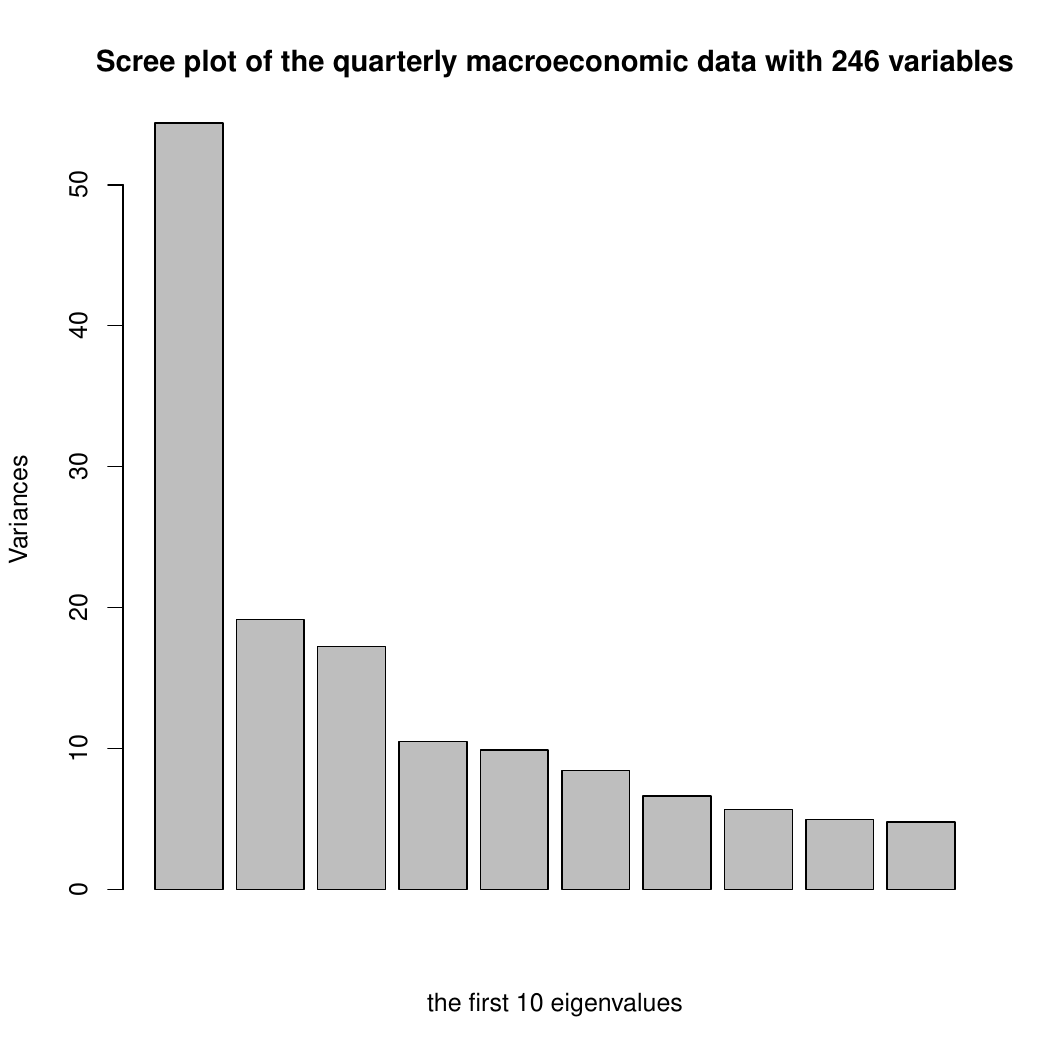}%
        	\caption{Scree plot of the first ten eigenvalues of the standardized data matrix $Y Y'/T$, where $Y$ is a $255 \times 246$ matrix.}%
        	\label{fig:screeplot}%
        \end{figure}

        \subsubsection{\Cref{fig:screeplot}: Scree plot of the macroeconomic data set}
        
\begin{figure}[ht]
			\centering
			\includegraphics[width=1.0\textwidth,keepaspectratio=TRUE]{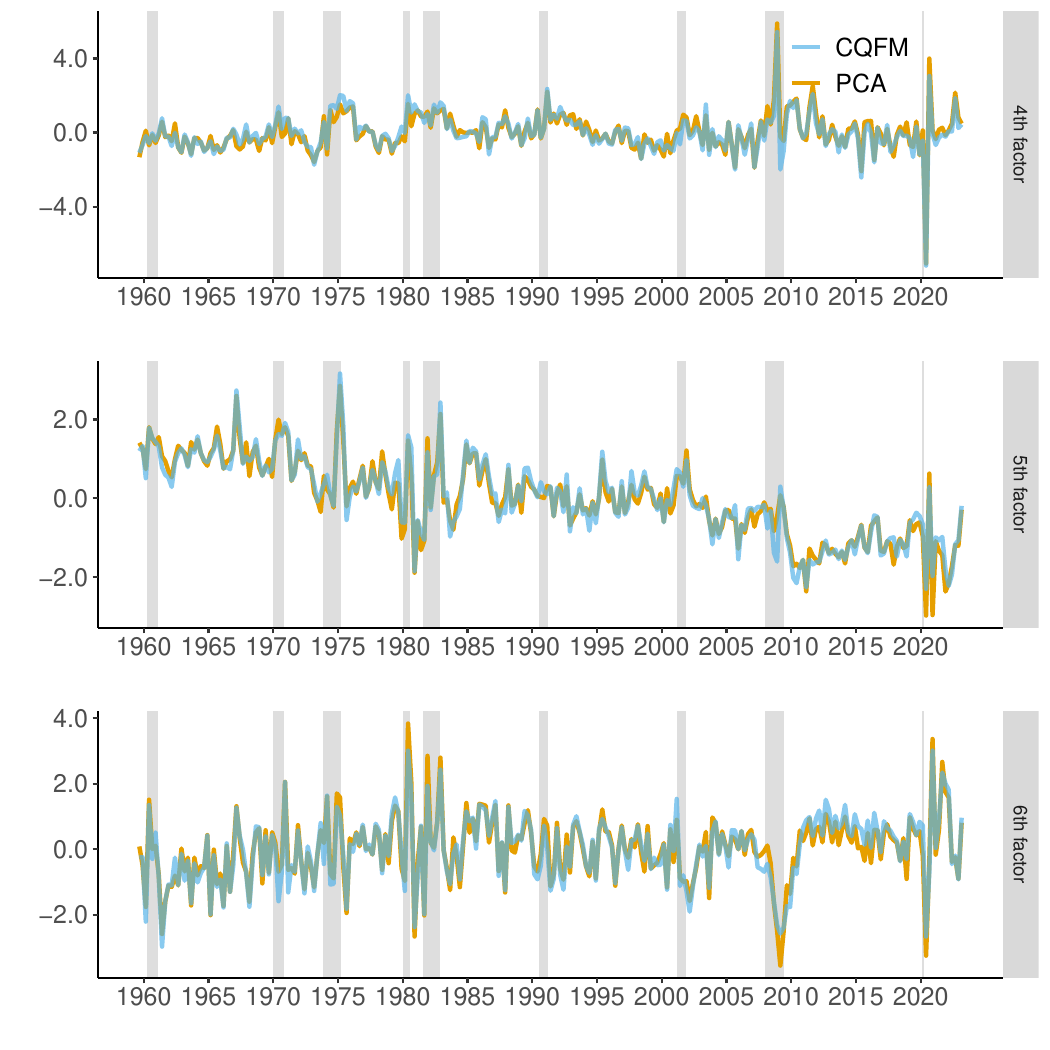}%
			\caption{The 4th, 5th, and 6th CQFM and PCA factors from 1959/3/1 to 2023/3/1}%
			\label{fig:factors4-6}%
		\end{figure}

		\subsubsection{\Cref{fig:factors4-6}: Time series plot of the 3rd, 4th, and 5th CQFM and PCA factors}
		
	\end{appendices}

\end{document}